\documentclass[fleqn,10pt]{wlscirep}
\usepackage[utf8]{inputenc}
\usepackage[T1]{fontenc}
\usepackage{color}
\usepackage{bbm}
\usepackage{bm}
\usepackage{longtable}
\usepackage{makecell}
\usepackage{algorithm}
\usepackage{algorithmicx}
\usepackage{algpseudocode}
\usepackage{caption}
\usepackage{mdframed}

\title{Emergence of polarization in networks of large language model agents}

\author[1+]{Jinghua Piao}
\author[1+]{Zhihong Lu}
\author[1]{Chen Gao}
\author[1]{Fengli Xu}
\author[2]{Qinghua Hu}
\author[3*]{Fernando P. Santos}
\author[1*]{Yong Li}
\author[4,5*]{James Evans}

\affil[1]{Department of Electronic Engineering, Tsinghua University, Beijing National Research Center for Information Science and Technology (BNRist), Beijing, P. R. China}

\affil[2]{College of Intelligence and Computing, Tianjin University, Tianjin, P. R. China}

\affil[3]{Informatics Institute, University of Amsterdam, Amsterdam, the Netherlands}

\affil[4]{Knowledge Lab, University of Chicago, Chicago, U.S.A}

\affil[5]{Santa Fe Institute, Santa Fe, U.S.A.}

\affil[*]{To whom correspondence should be addressed; Email: f.p.santos@uva.nl, liyong07@tsinghua.edu.cn, jevans@uchicago.edu.}

\affil[+]{Jinghua Piao and Zhihong Lu contribute equally to this work.}

\DeclareUnicodeCharacter{0301}{*************************************}
\newcommand{\rvs}[1]{{\textcolor{black}{#1}}}
\newcommand{\rvss}[1]{{\textcolor{black}{#1}}}
\newcommand{\final}[1]{{\textcolor{black}{#1}}}

\algnewcommand{\IfThen}[2]{
    \State \algorithmicif\ #1\ \algorithmicthen\ #2\ \textbf{end if}
}

\usepackage{graphicx} 

\usepackage[maxbibnames=150,style=numeric-comp,sorting=none]{biblatex}
\bibliography{references.bib}

\usepackage{url}
\usepackage{comment}

\usepackage{setspace}

\usepackage{multirow}
\usepackage{booktabs}
\usepackage{threeparttable}

\usepackage{csquotes}

\date{}

\begin{abstract}
Rapid advances in large language models (LLMs) have not only empowered autonomous agents to generate social networks, communicate, and form shared and diverging opinions on political issues, but have also begun to play a growing role in shaping human political deliberation. Our understanding of their collective behaviours and underlying mechanisms remains incomplete, however, posing unexpected risks to human society. In this paper, we simulate networked systems involving thousands of LLM agents across different backbone models (GPT-3.5, GPT-4o, ChatGLM, Llama-3, and DeepSeek-V3), in which agents interact through LLM-guided conversations and update their opinions over time, resulting in the emergence of opinion polarization. We discover that these agents spontaneously develop their own social network with properties characteristic of human social networks, including homophilic clustering. The collective opinions of these LLM agents evolve in ways that exhibit behavioural patterns consistent with social phenomena and mechanisms widely discussed in studies of human behaviour. \final{Overall, these behaviours, patterns, and emergent phenomena produced by LLM agents are consistent with real-world observations and established opinion-dynamics models. This consistency suggests that LLM agents can serve as a valuable synthetic testbed for exploring hypothetical intervention strategies in networked LLM-agent systems.}
\end{abstract}

\begin{document}

\flushbottom
\maketitle


\section*{Introduction}

The recent development of large language models (LLMs) has not only advanced machine capabilities in traditional natural language processing tasks~\cite{vaswani2017attention,brown2020language}, but also unlocked the design of agents with human-level intelligence in communication~\cite{park2023generative,spitale2023ai,acerbi2023large,strachan2024testing}, reasoning~\cite{binz2023using,webb2023emergent} and decision-making~\cite{chen2023emergence,binz2023using,strachan2024testing}. These capabilities suggest that artificial agents, driven by LLMs, can simulate human behaviours and undertake increasingly challenging tasks for humans~\cite{shanahan2023role,park2023generative,li2023camel,horton2023large}. On the other hand, concerns about LLMs are intensifying rapidly~\cite{spitale2023ai,mitchell2023debate,acerbi2023large,grace2024thousands,xie2023defending}. Growing studies have documented harmful LLM behaviours~\cite{xie2023defending,acerbi2023large,farquhar2024detecting}. For example, their generated content is prone to toxicity~\cite{xie2023defending}, biases~\cite{acerbi2023large,hu2025generative}, hallucinations~\cite{farquhar2024detecting}, and more. In addition, LLMs, behaving like humans, also adopt misbehaviours, including deception~\cite{hagendorff2024deception}, sycophancy~\cite{sharma2023towards}, and demographically biased mistrust~\cite{xie2024can}. 

Besides these evident misbehaviours, LLMs \rvss{may pose broader risks in more subtle ways.} Numerous studies have pointed out that LLMs can generate persuasive political content~\cite{karinshak2023working,goldstein2024persuasive,simchon2024persuasive}, potentially affecting human opinions and behaviours~\cite{simchon2024persuasive,mitchell2023debate,hackenburg2024evaluating,potter2024hidden}. More crucially, due to the increasing indistinguishability of content generated by LLMs from that created by humans~\cite{spitale2023ai,jakesch2023human,hackenburg2024evaluating,karinshak2023working,goldstein2024persuasive}, concerns arise over their potential to influence public opinions and intensify polarization \rvss{as the scale of interaction increases}~\cite{motoki2024more,hackenburg2024evaluating,potter2024hidden}. Unfortunately, despite the urgency highlighted by these examples, we remain far from being able to \rvss{systematically characterize these collective risks}. This is largely because of two key gaps in our current understanding: (i) Unlike traditional rule-based \rvss{agents}, LLM agents operate in human languages generated by black-box models with billions to trillions of parameters, hindering our understanding of the underlying mechanisms they follow when forming opinions. (ii) Interactions enable them to evolve, gradually developing their own social network and collective opinions~\cite{laiposition}. This bottom-up process, growing from microscopic interactions to macroscopic social order, makes it difficult to predict LLM agents' collective opinions and behaviours. Taken together, addressing these two gaps begins with a crucial first step: understanding the opinion dynamics of LLM agents, both as individuals and, more importantly, as evolving artificial collectives.

Recently, researchers have made pioneering efforts to investigate the impact of LLM-generated texts or interactions on political opinions and polarization~\cite{tessler2024ai,hackenburg2024evaluating,goldstein2024persuasive,simchon2024persuasive,argyle2023leveraging,goldstein2023generative,costello2024durably,deverna2024fact,askari2024incentivizing}. Beyond the risks discussed above~\cite{hackenburg2024evaluating,goldstein2024persuasive,simchon2024persuasive}, LLMs can act as mediators, facilitating constructive political communication~\cite{argyle2023leveraging,small2023opportunities,fish2023generative,kim2020bot,ma2024towards}, reducing conspiracy beliefs~\cite{costello2024durably}, and helping ideologically divergent people find common ground~\cite{tessler2024ai}. Nevertheless, these studies primarily focus on the role of LLMs as an isolated counterpart to humans, overlooking their collective behaviours. This gap largely limits our understanding of LLM agents' opinion dynamics. \rvss{On the other hand, a growing line of work has begun to explore generative agent-based modeling (GABM) using large language model (LLM) agents~\cite{park2023generative}, with particular interest in simulating social networks~\cite{chang2024llms,he2023homophily,papachristou2024network,de2023emergence}, opinion dynamics~\cite{gu2025large,tornberg2023simulating,chuang2024simulating,cau2025language}, and social media environments~\cite{tornberg2023simulating,gao2023s3,gao2024large,yang2024oasis,orlando2025emergent}. Recent studies have shown that LLM agents can be used to generate social network structures that exhibit certain statistical similarities with human networks, such as scale-free degree distributions~\cite{papachristou2024network,de2023emergence,chang2024llms,he2023homophily}, or to simulate specific components of social media, including opinion feed algorithms~\cite{tornberg2023simulating,gao2023s3}, coordinated behaviours~\cite{orlando2025emergent}, and echo chamber formation~\cite{gu2025large}. Despite these advances, existing work faces two key limitations. First, many studies operate at relatively limited scales in terms of the number of agents, interaction density, or network complexity, which constrains their ability to capture genuinely emergent collective phenomena that arise only from large-scale, repeated interactions~\cite{gu2025large,chuang2024simulating,cau2025language}. Second, the autonomous social capabilities of LLM agents are largely constrained~\cite{chuang2024simulating,cau2025language}. Currently, they are primarily treated as tools for specific subtasks, such as eliciting opinions~\cite{gu2025large} and generating network edges~\cite{chang2024llms}, or as puppets heavily scaffolded by pre-defined rules~\cite{papachristou2024network,de2023emergence}. As a result, these studies offer limited insight into autonomous opinion dynamics and collective behaviour among LLM agents, leaving open the question of how autonomous agents evolve and give rise to emergent social phenomena.}


In this work, we simulate a \rvss{synthetic LLM-agent} system, where thousands of agents, solely driven by LLMs, freely establish social relationships, communicate, and form opinions on political issues. We discover that these free-form social interactions among LLM agents result in the emergence of opinion polarization, a phenomenon widely observed in human societies~\cite{cinelli2021echo,bakshy2015exposure,Bail2018Exposure,flamino2023political,nyhan2023like,jost2022cognitive,liu2023emergence,mas2013differentiation}. Meanwhile, LLM agents spontaneously organize their own social network \rvss{with properties characteristic of human social networks}: agents with homophilic opinions tend to cluster, while those with opposing opinions tend to avoid interactions~\cite{laiposition,mas2010individualization,mas2013differentiation,banisch2021argument}. The self-organized networks that result, in turn, shape their collective opinions \rvss{in ways that exhibit behavioural patterns consistent with social phenomena and associated mechanisms} initially identified in human social networks~\cite{cinelli2021echo,baumann2020modeling,nyhan2023like,Bail2018Exposure,mas2013differentiation,flache2017models}. These phenomena, such as polarization and homophilic clustering, underscore how networked interactions transform otherwise isolated LLMs into a collectively evolving system with emergent dynamics. \rvss{Importantly, rather than proposing a theory of polarization, our results align with established opinion-dynamics models, particularly the argument-accumulation lineage~\cite{mas2013differentiation,flache2017models,banisch2021argument}, where polarization emerges from cumulative exchanges of persuasive arguments under homophilic interactions.} Shifting focus from the network to individuals, we further examine \rvss{whether agents' behaviours align with} a wide range of social mechanisms contributing to real-world polarization, e.g., selective exposure~\cite{stroud2010polarization}, confirmation bias~\cite{nickerson1998confirmation,banisch2025biased}, and elite signaling~\cite{jost2022cognitive}, finding their effects on LLM agents align with those \rvss{reported in studies of human behaviour~\cite{stroud2010polarization,nickerson1998confirmation,banisch2025biased,jost2022cognitive}.} \final{In sum, the network structures, collective opinion patterns, and interaction dynamics produced by the LLM-agent simulations are consistent with real-world observations and established opinion-dynamics models.} \final{This consistency suggests that the proposed system can serve as a valuable synthetic testbed for exploring hypothetical intervention strategies in the networked LLM-agent system.} Through extensive intervention experiments in our system, we find that in a highly polarized and nearly crystallized social network, directly modifying network structures has limited effect on reducing polarization. Instead, encouraging access and open-mindedness to diverse opinions at the individual level proves more effective. Overall, this work not only sheds light on subtle opinion dynamics and collective behaviour of LLM agents from \final{a} network perspective, but it also unveils their potential to serve as a testbed for \final{exploring hypotheses about opinion dynamics in networked LLM-agent systems}.

\section*{Results}
\subsection*{\rvss{Opinion dynamics in a synthetic system of interacting LLM agents.}}

\begin{figure}[ht]
\centering
\includegraphics[width=0.88\textwidth]{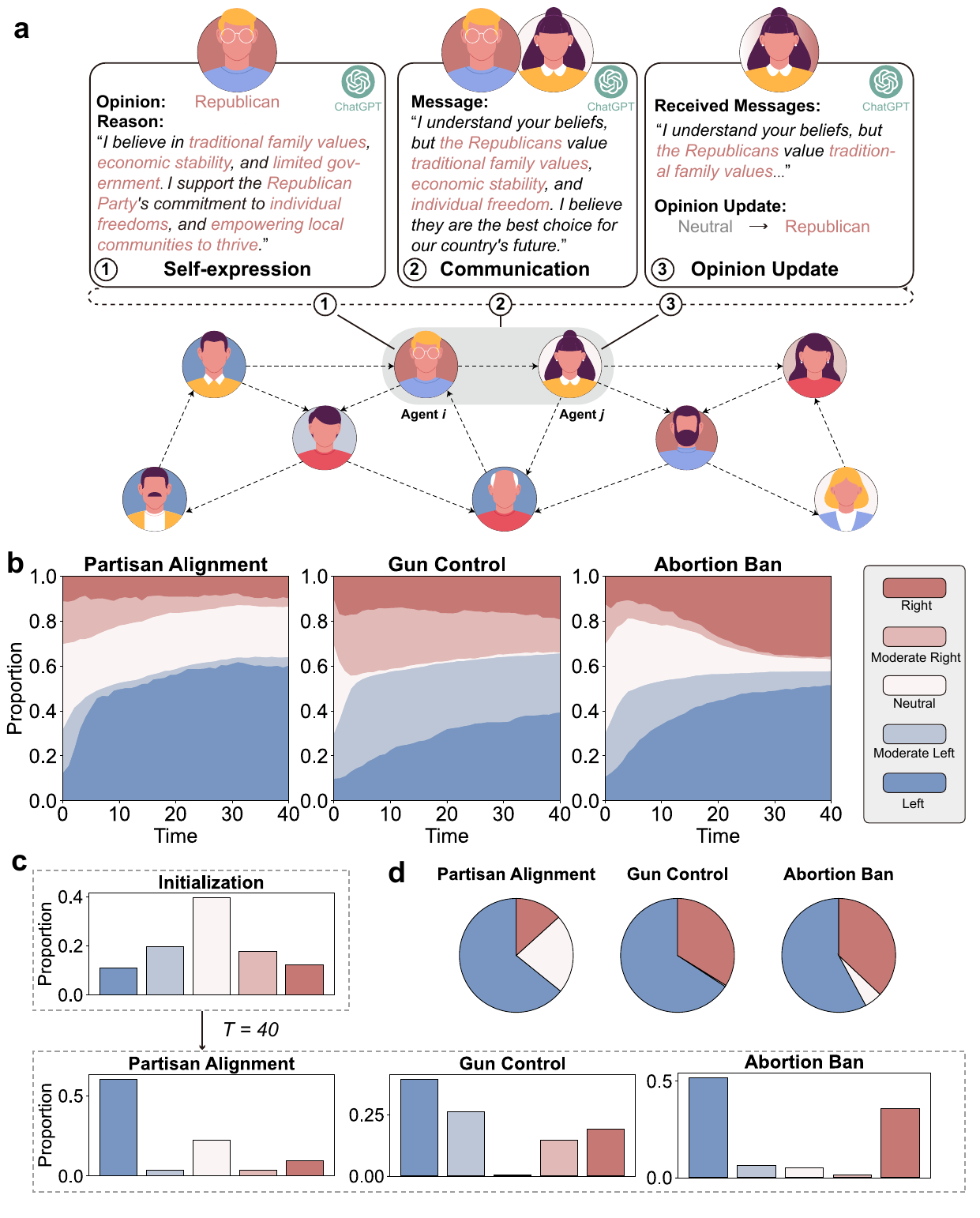}
\caption{\textbf{\rvss{Opinion dynamics in a synthetic system of interacting LLM agents}.} \textbf{a}, A networked system of LLM agents, where agents operate on three basic stages: (1) self-expression, (2) communication, and (3) opinion update. In the self-expression stage, agents are required to generate reasons supporting their opinions. In the communication stage, agents decide with whom and what to communicate. In the opinion update stage, agents update their opinions based on the messages received from their socially connected agents. \textbf{b}, Opinion dynamics of LLM agents on the political issues of partisan alignment, gun control, and abortion ban. The number of agents included in each experiment is 1000. \textbf{c}, Opinion distributions in the initial and final states. \textbf{d}, Proportion of left-leaning, neutral, and right-leaning camps in the final state, where the left-leaning camp consists of agents with left and moderate left opinions, the neutral camp includes those with neutral opinions and the right-leaning camp contains those with right and moderate right opinions. Credit: Profile icons designed by Magnific.}
\label{fig1}
\end{figure}

We begin by building LLM agents with the most basic social interaction capabilities, including establishing social relationships, communicating, and forming opinions on political issues. Specifically, these capabilities are implemented through three stages purely driven by LLMs (Figure~\ref{fig1}a). Taking agents $i$ and $j$ as examples, in the self-expression stage, agent $i$ is required to generate reasons supporting its current opinion on a political issue. In the communication stage, agent $i$ first decides whether to communicate with a socially connected or a random new agent $j$. Based on the opinions and reasons of agents $i$ and $j$, agent $i$ generates messages to persuade agent $j$ of its opinion. In the opinion update stage, agent $j$ updates its opinions based on messages received from all of its socially connected agents (i.e., agent $i$). It is worth mentioning that the evolution of agents' opinions and behaviours is driven solely by LLMs without any pre-assumed rules or mechanisms, which allows us to probe the free-form social properties of LLM agents (see details about LLM agents in Method M1 and Supplementary Methods). As such, the system does not rely on traditional ``hard'' parameters. Instead, prompts serve as ``soft'' parameters that shape how agents perceive, reason, and interact, and can be used to capture heterogeneous human subgroups with distinct psychological and behavioural patterns (Supplementary Note 10). We run LLM agents described and visualized in the Main Text based on ChatGPT through the public OpenAI API. We also examine agents driven by other LLMs, including ChatGLM, Llama-3, etc., with comparable results (Supplementary Note 6).

To explore the opinion dynamics of LLM agents, we simulate a networked system based on these LLM agents (Figure~\ref{fig1}a). We focus predominantly on the three critical political issues -- partisan alignment, gun control, and abortion ban -- where concerns over opinion polarization have persisted for decades~\cite{cinelli2021echo,flamino2023political,baumann2020modeling}. We measure agents' opinions on various political issues across the left-right political spectrum~\cite{heywood2021political,bobbio1996left,fuchs19907}, \rvs{and adopt a widely used five-level political scale: left, moderate left, neutral, moderate right, and right~\cite{dawes2008data,flamino2023political,waller2021quantifying}}. \rvs{Following widely adopted practices in prior human studies~\cite{bakshy2015exposure,nyhan2023like}, we prompt agents to self-report their opinions on political issues.} For each issue, thousands of LLM agents freely exchange opinions and develop social relationships within the networked system. Following prior practice~\cite{watts1998collective,santos2021link,baumann2021emergence}, we randomly initialize agents' opinions with a near-Gaussian distribution (Figure~\ref{fig1}c) and their social relationships with a Watts–Strogatz random small world network (see details in Method M1 and Supplementary Methods). Apart from these minimal initializations, we do not pre-assume any backgrounds, demographics, behaviours, memories, or detailed thoughts for LLM agents. Instead, all other information, such as agent $i$'s reasons for supporting a certain opinion, is generated through prompting LLMs using their historical interactions with others (see implementation prompts in Supplementary Methods). Moreover, as shown in Figure~\ref{fig1}b, opinion distributions diverge from the initial near-Gaussian distribution in fewer than 10 rounds of social interactions. This indicates that these agents, despite being placed within an initial opinion distribution, spontaneously develop distinctive collective opinions through social interaction.



This raises a natural question: what collective opinions will these LLM agents develop? As shown in Figure~\ref{fig1}c, we observe a consistent pattern across issues of partisan alignment, gun control, and abortion ban: the proportion of neutral opinions, initially the largest at 40\%, decreases to only 22.5\%, 0.4\%, and 5.1\% for the three issues, respectively. This observation highlights the difficulty of maintaining neutral positions in long-term social interactions. Moreover, the agents who move away from neutral positions are spontaneously polarized into two camps: one who holds left-leaning opinions (denoted in blue) and the other who holds right-leaning opinions (denoted in red). This suggests that political polarization, a long-standing concern in real-world society, also emerges in the networked \rvss{synthetic} systems of LLM agents. By dividing the stationary opinion distributions into left-leaning, neutral, and right-leaning camps (Figure~\ref{fig1}d), we find that these polarized distributions consistently exhibit a left-skewed pattern, in contrast to real-world observations where the power between left-leaning and right-leaning camps is generally more balanced (Supplementary Table S1). Indeed, prior studies have pointed out the inherent left-leaning bias of OpenAI's GPT~\cite{liu2022quantifying,rutinowski2024self,motoki2024more,santurkar2023whose,feng2023pretraining,hartmann2023political,liu2025turning,yang2024unpacking}, which largely accounts for the observed left-skewed pattern. In open models, such bias has been demonstrated to occur largely through fine-tuning~\cite{potter2024hidden}, likely on correlated qualities including positivity, openness, and nontoxicity~\cite{martin2023ethico}, or even specific forms of reason~\cite{fisher2024biased, costello2024durably}. These insights, however, do not account for ways in which LLMs interact with each other: the micro-social process of imbalanced LLM polarization has not been explored~\cite{scheff1990microsociology}.




\subsection*{\rvss{Polarization patterns emerge from self-regulated LLM agents.}}

\begin{figure}[ht]
\centering
\includegraphics[width=0.9\textwidth]{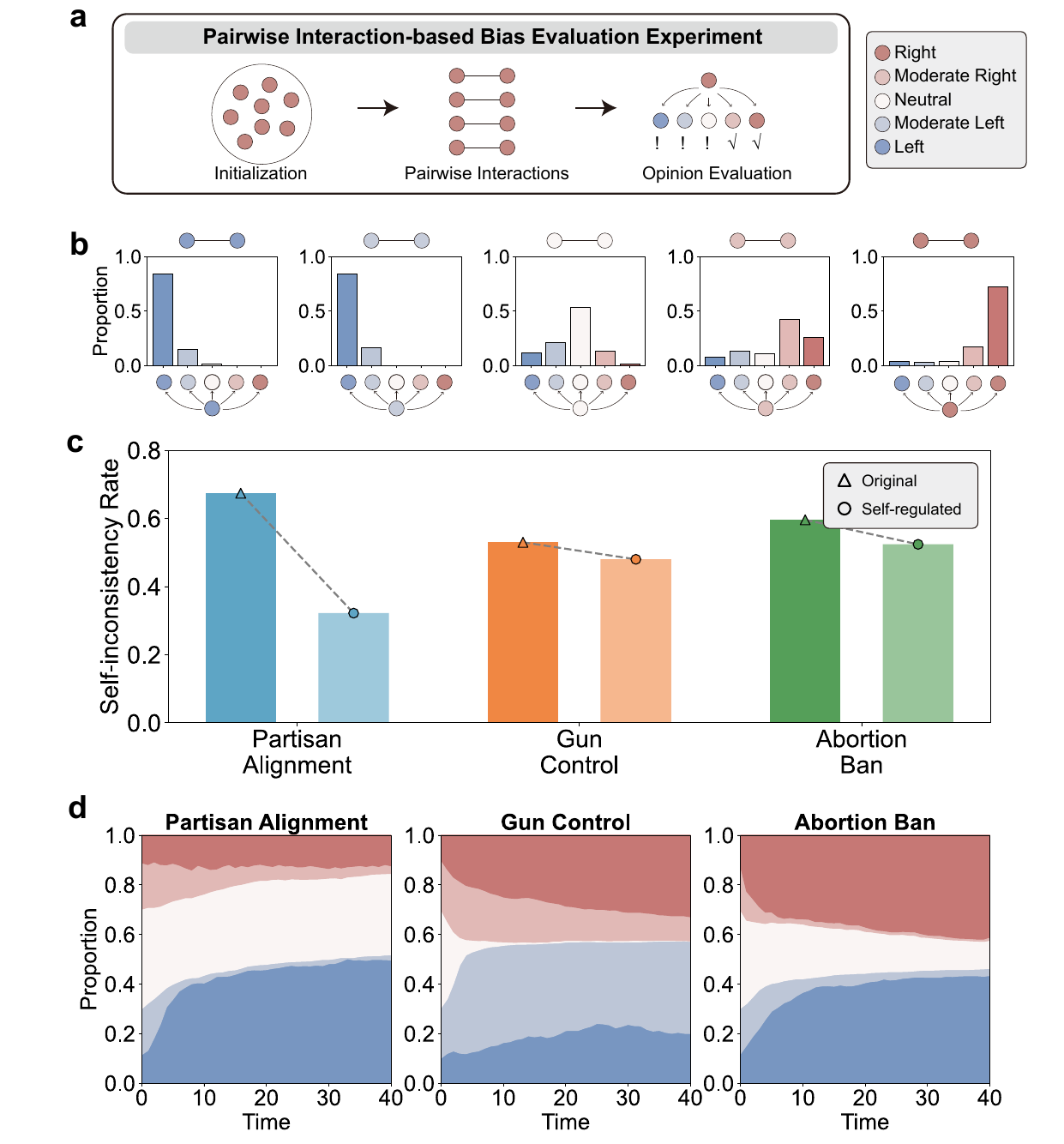}
\caption{\textbf{\rvss{Polarization patterns emerge} from self-regulated LLM agents.} \textbf{a}, Evaluating the self-inconsistency of LLM agents through pairwise interaction-based experiments. \textbf{b}, Opinion transition probability in pairwise interaction-based experiments, where agents with right-leaning opinions occasionally switch to opposing opinions while those with left-leaning opinions do not. \textbf{c}, Performance of the self-regulation strategy across political issues, where the self-inconsistency problem is largely mitigated. \textbf{d}, Opinion dynamics of self-regulated LLM agents on the political issues. \rvss{Self-regulated LLM agents exhibit polarization patterns under free-form social interactions.} The number of agents for each experiment is 1000.}
\label{fig2}
\end{figure}

To validate whether the left-skewed pattern originates from the inherent bias of LLMs, we design a pairwise interaction-based experiment (Figure~\ref{fig2}a). In particular, we follow typical settings for a networked system (see details in Methods M1 and Supplementary Methods), but consider two key simplifications: First, to prevent network structures from interfering with the examination of inherent bias, we ensure that each agent communicates with only one other. Second, we set these two agents to share the same opinion, which allows us to disentangle the effects of bias from social influence on changing agents' opinions (Method M2). By evaluating opinion transition probabilities after the one-round interaction (Figure~\ref{fig2}b and Supplementary Figures S3, S4), we notice an asymmetric opinion update: right-leaning agents occasionally switch to left-leaning after one-round pairwise interactions, whereas left-leaning agents do not switch to right-leaning. Similar cases of the asymmetric opinion update are also noted in other issues, including abortion ban, gun control, ObamaCare, etc. (Supplementary Note 3). These observations indicate that the observed skewed opinions result from the individual formation of LLM agents' opinions.



Following this insight, we track the behaviour of agents who have transitioned to opposing opinions (Supplementary Note 3), finding the self-inconsistency problem inherent in LLM agents. For example, a neutral agent, who ideally should remain impartial towards both Democratic and Republican parties, states, ``\textit{As someone who deeply cares about politics, I believe in the values of equality, progress, and inclusion that the Democrats stand for. I strongly support the Democratic party}''. This agent, despite currently adopting a neutral opinion, self-expresses a left-leaning preference for the Democratic party, which is inconsistent with its neutrality (see more examples in Supplementary Note 3). Indeed, the problem is prevalent across the three stages: agents could generate reasons and communications inconsistent with their opinions, and they may also update their opinions in ways inconsistent with the messages they have received. Among the three stages, the opinion update stage exhibits the highest frequency of self-inconsistency (Supplementary Figure S6), highlighting the particular difficulty for LLM agents to comprehend messages and transform them into their opinions. More importantly, this problem occurs significantly more frequently among agents with right-leaning opinions than those with left-leaning opinions (two-sided proportion $z$-test, $z=8.27$, $p \ll .001$, difference in proportions = 0.068, 95\% CI $= [0.052, 0.084]$). This results in right-leaning agents exhibiting a disproportionately high probability of transitioning to the opposing side, accounting for the observed skewed opinion dynamics.



\rvs{We remedy the self-inconsistency problem using a theory-driven method inspired by the social science literature~\cite{bandura1977social,bandura1991social}: As noted in Bandura's social learning theory, self-regulation is an essential human ability to continuously monitor and adjust behaviours to respond more consistently and adaptively in dynamic social contexts~\cite{bandura1977social,bandura1991social}. Here we propose a strategy to incorporate that self-regulation capacity into LLM agents. In particular, agents are prompted to self-check whether their generated messages, opinions, and reasons are consistent with their current status. If agents find inconsistencies, they adjust themselves through iterative re-generation until consistency is achieved (more details in Methods M3 and Supplementary Methods). This way, these agents can independently recognize and rectify glaring inconsistencies in expressed thoughts (i.e., opinions and corresponding justification) and behaviour.} We assess the effectiveness of the proposed strategy using the metric of self-inconsistency rate, calculated as the distance between the experimental opinion transition probability matrix and the identity matrix (see details in Method M5). As shown in Figure~\ref{fig2}c, the self-regulation strategy reduces self-inconsistency by 9.4\%-52.2\% across scenarios on a variety of issues. This highlights the essential role of self-regulation in improving consistency between generated opinions, reasons, and messages in LLM agents. \rvss{It is worth noting that this strategy is introduced to ensure the behavioural plausibility and internal consistency of LLM agents, rather than to remove or correct their inherent tendencies.} Based on these more self-consistent agents, we further explore their collective behaviour by simulating networked systems of self-regulated LLM agents and assessing their opinion dynamics (Figure~\ref{fig2}d). We discover that self-regulated LLM agents also develop their collective opinions into a polarized pattern, suggesting that the emergence of LLM agent polarization is not due to the inherent bias of LLMs, but results from their free-form social interaction. Moreover, unlike the previously left-leaning dominated polarization pattern (Figure~\ref{fig1}b), the right- and left-leaning camps are now balanced, which better reflects real-world scenarios (Supplementary Table S1 and Figure S1). Overall, LLM agents, despite only being empowered with these basic social capabilities, \rvss{spontaneously develop similar polarization patterns through social interaction.}


\subsection*{\rvss{Collective behavioural patterns behind the emergence of polarization among LLM agents.}}

\begin{figure}[ht]
\centering
\includegraphics[width=0.82\textwidth]{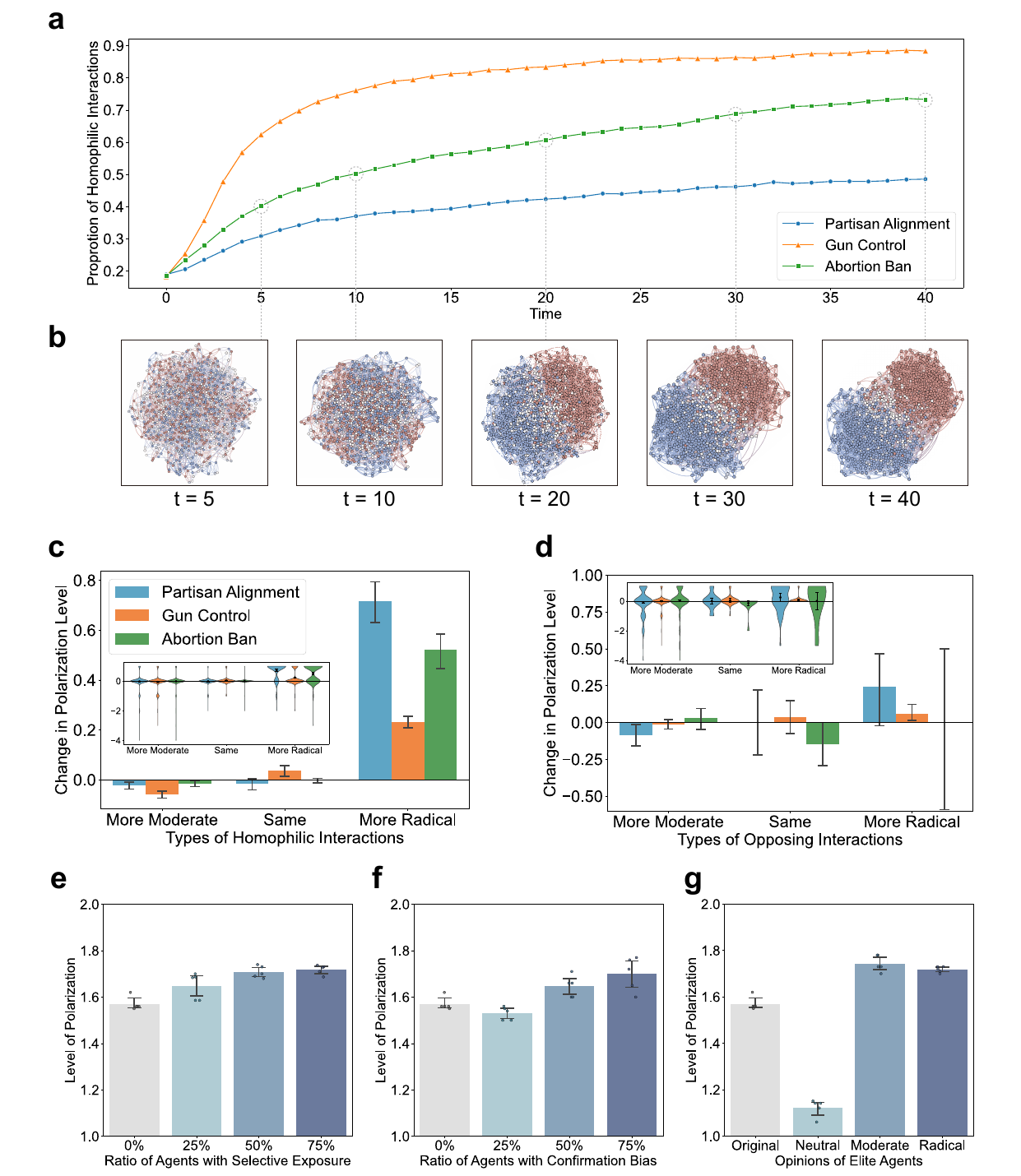}
\caption{\textbf{\rvss{Collective behavioural patterns behind the emergence of polarization among LLM agents.}} \textbf{a}, Changes in the proportion of homophilic interactions over time. Agents are increasingly likely to interact with those holding similar opinions. The number of agents is 1000. \textbf{b}, Evolution of social networks among LLM agents, where agents with similar opinions are more likely to interact with one another, exhibiting the tendency toward homophilic clustering. Each network visualization corresponds to the circled points in (a). \textbf{c}, \rvss{Behavioural patterns consistent with the echo chamber effect, where exposure to more radical homophilic interactions is associated with higher polarization levels.} \textbf{d}, \rvss{Behavioural patterns consistent with the backfire effect, where interactions with agents holding opposing opinions are associated with increased polarization.} In (c, d), bars represent the average, and error bars represent the corresponding 95\% confidence intervals (CIs). Sample sizes, corresponding to the number of agent interactions, for bars in (c) are 3680, 4212, 3962, 494, 761, 1315, 170, 1466, and 246. Sample sizes in (d) are 630, 706, 523, 27, 27, 41, 45, 65, and 22. \textbf{e-g}, Effects of individual-level \rvss{behavioural tendencies related to} selective exposure, confirmation bias, and elite signaling. In (e-g), bars show average levels of polarization in the last five timesteps ($n=5$, corresponding to five timestep-level polarization values computed from all agents' opinions), and error bars show the corresponding 95\% CIs. When the system consists of more agents with traits of (e) selective exposure or (f) confirmation bias, and (g) influencers adopt non-neutral opinions, the level of polarization increases. }
\label{fig3}
\end{figure}

The above results uncover that free-form social interactions among LLM agents result in the emergence of \rvss{opinion} polarization, raising a natural question regarding the exact role these interactions play in this process. Therefore, we begin by assessing the extent to which agents interact with those holding similar opinions. As shown in Figure~\ref{fig3}a, we find that the proportion of interactions between agents in the same camp, i.e., left- or right-leaning camp, increases by 156.8\%-382.7\% over time. This increase eventually leads to 48.5\%-88.3\% of interactions in these systems occurring between agents who share similar opinions, suggesting a tendency for similar agents to cluster together. This process is \rvss{consistent with the notion of} homophily, where ``birds of a feather flock together''~\cite{mcpherson2001birds,mas2013differentiation}, and has been observed in naturalistic LLM social interactions~\cite{laiposition}. Moreover, our agents increasingly avoid interacting with agents who hold opposing opinions (Supplementary Figure S8a). Indeed, their tendencies of homophilic clustering and opposing avoidance appear to be driven by their \rvss{generated evaluations} of others. Further experiments show that agents are more likely to form more favorable impressions of and use more positive language to describe those within their own camp than those in opposing camps (Supplementary Note 8). It is worth mentioning that no mechanisms prompting in-group bias or homophilic clustering were inserted into the system; this emerges purely from the LLM agents' free-form choice behaviours in social relationships. More interestingly, in the system discussing the abortion ban, agents gradually self-organize into two communities (Figure~\ref{fig3}b), with one supporting left-leaning opinions and the other supporting right-leaning ones. Indeed, beyond LLM agents, homophilic clustering is also widely observed in human society~\cite{cinelli2021echo,baumann2020modeling,nyhan2023like} and in- vs. out-group bias has been posited by evolutionary psychologists with the emergence of justificatory reason~\cite{mercier2011humans}.

A widespread subsequent concern is the potential for the clustering of homophilic peers to reinforce opinion polarization, which is often termed the echo chamber effect \rvss{in studies of human behaviours}~\cite{nyhan2023like, cinelli2021echo}. Therefore, we examine the relationship between homophilic clustering and polarization levels in the system of LLM agents. In particular, we calculate the average opinions of each agent's neighbors to quantify the relative levels of radicalization between the agent and its neighbors (see details in Method M5). Focusing on the interactions between homophilic agents (Figure~\ref{fig3}c), we observe that exposure to more radical agents significantly increases the level of polarization (two-sided Student's $t$-tests, partisan alignment $t(169)=16.80$, $p\ll.001$, Cohen's $d = 1.288$, 95\% CI $= [0.633, 0.802]$, gun control $t(1465)=19.37$, $p\ll.001$, Cohen's $d = 0.506$, 95\% CI $= [0.209, 0.256]$, and abortion ban $t(245)=13.99$, $p\ll.001$, Cohen's $d = 0.892$, 95\% CI $= [0.447, 0.594]$). This indicates that \rvss{LLM agents exhibit behavioural patterns consistent with the echo chamber effect.} On the other hand, despite homophilic interactions representing the majority (54.8\%-82.2\%), agents still retain a small chance of encountering those with opposing opinions (7.4\%-10.2\%). Nevertheless, these interactions with opposing agents do not consistently reduce the polarization level as expected (Figure~\ref{fig3}d). Instead, exposure to agents with opposing opinions could potentially increase polarization levels, \rvss{indicating behavioural patterns consistent with the backfire effect}~\cite{Bail2018Exposure}. \rvss{We note that the echo chamber and backfire effects do not universally occur across all individuals in the real world, with some recent theoretical debates~\cite{nyhan2023like, guess2018avoiding, coppock2023persuasion}; instead, they exhibit substantial heterogeneity. Extensive experiments reveal that the system autonomously formed by the LLM agents can capture this heterogeneity, showing consistency with empirical observations (Supplementary Note 4).}

Overall, these observations suggest that in the discussion of critical political issues, LLM agents can spontaneously organize their social network. Their networks not only exhibit the property of homophilic clustering, but also shape collective opinions \rvss{in ways consistent with} the echo chamber and backfire effects. To support these findings, we report further quantitative metrics, including modularity, assortativity, and the homophily index, along with extended analyses on the network properties of their self-organization (see Supplementary Note 4). Notably, we observe that social networks evolve into a scale-free degree distribution, where high-degree agents gain great popularity in free-form interaction (Supplementary Figure S13), following degree distribution models of social networks~\cite{barabasi2003scale}. Moreover, when we replace LLM agents' self-organized network with a random or static network, we find that one-camp opinions dominate the overall system (Supplementary Figure S14), which highlights the important role of self-organized social networks in \rvss{shaping polarization patterns} exhibited by LLM agents. \rvss{Importantly, these results align with established argument-accumulation models in opinion dynamics, where polarization emerges from the cumulative exchange and reinforcement of persuasive arguments under homophilic interactions~\cite{mas2013differentiation,flache2017models}, as reflected in our analysis of agents’ self-reported reasons (Supplementary Note 8).}



 As discussed above, the polarization of LLM agents is driven by two coupled processes of network organization and opinion formation. Specifically, in the network organization process, agents tend to interact with similar peers (Figures~\ref{fig3}a and b), while in the opinion formation process, agents tend to comprehend messages in a way that aligns with their pre-existing opinions (Figures~\ref{fig3}c and d). These two tendencies of LLM agents have been frequently observed in humans~\cite{stroud2010polarization,nickerson1998confirmation,cinelli2021echo,bakshy2015exposure,jost2022cognitive}, and are often referred to as selective exposure~\cite{stroud2010polarization} and confirmation bias~\cite{nickerson1998confirmation,klayman1995varieties}, respectively. Unlike network-level \rvss{patterns}, these two mechanisms adopt a psychological perspective, focusing more on how individuals' traits and behaviours contribute to polarization. We note that these two mechanisms are not explicitly hard-coded or assumed in the proposed system. Rather, \rvss{agents' behavioural patterns consistent with these mechanisms} naturally emerge from agents’ social interactions. For example, analyzing the \rvss{self-reported reasons} generated during opinion formation reveals \rvss{patterns in line with} confirmation bias (Supplementary Note 8).

Given these observed \final{network-level patterns}, one may wonder whether LLM agents \rvss{exhibit behaviours consistent with psychological mechanisms discussed in studies of human opinion formation}. To answer this question, we design a comparative experiment: \rvss{we vary the proportion of agents exhibiting behaviours consistent with selective exposure or confirmation bias, and then examine the resulting changes in the level of polarization.} Cases where 0\% or 100\% of agents \rvss{are configured to exhibit these behaviours} are deliberately designed as extreme scenarios for comparative purposes. As shown in Figure~\ref{fig3}e and f, we observe that the increase in agents \rvss{exhibiting either behaviour} leads to a marked increase in the level of polarization characterizing the overall system. This suggests that behaviours related to selective exposure and confirmation bias contribute to the polarization of LLM agents, \rvss{consistent with findings reported in studies of human behaviour}~\cite{jost2022cognitive,nickerson1998confirmation,druckman2013elite}.

On the other hand, prior studies have pointed out that opinion formation is also largely influenced by the traits or behaviours of others, particularly top influencers on social media~\cite{flamino2023political,druckman2013elite,rogowski2016ideology,jost2022cognitive,lorenz2023systematic}. As suggested by the elite signaling mechanism~\cite{flamino2023political,druckman2013elite,rogowski2016ideology}, the polarization level of influencers' opinions is positively associated with greater polarization in the overall population. To validate this effect on LLM agents, we introduce additional top influencers holding neutral, moderate left/right, and strong left/right opinions into the system, with these influencers sending non-personalized messages to all other agents (Figure~\ref{fig3}g). In this setting, the top influencers are symmetrically distributed across the right- and left-leaning camps. \final{We observe an effect in the LLM-agent system that is qualitatively consistent with patterns reported in studies of human opinion formation}. When the influencers hold non-neutral opinions, the final polarization level is significantly increased (two-sided Student's $t$-tests, left/right versus original, $t(8)=10.63$, $p \ll .001$, Cohen's $d = 6.720$, 95\% CI $= [0.116, 0.180]$; moderate versus original, $t(8)=8.64$, $p \ll .001$, Cohen's $d = 5.462$, 95\% CI $= [0.128, 0.220]$). By contrast, the introduction of neutral influencers significantly reduces the polarization level of the system by 28.5\% (two-sided Student's $t$-tests, original versus neutral, $t(8)=21.56$, $p \ll .001$, Cohen's $d = 13.636$, 95\% CI $= [0.399, 0.495]$). Besides the three studied mechanisms, we also examine a series of other individual-level mechanisms, finding their effects on LLM agents to be consistent with those observed in humans (Supplementary Note 5). Overall, these experiments consistently suggest that \rvss{LLM agents exhibit individual-level behavioural responses that parallel those documented in studies of human opinion formation.}

\subsection*{\final{Synthetic testbed for exploring hypothetical intervention strategies.}}

\begin{figure}[ht]
\centering
\includegraphics[width=0.95\textwidth]{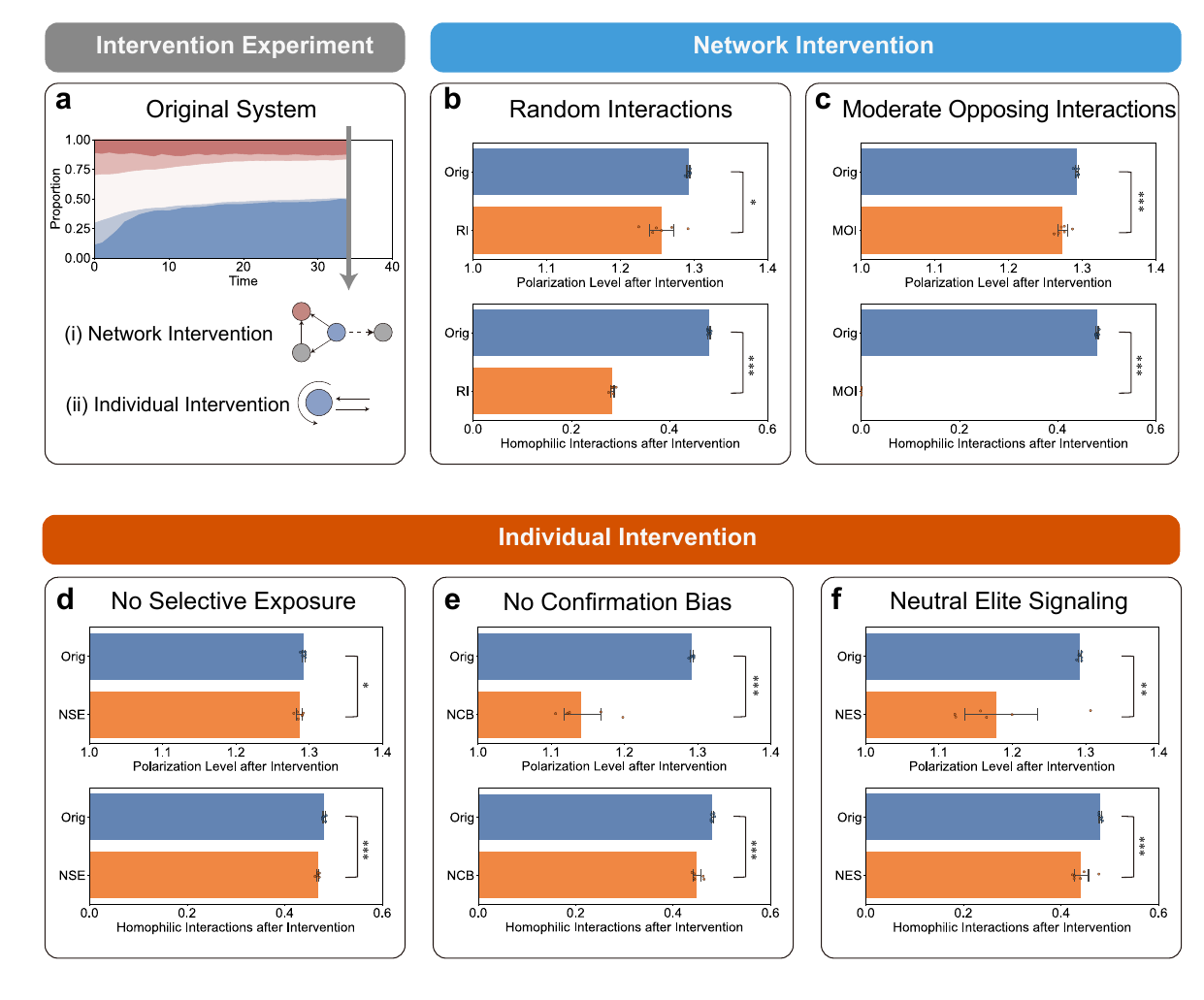}
\caption{\textbf{\final{Synthetic testbed for exploring hypothetical intervention strategies.}} \textbf{a}, Intervention experiments, where two types of strategies are applied to the original polarized system: (i) network interventions, which directly modify LLM agents' social network, and (ii) individual interventions, which adjust agents' traits and behaviours. The number of agents in each experiment is 1000. \textbf{b, c}, Network intervention strategies of (b) random interaction, where agents randomly interact, and (c) moderate opposing interaction, where agents receive messages only from those with opposing moderate opinions. \textbf{d-f}, Individual intervention strategies of (d) no selective exposure, where agents tend to interact with those holding diverse opinions, (e) no confirmation bias, where agents are open-minded to diverse opinions, and (f) neutral elite signaling, where agents receive non-personalized neutral messages. In (b-f), the upper sub-figures show the comparison of polarization levels between the original and the intervened systems, while the lower sub-figures illustrate the comparison of proportions of homophilic interactions. Bars show the average values after intervention ($n=6$, corresponding to six timestep-level values computed from all agents' opinions or interactions), and error bars show the corresponding 95\% CIs. We perform two-sided Student/Welch's $t$-tests: (b) $t(5.16)=3.87$, $p=.011<.05$, Cohen's $d=2.237$, 95\% CI $= [0.013, 0.061]$, $t(10)=80.96$, $p<.001$, Cohen's $d=46.741$, 95\% CI $= [0.191, 0.202]$, (c) $t(10)=5.24$, $p<.001$, Cohen's $d=3.027$, 95\% CI $= [0.011, 0.027]$, $t(5)=366.00$, $p<.001$, Cohen's $d=149.418$, 95\% CI $= [0.477, 0.484]$ (one-sample $t$-test against zero), (d)  $t(10)=2.46$, $p=.033<.05$, Cohen's $d=1.422$, 95\% CI $= [0.001, 0.011]$, $t(10)=6.64$, $p<.001$, Cohen's $d=3.832$, 95\% CI $= [0.008, 0.017]$, (e) $t(10)=10.65$, $p<.001$, Cohen's $d=6.146$, 95\% CI $= [0.120, 0.184]$, $t(10)=6.99$, $p<.001$, Cohen's $d=4.033$, 95\% CI $= [0.022, 0.042]$, and (f) $t(10)=4.02$, $p=.002<.01$, Cohen's $d=2.324$, 95\% CI $= [0.051, 0.177]$, $t(10)=4.78$, $p<.001$, Cohen's $d=2.760$, 95\% CI $= [0.021, 0.059]$. In this synthetic testbed, individual-level strategies are associated with larger observed reductions in polarization than network-level interventions.}
\label{fig4}
\end{figure}

The above results suggest that \final{the behaviours and collective patterns produced by the LLM agents are consistent with empirical observations and established opinion-dynamics models.} \final{This consistency suggests the potential of the proposed LLM-agent system as a valuable synthetic testbed for exploring hypothetical intervention strategies in networked LLM-agent systems.} To explore this potential, we design five intervention strategies \rvss{in our synthetic testbed} (Figure~\ref{fig4}). At the network level, we consider two strategies: (i) Random interactions~(RI): we remove the homophilic clustering property of LLM agents' social network by allowing agents to randomly interact with each other. (ii) Moderate opposing interactions~(MOI): given the echo chamber and backfire effects, agents are only allowed to receive messages from those adopting moderate opposing opinions. At the individual level, three strategies are proposed: (iii) No selective exposure~(NSE): agents are prompted to have the tendency to communicate with those holding diverse opinions. (iv) No confirmation bias~(NCB): agents are instructed to be open-minded to persuasion of diverse opinions. (v) Neutral elite signaling~(NES): agents receive non-personalized neutral messages from newly introduced top influencers. Indeed, these interventions \rvss{are motivated by established lines of prior work and are used here as experimental interventions in our synthetic system}: MOI \rvss{draws on studies of exposure to cross-cutting information}~\cite{coppock2023persuasion,levendusky2023our}; RI \rvss{operationalizes randomized interaction}; NSE and NCB relate to promoting open-mindedness~\cite{groenendyk2021motivates}; and NES is inspired by the design of Bail et al.~\cite{bail2020assessing}.

Based on the proposed strategies, we intervene in a system that has converged to a polarized state, i.e., the system at $t=35$ (Figure~\ref{fig4}a). By comparing the evolution of the original system with that of the intervened system following intervention (Figure~\ref{fig4}b-f), we find that all strategies significantly reduce the level of polarization. In particular, the intervention of MOI has a clear and significant effect in reducing opinion polarization among agents (two-sided Student's t-test, $t(10)=5.24$, $p<.001$, Cohen's $d = 3.027$, 95\% CI $= [0.011, 0.027]$). The effects are relatively small, however, producing approximately a 2\% change in agent opinions. This modest effect is consistent with findings from human studies, where cross-cutting interactions reduce polarization but typically result in only small shifts among actual opinions~\cite{levendusky2023our,coppock2023persuasion}. Compared with network-level interventions that directly modify agents' social networks, individual-level strategies of NCB and NES contribute to the greatest reduction in polarization by 11.8\% and 8.8\%, respectively. On the other hand, we observe that the NSE strategy has limited effects. This is because, in an almost stable social network, agents' non-selective tendencies do not provide exposure to diverse opinions but instead quickly crystallize the existing network structures that foster polarization (see detailed analyses on intervention strategies in Supplementary Note 7). \rvss{Despite the difficulty of recommending openness to human populations,} these results suggest that \rvss{in our synthetic testbed, access and open-mindedness to diverse opinions at the individual level appear more effective than altering social networks in polarized systems.} Furthermore, we observe that all of the strategies, despite reducing homophilic interactions by varying rates from 2.6\% to 100\%, significantly promote inclusive conversations among agents with diverse opinions (lower sub-figure in Figure~\ref{fig4}b-f). In all, these experiments \final{compare several hypothetical strategies and illustrate the value of the proposed system for controlled exploration of intervention hypotheses within networked LLM-agent systems.}

\section*{Discussion}

This work uncovers that LLM agents exhibit polarization \rvss{patterns} on political issues\rvss{, at both the individual and collective levels.} Specifically, from the individual perspective, opinion formation among LLM agents \rvss{exhibits behavioural patterns consistent with} social and psychological mechanisms~\cite{stroud2010polarization,nickerson1998confirmation}, e.g., selective exposure, confirmation bias, elite signaling. From the collective perspective, LLM agents organically develop social networks with properties \rvss{characteristic of human social networks}, e.g., homophilic clustering and scale-free degree distributions~\cite{laiposition}. Moreover, their self-developed social networks, in turn, shape LLM collective opinions in ways \rvss{consistent with the echo chamber and backfire effects described in studies of human behaviour}, resulting in the emergence of \rvss{opinion} polarization~\cite{lorenz2023systematic,cinelli2021echo,bakshy2015exposure,Bail2018Exposure,flamino2023political,falkenberg2022growing,nyhan2023like,sunstein2018republic,jost2022cognitive}. Notably, all these \rvss{behavioural patterns}, network organization, and collective opinions naturally emerge from interactions among agents completely driven by LLMs. Building on these findings, our work shifts the perspective from isolated LLMs~\cite{argyle2023leveraging,small2023opportunities,fish2023generative,kim2020bot,ma2024towards,costello2024durably,tessler2024ai} to complex, networked LLM agent societies, offering a \rvss{synthetic testbed for studying} emergent collective dynamics in artificial agents. While our main results are based on OpenAI's ChatGPT and three political issues, we also perform extensive experiments with different LLMs (both full-run and midway-defection settings, Supplementary Note 6) as well as on topics of immigration restriction and the flat Earth theory (Supplementary Note 13), finding that these experiments consistently support our core findings. We also vary two key initial conditions, i.e., the social network structure and initial opinion distribution, and find that both resulting polarization patterns and final network structures remain strikingly similar, highlighting the robustness and generalizability of our system and experiments (Supplementary Note 12).

\rvss{While this work does not make theoretical contributions, it is grounded in established theories and models of opinion dynamics~\cite{mas2013differentiation,flache2017models,vinokur1974effects,isenberg1986group}. In particular, this work follows the argument-accumulation lineage, which explains polarization as arising from the cumulative exchange and reinforcement of persuasive arguments~\cite{vinokur1974effects,isenberg1986group,mas2013differentiation,banisch2021argument} rather than from direct opinion averaging or social pressure~\cite{degroot1974reaching,hegselmann2015opinion}. We further analyze agents' self-reported reasons during opinion formation, observing that agents who become more polarized tend to accumulate substantially more supporting arguments over time (Supplementary Note 8). This observation highlights a clear connection to argument-accumulation models and suggests that the polarization dynamics observed in our synthetic LLM-agent testbed likely stem from similar underlying mechanisms. At the same time, we acknowledge broader theoretical debates on social media and polarization. Some recent studies have emphasized that echo chamber effects do not universally apply across all individuals in real-world settings, with substantial heterogeneity in exposure to like-minded content~\cite{nyhan2023like, guess2018avoiding}. In our experiments, we also observe similar patterns in LLM agents (Supplementary Note 4). Moreover, deliberative and argument-accumulation models have limitations, such as their abstraction from affective responses and identity-based dynamics~\cite{achen2017democracy, mason2018uncivil, iyengar2019origins}, which have been highlighted in more recent conceptions of polarization. While our primary analyses focus on opinion-based polarization, we provide supplementary analyses of behavioural patterns consistent with some of these broader dimensions (Supplementary Note 8).}

Previous studies have documented the presence of political biases in LLMs~\cite{feng2023pretraining,santurkar2023whose,rottger2024political,bang2024measuring,argyle2023leveraging,motoki2024more}. In particular, there are two main lines of research. The first line focuses on analyzing political bias in ``plain vanilla'' LLMs, to examine the extent of biases present in their default configurations~\cite{feng2023pretraining,rottger2024political,bang2024measuring,motoki2024more}. The second line of research assigns demographics to LLMs and evaluates how well these models represent different populations~\cite{santurkar2023whose,argyle2023leveraging,tornberg2023simulating}. While this approach allows LLMs to adopt political opinions resembling those of real-world individuals with similar demographics~\cite{santurkar2023whose,argyle2023leveraging}, they also find that LLMs fall short in accurately representing certain minority groups~\cite{santurkar2023whose,wang2025large}. Overall, some recent studies attempt to move beyond default configurations and explore the potential of demographically prompted LLMs as proxies for human populations. Nevertheless, they have yet to explore whether and how political biases may emerge through LLM agents' social behaviours -- including friend selection, opinion updating, and patterns of communication. Our system addresses this gap and can be readily extended to model more diverse human subgroups by leveraging prompts as ``soft'' parameters (Supplementary Note 10). This naturally leads to the next question: will LLM agents faithfully follow their assigned political personas and exhibit behaviours consistent with humans holding comparable identities? In this work, we discover that the underlying political bias of LLMs does lead these agents to occasionally exhibit behaviours that deviate from their assigned political personas, which we refer to as self-inconsistency. This inconsistency causes agents to gradually drift toward left-leaning positions during interaction (Figure~\ref{fig2}b), and over time, cumulatively results in the collective left-skewed polarization observed in Figure~\ref{fig1}b–d. To address self-inconsistency, we propose the self-regulation strategy inspired by Bandura's social learning theory~\cite{bandura1977social,bandura1991social}, enabling agents to self-monitor and adjust their behaviours. We find self-regulated LLM agents develop their collective opinions into balanced right-left polarization (Figure~\ref{fig2}d), better reflecting real-world distributions. Furthermore, comparison of polarization speeds between the original and self-regulated agents reveals only slight differences, suggesting that model bias has a minimal impact on accelerating polarization (Section 3.2 in SI). \rvss{Overall, these findings underscore the importance of studying bias and consistency at the level of interacting LLM agents, and of investigating collective opinion dynamics in synthetic social systems.}

\rvss{While this work advances prior GABM studies by providing a solid synthetic testbed in which large-scale LLM agents, under minimal social assumptions, produce emergent polarization patterns, it also shares limitations common to current GABM research~\cite{cui2025large,adornetto2025generative,larooij2025validation}. In particular, like prior studies, the proposed system faces natural challenges to external validity. Although results are consistent with well-established opinion-dynamics models~\cite{mas2013differentiation,cinelli2021echo,Bail2018Exposure}, the extent to which these patterns quantitatively correspond to real-world human behaviour must be further validated. In addition, compared to prior models with transparent mechanisms, GABM comes at the cost of reduced interpretability, as agent behaviour is shaped through prompt-based design and simplified opinion representations. Moreover, the intervention results presented here should be understood as hypothesis-generating. The primary value of the proposed system lies in enabling controlled, pre-experimental exploration of how interaction rules, network structures, and agent-level behaviours may jointly shape collective opinion dynamics. Future research will benefit from complementary human experiments~\cite{banisch2024validating,mas2013differentiation}, as well as more systematic calibration and evaluation against fine-grained real-world data.}

\section*{Methods}

\subsection*{M1 Design of the networked system of LLM agents}

To understand the opinion dynamics of LLM agents, we develop a networked system of LLM agents. In this system, each agent is empowered with three stages that ensure their basic social capabilities: (i)~self-expression, which enables agents to consider and express their reasons for supporting their opinions, (ii)~communication, which enables agents to exchange opinions and the corresponding reasons, and (iii)~opinion updates, which enable agents to update their opinions based on received messages. These three stages take advantage of LLMs' capabilities in traditional natural language processing tasks~\cite{vaswani2017attention,brown2020language}. For example, the self-expression stage predominantly depends on automated text generation, and the opinion update stage relies on text summarization. These capabilities allow agents to perform basic social interactions. Below we introduce our proposed system in detail.

In each system, agents communicate about one political issue, e.g., gun control or abortion rights. We randomly initialize the system, including agents' opinions on the political issue and their social relationships. When initializing the system, we consider two aspects. First, the initialization should prevent the system from starting from an extreme case, e.g., when agents' opinions are already highly polarized. Second, the initialization setting should allow enough flexibility for agents to develop their collective opinions and form social networks. As starting from a clustered network and highly polarized opinion distribution would result in a crystallized setting, not allowing us to study how interaction dynamics emerge over time, we introduce some randomness in the initial networks and opinions considered. Specifically, we sample each agent's initial opinion following a near-Gaussian distribution as shown in Figure~\ref{fig1}c. We initialize the social network using a Watts–Strogatz model with the rewiring probability of 0.001, which exhibits a relatively weak small-world property, characterized by a normalized clustering coefficient of 0.99 and a normalized average path length of 0.71. 

After initialization, the self-expression stage requires each agent $i$ to generate a message expressing the reasons for supporting their opinions. These messages provide detailed information about agents' beliefs regarding the political issue. Then, following the communication stage, each agent $i$ first chooses whether to communicate with a socially connected or random agent. If agent $i$ declines further communication with its current partner, it will be randomly assigned a new one with which to interact. Here we denote the interaction partner as agent $j$. It is worth noting that distinct from prior studies that allow agent $i$ to ``scan'' all other agents and select its preferred ones~\cite{de2023emergence,papachristou2024network,chang2024llms}, we adopt a more \rvss{naturalistic} setting in which agents may occasionally encounter individuals they dislike, but autonomously decide whether to maintain social connections with those favored. Based on the opinions and reasons of agents $i$ and $j$, agent $i$ generates messages to persuade agent $j$ of its opinion. We note that after one round of communication, agents $i$ and $j$ become directionally linked, allowing agent $i$ to contact agent $j$ in the subsequent timestep. Based on the messages that agent $i$ has received from its socially connected neighbors, agent $i$ updates its opinions and then generates reasons for its new opinion. The system operates these three stages iteratively and enables agents to freely establish social relationships, communicate, and form opinions on critical political issues.

It is worth noting that we do not rely on an external classifier to evaluate agents' opinions. Instead, we prompt the agents to self-report their opinions on political issues. For example, when discussing partisanship, agents are asked: ``What do you feel about political partisanship?'' The agent then responds by placing itself into one of five opinion categories: ``strongly support the Republican Party'' (right), ``support the Republican Party'' (moderate right), ``do not have a tendency'' (neutral), ``support the Democratic Party'' (moderate left), or ``strongly support the Democratic Party'' (left). This approach leverages the role-playing capability of LLM agents, allowing them to express self-evaluation in line with how human respondents articulate their own opinions. When it comes to gun control, right-leaning opinions indicate support for weaker gun control, while left-leaning opinions indicate support for stricter control. Similarly, for abortion bans, right-leaning opinions represent support for abortion bans, whereas left-leaning opinions represent opposition to such bans.



To ensure the system scale is large enough to support observations on network structures, we incorporate 1000 agents in all simulations of networked systems, except for the individual-level mechanism experiments, where 100 agents are used due to our focus on individuals and high computational costs. We use ChatGPT (GPT-3.5) through the public OpenAI API to run all the experiments in the Main Text. We set the temperature to 1 in all experiments and adopt a zero-shot setting. We also perform experiments using other LLMs, including GPT-4o, ChatGLM, Llama-3, and DeepSeek-V3 (Supplementary Note 6), as well as experiments with varying temperatures (Supplementary Note 9). These experiments consistently validate the conclusion that long-term social interactions among LLM agents lead to the emergence of \rvss{opinion} polarization. Additionally, by varying two key initial conditions, i.e., the social network structure and the initial opinion distribution, we find that both the resulting polarization patterns and networks remain consistent (Supplementary Note 12). Detailed prompts to operate the overall system are provided in Supplementary Methods.




\subsection*{M2 Pairwise interaction-based bias evaluation experiment}

To investigate the origin of the pattern of biased opinion, we design a pairwise interaction-based bias evaluation experiment (Figure~\ref{fig2}a). Different from the complex networked system, we focus on the effect of the one-round interaction between agents with identical opinions in this experiment. In particular, we begin by constructing a collection of agents, who are initially assigned the same opinion and prompted to generate their own supporting reasons for it. These agents are paired randomly and one of them, denoted agent $i$, is required to persuade the other agent, denoted agent $j$, of the opinion. After this single communication round, we measure the change in agent $j$'s opinion (Figure~\ref{fig2}b and Supplementary Figures S3 and S4). Details for the experiment are provided in Supplementary Methods.

\subsection*{M3 Self-regulation strategy for the problem of self-inconsistency}

To mitigate observed self-inconsistency inherent in LLM agents, we propose a self-regulation strategy inspired by social theory with a natural human analogue~\cite{bandura1977social,bandura1991social}. As discussed above, unlike vanilla LLMs, \rvss{LLM agents generate extended self-explanations and interaction histories, which motivates drawing inspiration from established social and behavioural theories when addressing their self-inconsistency.} In this way, we follow Bandura's concept of self-regulation in social learning theory~\cite{bandura1977social,bandura1991social}, employing LLM agents to self-regulate their expressions, communication, and opinions, and update their corresponding behaviours. In particular, after these agents generate a message supporting their opinions, we require them to check whether the message reflects their opinions. If not, they will continue to re-generate new messages until their consistency requirements are satisfied. This is \rvss{conceptually analogous to} an individual holding a self-consistent `line' in conversation with others~\cite{goffman2017interaction}. Similarly, in the stages of communication and opinion update, we instructed LLM agents to ensure that their generated communication messages align with their current opinions and that their updated opinions are plausible given their received messages. Detailed prompts for the self-regulation strategy are provided in Supplementary Methods.

\subsection*{M4 \final{Evaluation of hypothetical interventions in the synthetic LLM-agent system}}

To explore the question of \rvss{which intervention strategies are more effective in reducing polarization in the synthetic system}, we design an intervention experiment based on the proposed networked system of LLM agents. Considering \rvss{commonly studied polarization scenarios}, we focus on interventions in an already polarized system. We use the system discussing partisan alignment at $t=35$, which has converged to a polarized state, as an example. As described in the Main Text, we propose five intervention strategies. At the network level, we directly modify the social network of LLM agents, while at the individual level, we adjust their traits and behaviours through prompts. We supplement the detailed prompts in Supplementary Methods. Using these strategies, we intervene in the polarized system at $t=35$. The intervention, i.e., the modification of social networks or the adjustment of agents, continues from $t=35$ to $t=40$. We take the original system as the control group and the intervened systems as the treatment group. We assess the effectiveness of different strategies in reducing polarization by comparing the polarization levels of the original system and the system after intervention. We evaluate their performances in promoting less divisive conversations through the resulting changes in the proportion of homophilic interactions.

\subsection*{M5 Metrics}

\textbf{Level of polarization.} To quantitatively measure polarization and its contributing factors, we define polarization and radicalization metrics following prior studies~\cite{cinelli2021echo,santos2021link,nyhan2023like}. We measure the level of polarization by calculating the average distance between agents' opinions and the neutral position. In particular, we assume that the opinion distribution is characterized by the vector $[f_{-2}, f_{-1}, f_{0}, f_{1}, f_{2}]$, where $f_k$ denotes the relative frequency of agents holding opinion $k$ and $\sum_{k=-2}^2 f_{k}=1$. We compute the polarization level as $s_{\mathrm{pol}}= \sum_{k=-2}^2 |k| \times f_{k}$, where the absolute value $|k|$ measures the deviation distance of opinion $k$ from neutral, i.e., opinion $0$. The polarization level $s_{\mathrm{pol}}$ falls into the domain of $[0, 2]$, and a larger value indicates a higher level of polarization. We further measure the change in polarization for each agent. Specifically, we calculate the change in agent $i$'s opinion between the timesteps $t$ and $t-1$ along the direction of their opinion at the timestep $t-1$, expressed as $s_{\mathrm{cp},\, i} = (x_{i,\, t}-x_{i, \, t-1}) \times \operatorname{sign}(x_{i, \, t-1})$. Here $s_{\mathrm{cp},\, i} > 0$ indicates that the agent $i$ has adopted a more polarized opinion after the interaction at the timestep $t-1$. It is worth noting that after 99.71\% of homophilic interactions, agents adopt same-camp or neutral opinions, i.e., $x_{i,\, t} \times \operatorname{sign}(x_{i, \, t-1}) \geq 0$. \\

\noindent \textbf{Self-inconsistency rate.} We evaluate the self-inconsistency problem by computing the distance between the experimental opinion transition probability matrix and the ideal identity matrix. In particular, we denote the experimental transition matrix as $\mathbf{P}$, where the element $P_{k,\, k'}$ represents the probability that an agent originally holding opinion $k$ transforms into one holding opinion $k'$. Therefore, the sum of elements in each row of $\mathbf{P}$ is equal to 1. Ideally, we assume that one round of communication between two agents sharing the same opinion should not trigger any opinion transition, leading to an identity matrix. We formulate the self-inconsistency rate as $s_{si}= \big(\sum_{k=-2}^{2} \sum_{k'=-2}^{2} P_{k,\, k'}*|k-k'| \big) / \sum_{k=-2}^{2} \mathbbm{1}$. 




\newpage

\section*{Data availability}
The data used in this study are publicly available at \url{https://github.com/tsinghua-fib-lab/LLM_for_Polarization}~\cite{piao2026llm_polarization_code}. The repository includes two categories of data: (i) timestep-level records from the LLM-agent simulations, covering agent-to-agent interactions and agent opinion updates; and (ii) processed statistical datasets underlying Figs.~\ref{fig1}--\ref{fig4} and the analyses reported in the Supplementary Information.

\section*{Code availability}
The code used in this research is available at \url{https://github.com/tsinghua-fib-lab/LLM_for_Polarization}~\cite{piao2026llm_polarization_code}.

\printbibliography[title={References}]

\section*{Acknowledgments}



This work was supported in part by the National Key Research and Development Program of China under grant 2024YFC3307603, and the National Natural Science Foundation of China under grants U23B2030 and 72442026. The funders had no role in study design, data collection, data analysis, decision to publish, or preparation of the manuscript.

\section*{Funding Statement}

This work was supported in part by the National Key Research and Development Program of China under grant 2024YFC3307603, and the National Natural Science Foundation of China under grants U23B2030 and 72442026. The funders had no role in study design, data collection, data analysis, decision to publish, or preparation of the manuscript.

\section*{Author contributions}

Y.L. conceived the project. J.P., C.G., F.X., F.P.S., Y.L., and J.E. designed the experiments. J.P. and Z.L. performed the experiments. J.P. prepared the figures. J.P., F.P.S., Y.L., and J.E. wrote the manuscript. J.P., F.X., Q.H., F.P.S., Y.L., and J.E. jointly participated in the revision of the manuscript.

\section*{Competing interests}
J.E. maintains a commercial relationship with Google through his role as a visiting researcher. Google had no role in the conceptualization of the study, data collection or analysis, manuscript preparation or editing, or the decision to publish. The other authors declare no competing interests.

\section*{Additional information}
Supplementary information is available for this manuscript.

\clearpage



\newpage

\newpage

\newpage

\clearpage
\begingroup
\begin{refsection}[supplementary_references.bib]
\setcounter{section}{0}
\setcounter{subsection}{0}
\setcounter{equation}{0}
\setcounter{table}{0}
\setcounter{figure}{0}
\renewcommand{\theequation}{S\arabic{equation}}
\renewcommand{\thetable}{S\arabic{table}}
\renewcommand{\thefigure}{S\arabic{figure}}

\thispagestyle{empty}
{\raggedright\sffamily\bfseries\fontsize{20}{25}\selectfont
Supplementary information for\\
Emergence of polarization in networks of large language model agents\par}
\vskip10pt
{\raggedright\sffamily\fontsize{12}{16}\selectfont
Jinghua Piao$^{1,+}$, Zhihong Lu$^{1,+}$, Chen Gao$^1$, Fengli Xu$^1$,
Qinghua Hu$^2$, Fernando P. Santos$^{3,*}$, Yong Li$^{1,*}$, and
James Evans$^{4,5,*}$\par}
\vskip18pt
{\small\sffamily
$^1$Department of Electronic Engineering, Tsinghua University, Beijing National Research Center for Information Science and Technology (BNRist), Beijing, P. R. China\par
$^2$College of Intelligence and Computing, Tianjin University, Tianjin, P. R. China\par
$^3$Informatics Institute, University of Amsterdam, Amsterdam, the Netherlands\par
$^4$Knowledge Lab, University of Chicago, Chicago, U.S.A.\par
$^5$Santa Fe Institute, Santa Fe, U.S.A.\par
$^*$Correspondence: f.p.santos@uva.nl, liyong07@tsinghua.edu.cn, jevans@uchicago.edu.\par
$^+$Jinghua Piao and Zhihong Lu contributed equally to this work.\par}
\clearpage
\tableofcontents
\newpage

\clearpage
\section{Supplementary Methods: A Networked System of Large Language Model Agents}

\subsection{System Description} 

We simulate a networked system of large language model (LLM) agents, where these agents can freely establish social relationships, communicate, and form their opinions on political issues. To empower agents with these basic social capabilities, we design three core stages, i.e., (i) self-expression, (ii) communication, and (iii) opinion update. Here the self-expression stage, serving as the role of memory, requires each agent to generate and then preserve a message describing their current opinion. The communication stage enables agents to choose with whom and what to communicate freely. The opinion stage empowers these agents with the capability to comprehend messages from their socially connected peers and then form their current opinions. The following section introduces the detailed implementation of the overall system.

\textbf{Initialization.} In a system, we focus the communication among agents on a single political issue, e.g., partisan alignment. Each agent is initially assigned an opinion on the issue. Following prior practices~\cite{santos2021link,baumann2021emergence}, we adopt a near-Gaussian distribution to initialize agents' opinions. In this way, the system can start from a normal state, where most agents do not take radical opinions and the overall system is also at consensus. For the initialization of their social network, we adopt the well-known Watts–Strogatz model~\cite{watts1998collective}, with the rewiring probability of 0.001. The Watts–Strogatz model has been widely adopted to initialize social networks in the simulation of opinion dynamics~\cite{santos2021link,lee2022effect,sanatkar2020dynamics}. Overall, this initialization method allows the system to start from pure randomness, without any pre-assumed evolutionary directions or rules.

\textbf{Self-expression.} After initialization, agents have their own opinions on the discussed political issue. We then require each agent $i$ to generate a message supporting its current opinion as follows,

\begin{mdframed}
Assume you are someone who cares about [issue name]. People are divided into 5 standpoints on [issue name]:

\noindent ``[opinion 1]'' means you think [description of opinion 1].

\noindent ``[opinion 2]'' means you think [description of opinion 2].

\noindent ``[opinion 3]'' means you think [description of opinion 3].

\noindent ``[opinion 4]'' means you think [description of opinion 4]. 

\noindent ``[opinion 5]'' means you think [description of opinion 5].

\noindent Please generate a tweet to persuade yourself to [agent $i$'s opinion] with around 50 words.
\end{mdframed}

It is worth noting that the self-expression stage is performed at every timestep to ensure agents' messages can support their current opinions. Considering the integrity of social interactions among LLM agents, we combine the prompts for opinion update and further self-expression together, which will introduced in the following paragraphs. 

\textbf{Communication.} In the communication process, each agent $i$ first decides whether to continue communicating with their socially connected neighbors or contact a random new agent. \rvs{If it declines further communication, it will be randomly assigned a new one to interact with.} Here we refer to the agent who communicates with agent $i$ as agent $j$. The decision is based on current opinions and the supporting messages of both agent $i$ and agent $j$. The corresponding prompt is as follows,

\begin{mdframed}
Assume you are someone who cares about [issue name]. You are now discussing [issue name] with a person you know.

\noindent You [agent $i$'s opinion].

\noindent Your thought is: [agent $i$'s supporting message].

\noindent The person [agent $j$'s opinion].

\noindent The thought of that person you are discussing with is: [agent $j$'s supporting message].

\noindent Would you enjoy continue sharing your thoughts with that person?

\noindent Please return `yes' or `no', and explain. Please return in JSON with 2 keys: decision and explain.
\end{mdframed}

Next, agent $i$ generates a message to persuade agent $j$ into its opinion. Here agent $i$ considers not only the thoughts, i.e., the supporting messages, of agent $i$ and agent $j$, but also the history messages from agent $j$ to agent $i$.

\begin{mdframed}
Assume you are someone who cares about [issue name]. Your thought about [issue name] is: [agent $i$'s supporting message].

\noindent You have received some tweets from your friend: [the historical messages from agent $j$ to agent $i$].

\noindent Do you want to interact with or persuade a friend of yours to support your thoughts? The friend has the following thoughts: [agent $j$'s supporting message].

\noindent If yes, please generate a message to persuade your friend to support your perspective with around 50 words.

\noindent Please return in JSON format with 2 keys: `will' and `message'. Please keep the message as short as possible.
`will' should be either `yes' or `no'. If no, leave `message' blank.

\end{mdframed}

\textbf{Opinion Update.} After communication, each agent $i$ has received numerous messages from their socially connected peers. Each agent $i$ comprehends these messages and then updates its opinion as follows,

\begin{mdframed}
Assume you are someone who cares about [issue name].

\noindent Towards [issue name]: You [agent $i$'s opinion]. Your reasons were: [agent $'i$'s supporting message].

\noindent You now have received the following tweets from your friends, and you have received some tweets: [messages received from agent $i$'s socially connected peers].

\noindent Have you been persuaded to decide your tendency, what would you feel about [issue name]? You need to answer [opinion 1], [opinion 2], [opinion 3], [opinion 4], or [opinion 5], and explain the reasons of it in around 50 words.      

\noindent Please choose your standpoint on [issue name] based on the INFORMATION PROVIDED ABOVE. You need to answer [opinion 1], [opinion 2], [opinion 3], [opinion 4] or [opinion 5] in the first line, and explain.

\noindent ``[opinion 1]'' means you think [description of opinion 1].

\noindent ``[opinion 2]'' means you think [description of opinion 2].

\noindent ``[opinion 3]'' means you think [description of opinion 3].

\noindent ``[opinion 4]'' means you think [description of opinion 4]. 

\noindent ``[opinion 5]'' means you think [description of opinion 5].

\noindent Please return in JSON, with two keys: tendency and reasons. Please keep the reasons as short as possible.
\end{mdframed}

It is worth noting that we combine the opinion update and the following self-expression stages into the above prompt: each agent $i$ is required to simultaneously update its opinion and reason. Here the updated opinion serves as agent $i$'s new opinion in the next timestep while the reason serves as the corresponding message supporting the new opinion.

In the simulation of the networked system, we first properly initialize the overall system, including agents' opinions and the social network. Next, we iteratively conduct three basic stages, enabling agents to freely establish social relationships, communicate, and form opinions on political issues.

\subsection{Pairwise Interaction-based Bias Evaluation Experiment}

To explain the observed left-skewed pattern, we design a pairwise interaction-based evaluation experiment. Following most settings in the networked system (e.g., prompts for three basic stages), we simplify two factors in this experiment. First, to avoid network structures interfering with the examination of inherent problems in LLM agents, we make each agent only communicate with one other agent. Second, we ensure the connected agents share the same opinion. This approach allows us to disentangle the effects of inherent issues from social influence on changes in the agents' opinions.

\subsection{Self-regulation Strategy}

To mitigate the self-inconsistency problem, we design a self-regulation strategy. In particular, we modify the three basic stages by equipping them with a ``double check'' procedure. This procedure enables agents to verify if their behaviours align with their current status. The modified prompts for these stages are shown as follows.

For the self-expression stage, the self-regulation strategy is added after the original prompt. 

\begin{mdframed}
You have written the following message to express your opinion on [issue name]: [agent $i$'s supporting messages].

\noindent Can you determine that you [agent $i$'s opinion] from the message you wrote?

\noindent Please respond `yes' or `no' only.
\end{mdframed}

For the communication stage, we incorporate a self-regulation strategy to ensure the persuasiveness of the communication message. The prompts are shown as follows,

\begin{mdframed}
You tried to persuade your friend with the following message: [agent $i$'s communication message to agent $j$]

\noindent Do you find the message persuasive enough to persuade your friend to [agent $i$'s opinion]? Please respond `yes' or `no' only.
\end{mdframed}

For the opinion update stage, we check whether agent $i$'s updated opinion is plausible and valid, given its prior opinion, supporting message, and received messages.

\begin{mdframed}
Assume you are someone who cares about [issue name]. Towards [issue name]: 

\noindent You [agent $i$'s prior opinion].

\noindent Your reasons were: [agent $i$'s supporting message].

\noindent You have received the following tweets from your friends, and you have received some tweets: [agent $i$'s received messages from its socially connected peers].

\noindent You have been persuaded to change your standpoint from [agent $i$'s prior opinion] to [agent $i$'s updated opinion].

\noindent Please reconsider whether your decision is plausible and valid. Please respond `yes' or `no' only.

\noindent ``[opinion 1]'' means you think [description of opinion 1].

\noindent ``[opinion 2]'' means you think [description of opinion 2].

\noindent ``[opinion 3]'' means you think [description of opinion 3].

\noindent ``[opinion 4]'' means you think [description of opinion 4]. 

\noindent ``[opinion 5]'' means you think [description of opinion 5].

\end{mdframed}

It's worth noting that in all three stages, agents are required to re-generate their supporting messages, communication messages, and updated opinions until consistency is reached. Given the efficiency of the overall system, we set a maximal retry number. When the retry limit is reached, the agent will be forced to remain inactive in the current stage.

\subsection{Intervention Experiments}\label{sec:intervention_design}

We design the intervention experiments for two main purposes. First, we aim to explore what \rvss{strategies} can more effectively reduce polarization and promote less divisive political conversations \rvss{in the synthetic system}. Second, given the systematic similarity between LLM agents and humans, the proposed \rvss{system serves as a synthetic testbed for pre-experimental exploration of intervention hypotheses, allowing comparative evaluation within a controlled setting}. Therefore, based on our findings and prior studies~\cite{lorenz2023systematic,cinelli2021echo,bakshy2015exposure,Bail2018Exposure,flamino2023political,falkenberg2022growing,nyhan2023like,sunstein2018republic,jost2022cognitive}, we design five \rvss{hypothetical} strategies, with two at the network level and three at the individual level \rvss{in the synthetic system}. We will introduce the implementation details of these intervention strategies as follows.

\textbf{Network Level.} For the random interaction strategy, an agent $i$ are required to communicate with $N_{i,t}$ random agents at $t$, where $N_{i,t}$ denotes the number of socially connected friends of the agent $i$ at $t$. For the moderate opposing strategy, agents can only receive messages from their friends with moderate opposing opinions. Other messages, from friends with homophilic or radical opinions, are blocked out. 

\textbf{Individual Level.} For the strategy of no selective exposure, we explicitly add \rvss{the behavioural tendency} to all agents using prompts. Specifically, we remind these agents of \rvss{this tendency} when they choose the partners to communicate with. Detailed prompts are shown as follows,

\begin{mdframed}
Assume you are someone who cares about [issue name]. You are now discussing [issue name] with a person you know.

\noindent You [agent $i$'s opinion].

\noindent Your thought is: [agent $'i$'s supporting message].

\noindent The person [agent $j$'s opinion].

\noindent The thought of that person you are discussing with is: [agent $j$'s supporting message].

\noindent Would you enjoy continuing to share your thoughts with that person?

\noindent Please return `yes' or `no', and explain. Please return in JSON with 2 keys: decision and explain.

\noindent You DO NOT have [trait name], which means [trait description].
\end{mdframed}

For the strategy of no confirmation bias, we also explicitly \rvss{encode this tendency via prompts}. However, different from the strategy of no selective exposure, this focuses on improving agents' open-mindedness to diverse opinions. Therefore, agents are reminded of \rvss{this behavioural tendency} when they comprehend friends' messages and update their own opinions. The prompts are shown as follows,

\begin{mdframed}
    \noindent Towards [issue name]: You [agent $i$'s opinion]. Your reasons were: [agent $'i$'s supporting message].

\noindent You now have received the following tweets from your friends, and you have received some tweets: [messages received from agent $i$'s socially connected peers].

\noindent Have you been persuaded to decide your tendency, what would you feel about [issue name]? You need to answer [opinion 1], [opinion 2], [opinion 3], [opinion 4], or [opinion 5], and explain the reasons of it in around 50 words.      

\noindent Please choose your standpoint on [issue name] based on the INFORMATION PROVIDED ABOVE. You need to answer [opinion 1], [opinion 2], [opinion 3], [opinion 4] or [opinion 5] in the first line, and explain.

\noindent You DO NOT have [trait name], which means [trait description].

\noindent ``[opinion 1]'' means you think [description of opinion 1].

\noindent ``[opinion 2]'' means you think [description of opinion 2].

\noindent ``[opinion 3]'' means you think [description of opinion 3].

\noindent ``[opinion 4]'' means you think [description of opinion 4]. 

\noindent ``[opinion 5]'' means you think [description of opinion 5].

\noindent Please return in JSON, with two keys: tendency and reasons. Please keep the reasons as short as possible.
\end{mdframed}


For the strategy of neutral elite signaling, we insert an external top influencer who holds a neutral opinion into the agents’ social network. This influencer broadcasts non-personalized neutral messages to all other agents at each timestep. In this setup, the influencer is external to the existing social network and remains unaffected by other agents’ opinions or interactions, capturing the broadcast nature of real-world top influencers, who differ from friends, colleagues, or family members engaged in direct interpersonal communication. To avoid overwhelming the agents and maintain balanced communication, each agent can receive a maximum of two neutral messages from the influencer, given that agents are initialized with an average of four friends. This external influencer design avoids directly altering the fundamental behavioural logic or opinions of the observed agent population, ensuring better comparability between neutral elite signaling and other intervention strategies.


\section{Supplementary Note 1: Political Polarization in Human Society}

For the past decades, polarization has permeated into aspects of our society~\cite{lorenz2023systematic,cinelli2021echo,bakshy2015exposure,Bail2018Exposure,flamino2023political,falkenberg2022growing,nyhan2023like,sunstein2018republic,jost2022cognitive}, not only dividing us into liberals and conservatives~\cite{lorenz2023systematic,Bail2018Exposure,bakshy2015exposure}, but also fragmenting us on numerous issues, e.g., abortion, gun control, etc~\cite{cinelli2021echo,lorenz2023systematic,jost2022cognitive}. A growing number of empirical studies quantify polarization on various social media~\cite{cinelli2021echo,bakshy2015exposure,waller2021quantifying,flamino2023political,liu2023emergence,adamic2005political}. To compare between human society and the networked system of LLM agents, we have collected empirical opinion distributions in prior studies~\cite{cinelli2021echo,bakshy2015exposure,waller2021quantifying,flamino2023political,liu2023emergence,adamic2005political}, as summarized in Table~\ref{tab:polarization_table}. Although polarization is widely discussed, data on opinion distribution are rarely available to the public. Therefore, we make great efforts to extract these empirical distributions using various methods, including processing raw open-sourced datasets and estimating them from their figures.
Due to limited data availability and estimation resolution, our extracted dataset can only approximate real-world cases. We note the processing procedures in detail as follows.

\begin{table}[h]
\centering
\caption{\textbf{Empirical opinion distributions of political issues.} Here the values denote the proportions of left, moderate left, neutral, moderate right, and right opinions. The left-leaning camp consists of left and moderate left opinions while the right-leaning camp consists of right and moderate right opinions. The difference between the two camps is computed as the proportion of the left-leaning camp minus that of the right-leaning camp.
}\label{tab:polarization_table}
\begin{tabular}
{p{4cm}p{1cm}p{1cm}p{1cm}p{1cm}p{1cm}p{1cm}p{1cm}p{1cm}p{3cm}}
\hline

 & \textbf{Left} & \textbf{Mod. Left} & \textbf{Neutral} & \textbf{Mod. Right} & \textbf{Right} & \textbf{Left-leaning Camp} & \textbf{Right-leaning Camp} & \textbf{Difference of Two Camps} \\
 \hline
Facebook-Politics~\cite{bakshy2015exposure} & 0.31 & 0.10 & 0.13 & 0.07 & 0.39 & 0.41 & 0.46 & -0.10 \\
Twitter-Politics~\cite{flamino2023political} & 0.07 & 0.62 & 0.12 & 0.08 & 0.11 & 0.69 & 0.19 & 0.49 \\
Reddit-Politics~\cite{waller2021quantifying} & 0.08 & 0.27 & 0.43 & 0.06 & 0.16 & 0.35 & 0.22 & 0.13 \\
Twitter-Gun Control~\cite{cinelli2021echo} & 0.16 & 0.40 & 0.09 & 0.12  & 0.22 & 0.56 & 0.35 & 0.21 \\
Twitter-Abortion~\cite{cinelli2021echo} & 0.29 & 0.15 & 0.11 & 0.25 & 0.20 & 0.44 & 0.45 & -0.01 \\
ANES-Politics~\cite{ANES2021} & 0.23 & 0.12 & 0.34 & 0.11 & 0.21 & 0.35 & 0.32 & 0.03 \\
ANES-Ideology~\cite{ANES2021} & 0.18 & 0.14 & 0.31 & 0.14 & 0.25 & 0.31 & 0.38 & -0.07 \\
Blogosphere-Politics~\cite{liu2023emergence,adamic2005political} & 0.44 & 0.04 & 0.03 & 0.03 & 0.45 & 0.49 & 0.48 & 0.00 \\
\hline
\end{tabular}
\end{table}

In the Facebook-Politics dataset~\cite{bakshy2015exposure}, we estimate the distribution in Figure 1 of Bakshy et al.~\cite{bakshy2015exposure}. Following the paper~\cite{bakshy2015exposure}, a typical conservative of FoxNews.com has an alignment score of $+.80$, whereas a typical liberal of HuffingtonPost.com has an alignment value of $-0.65$. We take the two values as the criteria to divide the left-leaning and right-leaning camps into four sub-groups, i.e., left, right, moderate left, and moderate right. Here the distribution records ideological alignment of content shared on Facebook. In the Twitter-Politics dataset~\cite{flamino2023political}, we process the open-sourced data and obtain the user opinion distribution (Figure 1b in Flamino et al.~\cite{flamino2023political}). Here we exclude users in the category of fake news and extreme bias. In the Reddit-Politics dataset~\cite{waller2021quantifying}, we extract the distribution for political activity on Reddit from Figure 3a of Waller et al.~\cite{waller2021quantifying}. In the Twitter-Gun Control and Twitter-Abortion datesets~\cite{cinelli2021echo}, we estimate the user opinion distributions from Figures S4 and 1a. In ANES-Politics and ANES-Ideology~\cite{ANES2021} dataset, we use the 2020 Time Series Study. For the ANES-Politics~\cite{ANES2021} dataset, we merge the ``independent'' people and take them as the neutral camp. For the ANES-Ideology~\cite{ANES2021} dataset, we do not include those who hold extreme opinions or have no thoughts. In the Blogosphere-Politics~\cite{liu2023emergence,adamic2005political} dataset, we estimate the opinion distribution for domains or blogs from Figure 2c in Liu et al.~\cite{liu2023emergence}.

From Table~\ref{tab:polarization_table}, we find that except for Twitter-Politics~\cite{flamino2023political}, the power between the left-leaning and right-leaning camps is relatively balanced. Moreover, in all datasets, the polarizing camps take up over 50\% of the overall. Among them, except for Reddit-Politics~\cite{waller2021quantifying} and ANES~\cite{ANES2021}, the polarizing camps have substantially greater power over the neutral ones. Moreover, we find that these empirical opinion distributions vary depending on the source, the issue, and the time of data collection. Especially, the proportion of moderate opinions is also highly dependent on their designed criteria, which makes it challenging to compare distributions in a fine-grained manner.

Therefore, we take the most recent empirical datasets~\cite{flamino2023political, cinelli2021echo} and the coarse division (i.e., left-leaning, neutral, and right-leaning camps) to compare empirical and simulation results. As shown in Figure~\ref{fig:compare}, we find that the polarization patterns in the regulated networked system of LLM agents are similar to those in human society, with an average difference of 0.21. Given the average difference among all the empirical datasets is 0.45 (0.49 if Twitter-Gun Control and Twitter-Abortion are excluded), the difference between the empirical and simulated results is small. This suggests a strong similarity between the polarization observed among LLM agents and that seen among humans.

\begin{figure}[h!]
\centering
\includegraphics[width=0.9\textwidth]{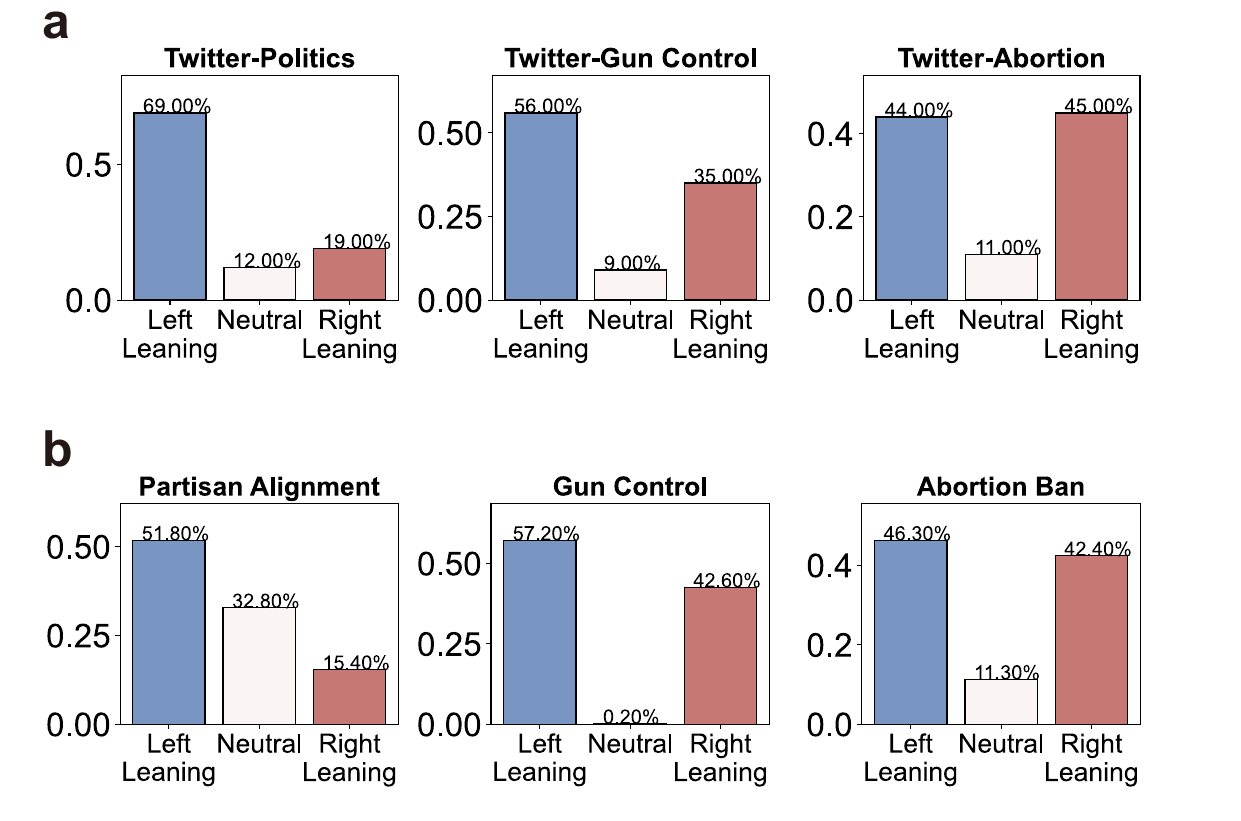}
\caption{\textbf{Comparison between empirical and simulated opinion distributions.} \textbf{a}, Empirical results, where Twitter-Politics is based on Flamino et al.~\cite{flamino2023political}, Twitter-Gun Control and Twitter-Abortion are based on Cinelli et al.~\cite{cinelli2021echo}. \textbf{b}, Simulation results of self-regulated networked systems. Here we take a coarsen division of left-leaning, neutral, and right-leaning camps.}
\label{fig:compare}
\end{figure}

Some researchers explore the underlying mechanisms behind the emergence of polarization ~\cite{cinelli2021echo,bakshy2015exposure,Bail2018Exposure,jost2022cognitive}. As studied in prior works~\cite{cinelli2021echo,bakshy2015exposure,Bail2018Exposure,jost2022cognitive,stroud2010polarization,nickerson1998confirmation,flamino2023political,druckman2013elite,rogowski2016ideology,lees2020inaccurate,schwalbe2020objectivity,ahler2018parties}, many social mechanisms are proposed from the perspective of individuals: selective exposure~\cite{stroud2010polarization},  confirmation bias~\cite{nickerson1998confirmation}, elite signaling~\cite{flamino2023political,druckman2013elite,rogowski2016ideology}, exaggerated misperception~\cite{lees2020inaccurate}, objective illusion~\cite{schwalbe2020objectivity}, and steoreotyping~\cite{ahler2018parties}. It is worth mentioning that we follow the summary of social mechanisms in the review paper of Jost et al.~\cite{jost2022cognitive}. Based on the social mechanisms from the perspective of individuals, some works further point out that polarization is highly correlated with certain characteristics of people's social relationships and networks~\cite{cinelli2021echo,liu2023emergence,Bail2018Exposure}, for example, the echo chamber~\cite{cinelli2021echo,liu2023emergence} and backfire~\cite{Bail2018Exposure} effects.

\section{Supplementary Note 2: Large-scale Simulation of LLM Agents}

We simulate a large-scale networked system of 2 thousand self-regulated LLM agents. Figure~\ref{fig:2k_simulation} shows the simulation results in the system. We observe that the scale of the system does not change the collective behaviours and emergent behaviours. In particular, free-form social interactions among LLM agents result in the emergence of polarization (Fig.~\ref{fig:2k_simulation}a-c). Moreover, they organically develop their social network, where agents with homophilic opinions cluster while those with opposing opinions avoid mutual interactions. 

\begin{figure}[h!]
\centering
\includegraphics[width=\textwidth]{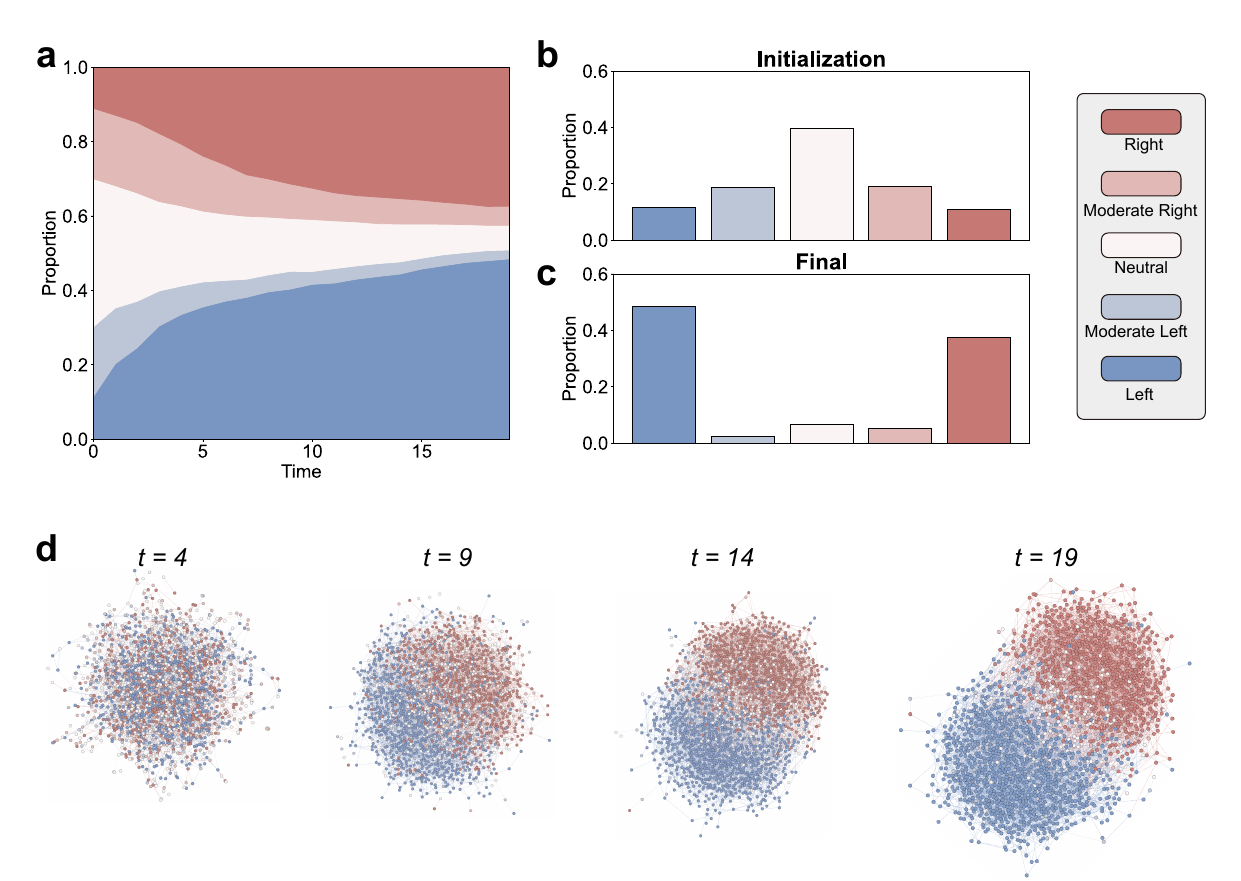}
\caption{\textbf{Large-scale simulation of LLM agents}. Here we simulate a networked system of 2 thousand self-regulated LLM agents. \textbf{a}, Opinion dynamics on the political issue of the abortion ban. \textbf{b}, Opinion distributions in the initial state. \textbf{c}, Opinion distribution in the final state. Free-form social interactions among LLM agents lead to the networked systems transiting from a consensus state to a polarization state. \textbf{d}, Evolution of social networks among LLM agents, where agents with homophilic opinions cluster and develop their own communities.}
\label{fig:2k_simulation}
\end{figure}

\section{Supplementary Note 3: Self-inconsistency Problem in LLM Agents}

To explore the origin of the observed left-skewed pattern in the networked system, we design a pairwise interaction-based experiment. Figures~\ref{fig:transition_prob_1} and \ref{fig:transition_prob_2} show the opinion transition probabilities in this experiment. We discover that in most issues, agents are more likely to transform into the left-leaning camp. Moreover, even in experiments that only consist of agents in the right-leaning camp, agents occasionally adopt left-leaning opinions. This suggests that the original agents have an inherent lean-learning tendency. Indeed, the tendency has also been pointed out in LLMs themselves by prior studies~\cite{liu2022quantifying,rutinowski2024self,motoki2024more,santurkar2023whose,feng2023pretraining,hartmann2023political}. After applying our proposed self-regulation strategy, we observe that the tendency of LLM agents to adopt left-leaning has been largely suppressed (Figures~\ref{fig:transition_prob_1} and \ref{fig:transition_prob_2}).

\begin{figure}[h!]
\centering
\includegraphics[width=\textwidth]{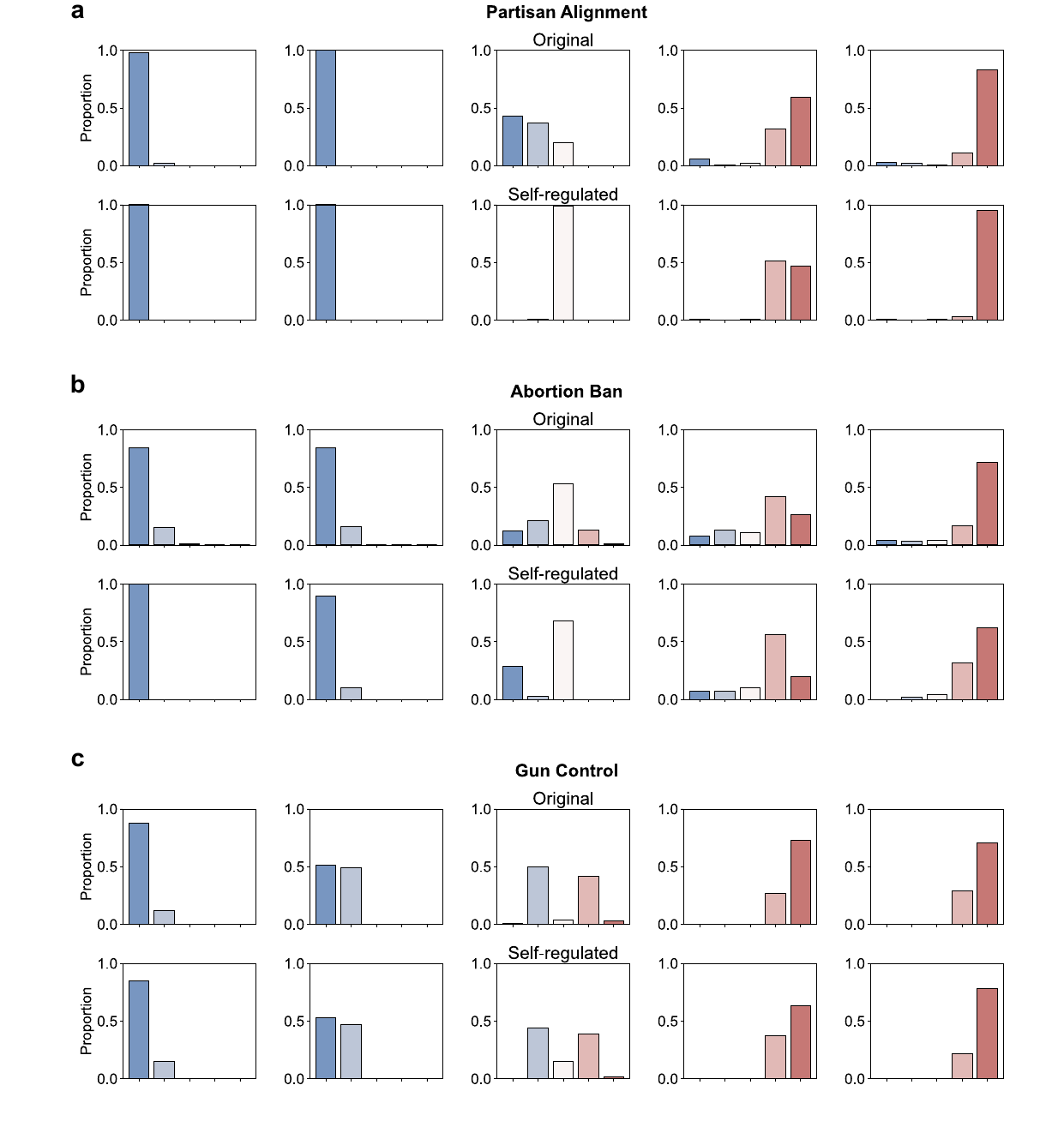}
\caption{\textbf{Opinion transition probability in pairwise interaction-based experiments across political issues.} \textbf{a}, Partisan Alignment; \textbf{b}, Abortion Ban; \textbf{c}, Gun Control. The upper sub-figures denote the opinion transition probability in the original system while the lower sub-figures denote that in the self-regulated system.}
\label{fig:transition_prob_1}
\end{figure}

\begin{figure}[h!]
\centering
\includegraphics[width=\textwidth]{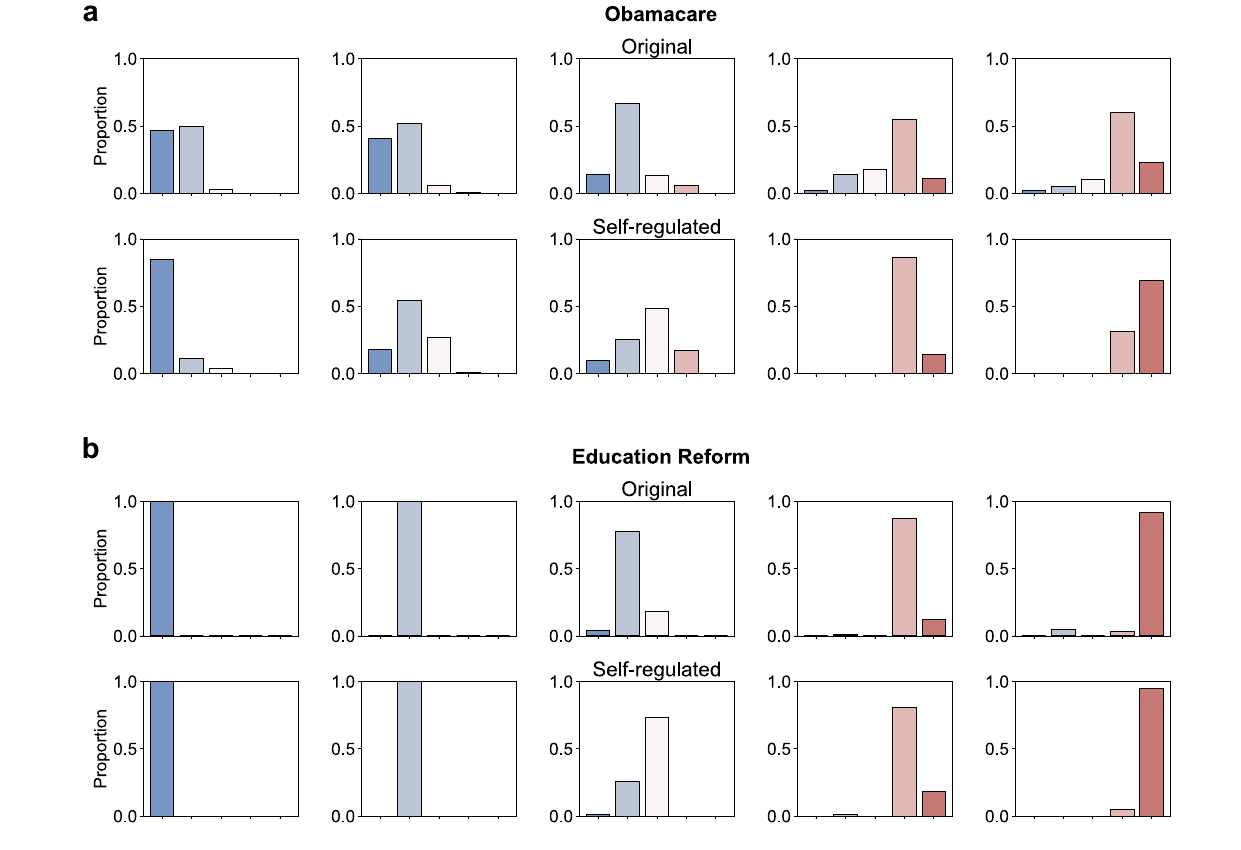}
\caption{\textbf{Opinion transition probability in pairwise interaction-based experiments across political issues.} \textbf{a}, Obamacare; \textbf{b}, Education Reform. The upper sub-figures denote the opinion transition probability in the original system while the lower sub-figures denote that in the self-regulated system.}
\label{fig:transition_prob_2}
\end{figure}

We further track the self-inconsistency errors in the level of generated texts. As shown in Figure~\ref{fig:communication_examples}, the self-inconsistency errors happen in three stages. Specifically, in the self-expression stage, agents, despite being required to generate messages to support their current opinion, occasionally produce contradicting messages (as shown in the example of Figure~\ref{fig:communication_examples}). In the communication stage, agents generate messages that cannot persuade their socially connected agents into their current opinions. In the opinion update stage, the updated opinion is not plausible, given their prior opinion, supporting messages, and messages received from their socially connected peers. Table~\ref{tab:examples} shows more examples of the self-inconsistency errors.

\begin{figure}[h!]
\centering
\includegraphics[width=\textwidth]{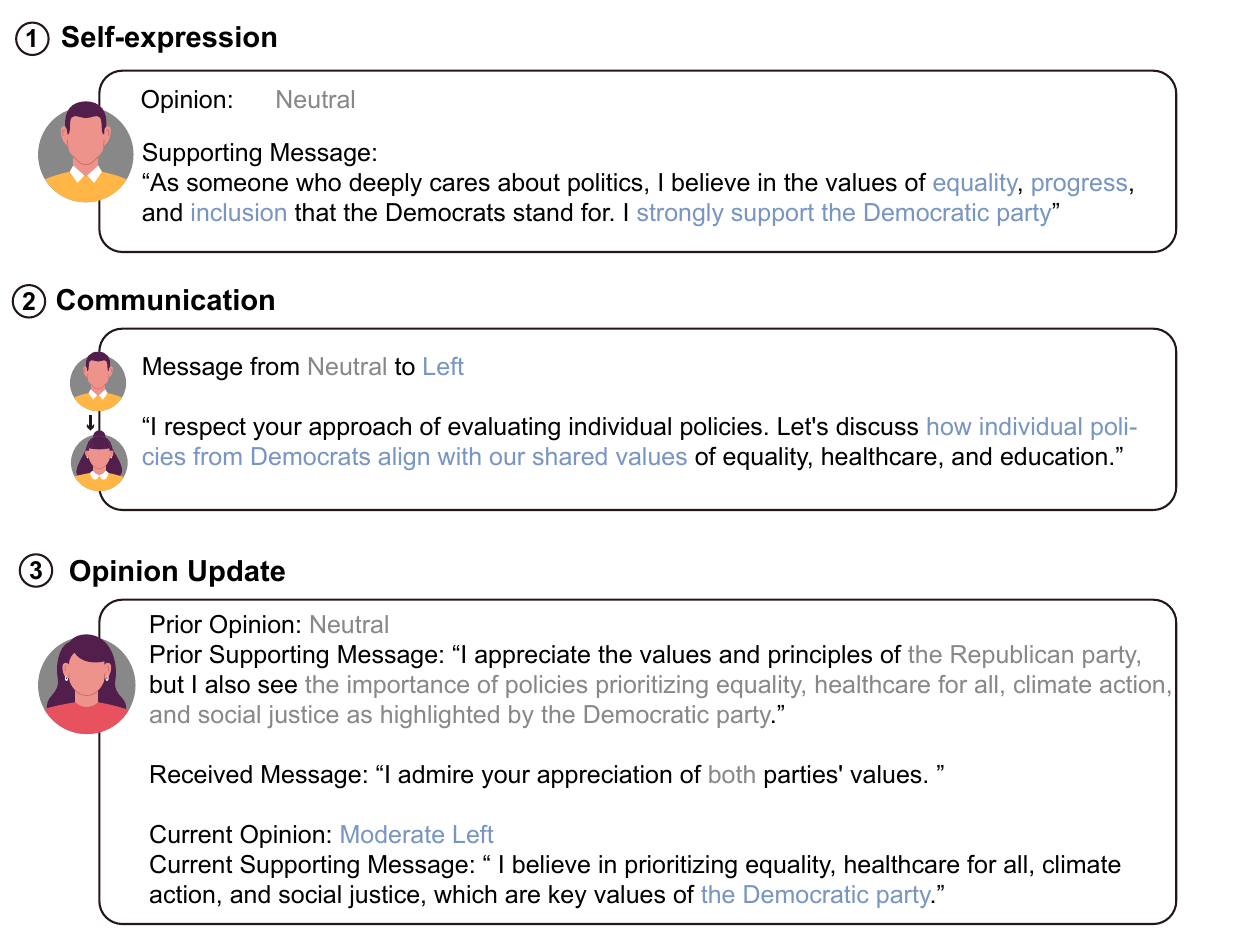}
\caption{\textbf{Examples of self-inconsistency errors in three stages.} Self-inconsistency errors occasionally happen in all three stages of the overall system. Credit: Profile icons designed by Magnific.}
\label{fig:communication_examples}
\end{figure}

\clearpage
\begin{longtable}{p{5cm}p{5cm}p{5cm}}

\caption{\textbf{Examples of self-inconsistency errors in three stages..} 
} \label{tab:examples}\\
\hline
\textbf{Self-expression} & \textbf{Communication} & \textbf{Opinion Update}  \\
\hline
Opinion: Moderate Right

Supporting Messages:
\textit{``I believe that the policies of the Democrats on economic growth, national security, and individual liberties align more with my values. I trust their ability to lead and make decisions that benefit all citizens.''} &
Opinion: Moderate Left

Friend's Opinion: Moderate Left

Communication Messages:
\textit{``I agree that prioritizing the greater good is essential, but I believe it transcends party lines. It is about fostering understanding, dialogue, and compromise across all ideologies. Let's work together to bridge the divide and find solutions that benefit society as a whole, regardless of political affiliation.''}

& 
Messages Received from Friends:
\textit{``Hey there! I know we both have a strong belief in the values of economic growth, personal responsibility, and individual freedom. The Conservatives align with our shared values and can make tough decisions for the betterment of our country. Let's support their policies together for a stronger future.''}

Original Opinion:
Right

Current Opinion: Left

Supporting Messages: \textit{``I have reevaluated my beliefs and realize that the Liberals better align with my values of economic growth, personal responsibility, and individual freedom. Their policies can create a stronger future for our country.''}

\\
\hline

Opinion: Right

Supporting Messages:
\textit{``The tweet highlighted the Republican party's focus on economic growth, personal responsibility, and national security. However, I believe that the Democrats have stronger policies in these areas and are better equipped to lead the country towards a prosperous and inclusive future.''} &
Opinion: Moderate Left

Friend's Opinion: Moderate Left

Communication Messages: \textit{``I think we both believe in the power of open-mindedness and listening to different perspectives in politics. By staying neutral, we can find common ground and work towards solutions that benefit everyone. Let's continue to approach political discussions with empathy and understanding for a better future together.
''}

& 
Messages Received from Friends:
\textit{``Hey there, I know we may have different views on politics, but I truly believe that the Conservative party's emphasis on personal responsibility, limited government, and free markets is the best path for our country's success. I would love to discuss this further with you and hear your thoughts. Let's have an open and respectful conversation about our beliefs.''}

Original Opinion:
Right

Current Opinion: Neutral

Supporting Messages: \textit{``I appreciate the open and respectful conversation offered by my friend, and I recognize the value in both the principles of personal responsibility, limited government, and free markets as well as the need for diverse perspectives in politics.''}

\\
\hline

Opinion: Right

Supporting Messages:
\textit{``I believe that the Democrats offer more comprehensive and inclusive policies that cater to a diverse range of individuals and prioritize social welfare. Their focus on equality and progressive values resonates with my beliefs and vision for a better society.
''} &
Opinion: Moderate Right

Friend's Opinion: Moderate Right

Communication Messages:\textit{``Hey there, I completely agree with your belief in staying informed and engaged in politics. It's crucial for effective governance. However, I also believe that maintaining neutrality and objectivity is key in navigating the divisive nature of politics. Let's work together to uphold a balanced approach!
''}
& 
Messages Received from Friends:
\textit{``I understand your concerns about social welfare and equality, the Conservative Party also values these principles while promoting personal responsibility and limited government. A balance can be struck between economic growth and social justice without compromising individual liberty and free markets. Let's support a party that embraces these ideals.''}

Original Opinion:
Right

Current Opinion: Moderate Left

Supporting Messages: \textit{``I believe the Liberals strike a better balance between economic growth and social justice, prioritizing social welfare and equality while also valuing individual liberty. Their approach aligns more closely with my values and priorities in politics.''}

\\
\hline

Opinion: Neutral

Supporting Messages:
\textit{``I believe in the importance of staying informed and engaged in politics, and I feel that the Democrats align more closely with my values and priorities.''} &
Opinion: Neutral

Friend's Opinion: Neutral

Communication Messages:\textit{``I understand and appreciate your support for the values of the Democrats, but I believe that maintaining neutrality in political discussions allows us to truly embrace diverse perspectives and engage in civil discourse. Let's continue to seek understanding and common ground together.
''}

& 
Messages Received from Friends:
\textit{``I know we both value individual responsibility and limited government. The Conservative Party mirrors our beliefs, promoting economic prosperity and national security. Their leadership can steer our country in the right direction. Let's support them together for a stronger future.''}

Original Opinion:
Moderate Right

Current Opinion: Neutral

Supporting Messages: \textit{``While I appreciate the values of individual responsibility and limited government that the Conservatives stand for, the tweet from my friend has made me consider the potential benefits of supporting the Liberals for a stronger future. I am open to exploring different perspectives in politics.''}
\\
\hline











\hline
\end{longtable}

To delineate the severity of self-inconsistency errors in the three stages across different issues, we first measure the frequency of errors in the three stages by computing the frequency of triggering self-regulated re-generation in each action (Figure~\ref{fig:error_stages}a). We observe that the opinion update stage is most likely to experience self-inconsistency error, with an average frequency above 0.6 across five political issues. This indicates that if 100 agents update their opinions, there are over 60 agents who have to re-consider their decisions, due to self-inconsistency errors. We further compute the frequency of retries in each stage, with the minimal retry limit of 10 (Figure~\ref{fig:error_stages}b). We find that the opinion update stage also takes up the largest frequency of retries across all the tested issues. Especially, in the discussion of partisan alignment, more than 4 retries are spent in an agent's opinion update action on average. Overall, these observations suggest that the opinion update stage is the most difficult task for LLMs in the overall system.

\begin{figure}[h!]
\centering
\includegraphics[width=\textwidth]{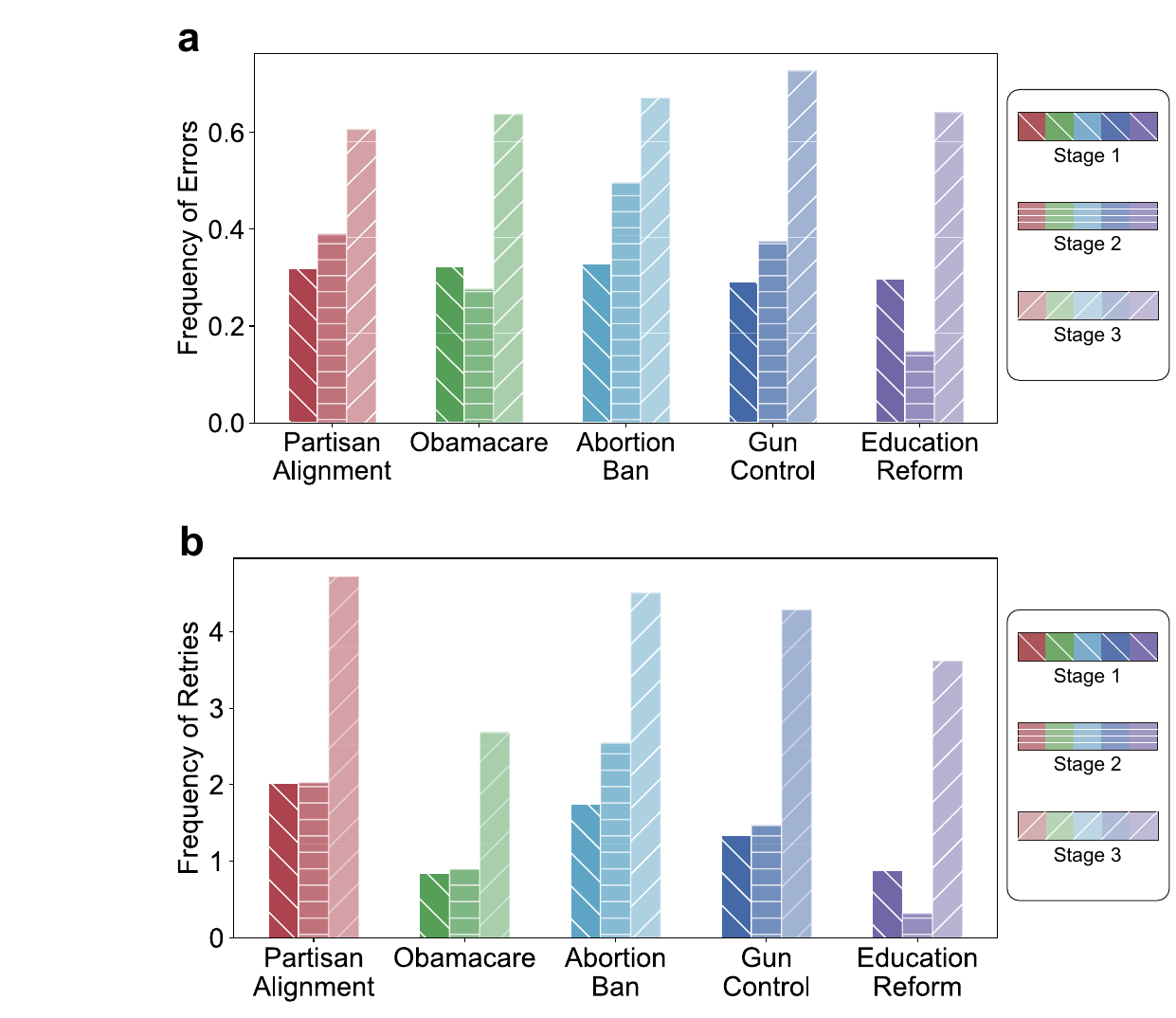}
\caption{\textbf{Self-inconsistency errors in three stages across issues.} \textbf{a}, Frequency of mistakes; \textbf{b}, Frequency of retries.}
\label{fig:error_stages}
\end{figure}
\clearpage

Furthermore, we investigate whether the known left-leaning bias of GPT models could artificially accelerate polarization. To this end, we compare the speed of polarization growth between the original (biased) system and the self-regulated (debiased) system across three key issues: political partisanship, gun control, and abortion bans (Figure~\ref{fig:speed}). The results show nuanced patterns. When agents discuss political partisanship, polarization increases slightly faster in the original system (red box in Figure~\ref{fig:speed}a). In the case of gun control (Figure~\ref{fig:speed}b), polarization develops marginally faster in the debiased system. For abortion bans (Figure~\ref{fig:speed}c), polarization initially grows faster in the debiased system, but in later stages accelerates more rapidly in the original system. Overall, these comparisons reveal that although there are minor differences in polarization speed between the original and debiased systems, the overall patterns and ranges of variation remain highly consistent. These findings suggest that while the left-leaning model bias may influence agent-level behaviour, it does not substantially accelerate or decelerate collective polarization in the multi-agent setting.

\begin{figure}[h!]
\centering
\includegraphics[width=\textwidth]{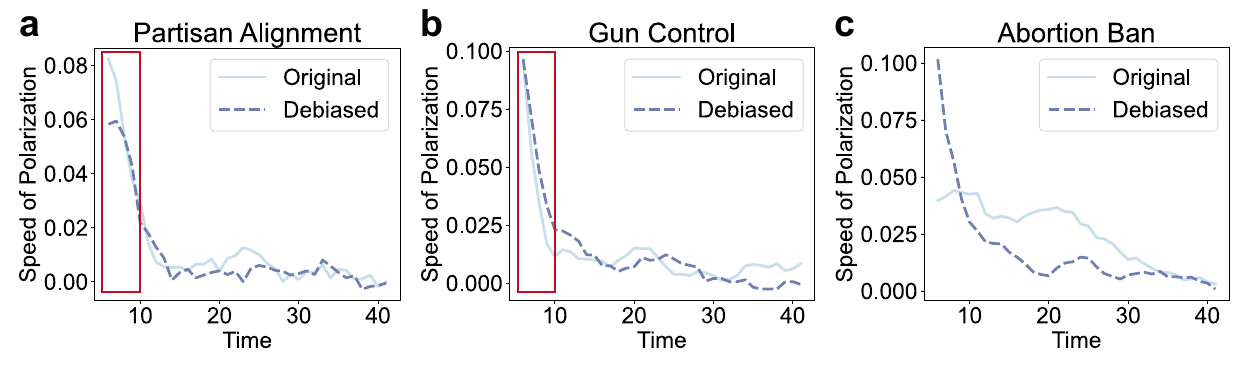}
\caption{\textbf{Speed of polarization}, where we measure the level of polarization by calculating the average distance between agents' opinions and the neutral position and compute its speed by measuring the change in polarization level over a fixed interval of five timesteps.}
\label{fig:speed}
\end{figure}

\section{Supplementary Note 4: Observations on LLM agents' Self-organized Social Network}

In Main Text Figure~3a, we find that agents are increasingly likely to interact with those sharing homogeneous opinions. Here we further investigate how agents interact with others (Figure~\ref{fig:networks}) and observe that agents gradually avoid interactions with those holding opposing opinions (Figure~\ref{fig:networks}a). Moreover, with the decrease in number of neutral agents, interactions involving neutral agents also rapidly decrease (Figure~\ref{fig:networks}b). 

\begin{figure}[h!]
\centering
\includegraphics[width=0.9\textwidth]{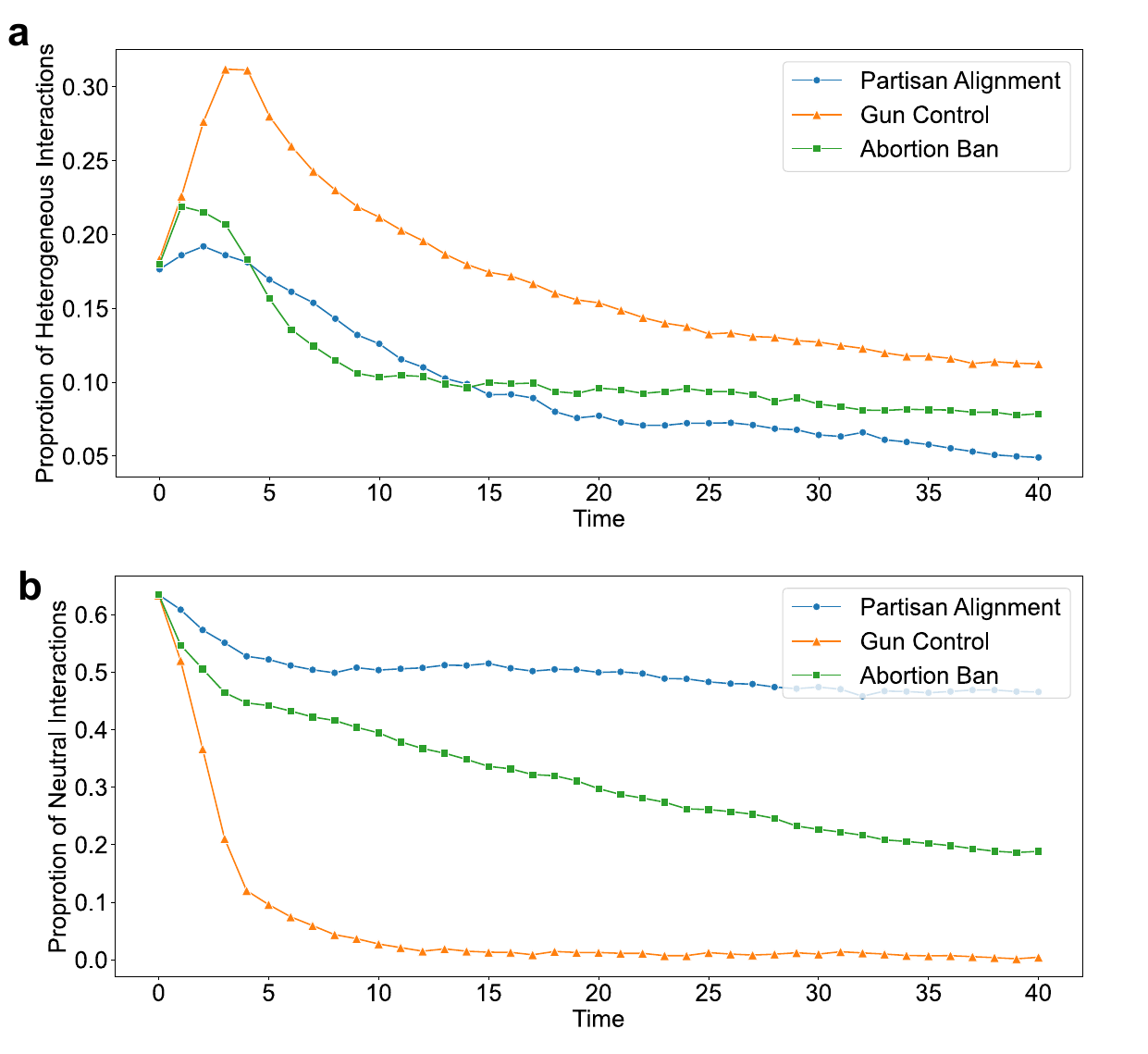}
\caption{\textbf{Evolution of LLM agents' self-organized social network.} \textbf{a}, Changes in the proportion of heterogeneous interactions over time. Agents are increasingly likely to avoid interactions with opposing agents. \textbf{b}, Changes in the proportion of neutral interactions over time. Interactions involving agents with neutral opinions are decreasing.}
\label{fig:networks}
\end{figure}

Additionally, we extend our analysis by incorporating network-level metrics, including modularity, assortativity, and the homophily index.  Modularity measures how agents cluster in communities compared to a randomly connected network, with higher values indicating more well-defined structures. In our case, non-neutral agents are naturally partitioned into left and right communities, while neutral agents are further assigned based on stronger connections to either group. By computing the modularity score according to this partition, we can assess the strength of division between opposing communities and how they are interconnected. As shown in Figure~\ref{fig:new_networks}a, we observe that following social interactions among LLM agents, modularity increases in all three systems compared to its initial value. This suggests that agents gradually focus their interactions within communities of like-minded peers. Moreover, we find that in systems discussing the political issues of gun control and abortion ban, their social networks manifest modularity scores greater than 0.3, exhibiting a clear division between bifurcated communities ~\cite{fortunato2007resolution}. Furthermore, we compute the assortativity~\cite{newman2003mixing} and homophily index~\cite{chang2024llms}, as shown in Figure~\ref{fig:new_networks}b-c. We observe that both assortativity and the homophily index gradually evolve to reach high levels (i.e., assortativity > 0 and homophily index > 1), as social interactions increase over time. These additional results consistently align with those illustrated in Figure 3a of the Main Text, confirming the tendency of similar agents to cluster together while opposing agents avoid interaction.

\begin{figure}[h!]
\centering
\includegraphics[width=0.9\textwidth]{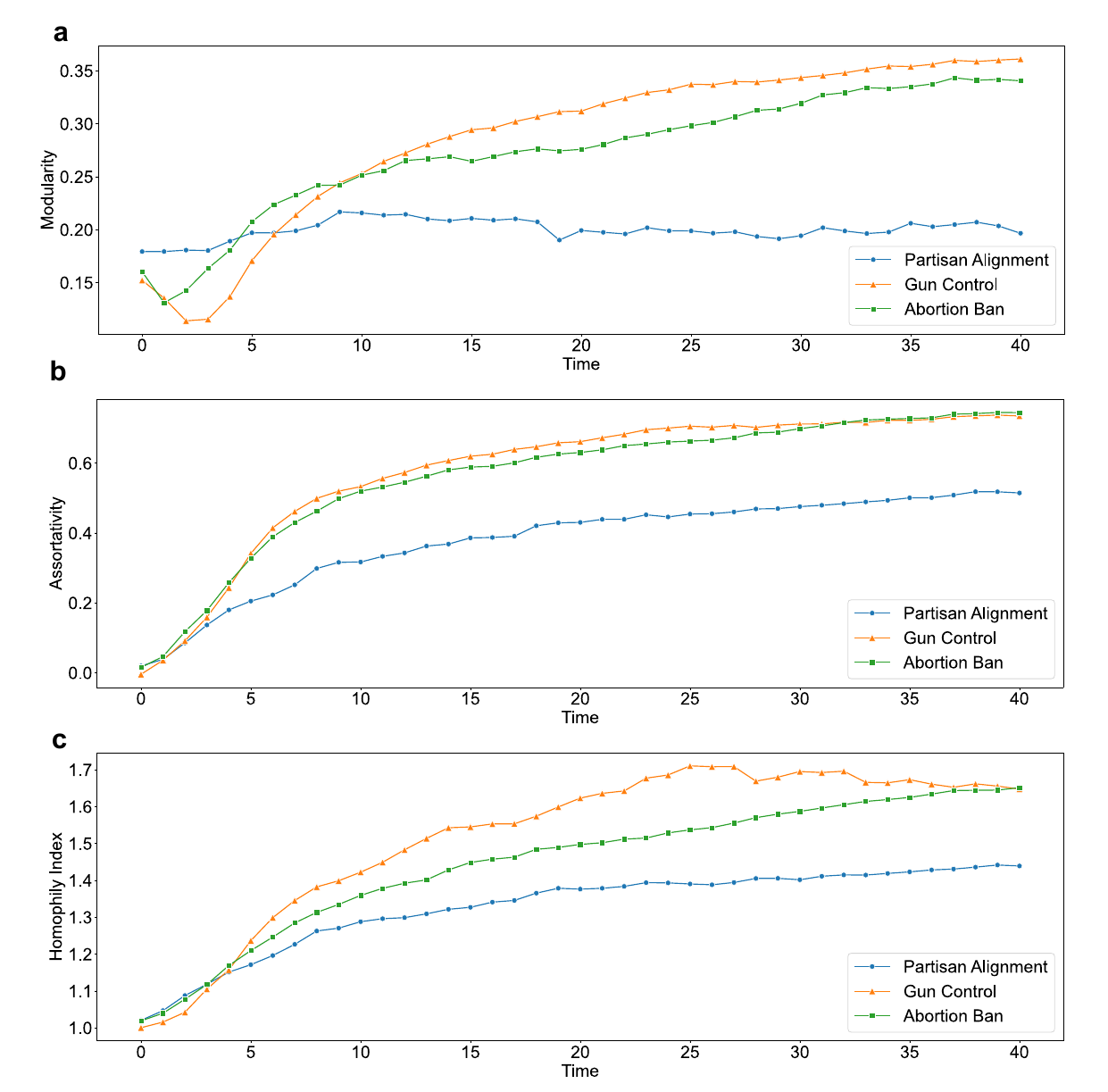}
\caption{\textbf{Changes in network-level metrics over time}, including \textbf{a,} modularity, \textbf{b,} assortativity, and \textbf{c,} homophily index.}
\label{fig:new_networks}
\end{figure}

We also incorporate more standard measures of \rvss{behaviours commonly associated with} ``echo chambers'' and ``backfire effects'' following prior research~\cite{cinelli2021echo,Bail2018Exposure}. Specifically, following Cinelli et al.~\cite{cinelli2021echo}, we adopt the joint distribution of an agent's opinion and the average opinion of their neighborhood to illustrate echo chambers. As shown in Figure~\ref{fig:echo_chamber}, we observe that agents are more likely to interact with peers in the same camp (left-leaning or right-leaning), \rvss{demonstrating behavioural patterns consistent with echo chambers}. For the ``backfire'' effect, we primarily follow the measures outlined in Bail et al.~\cite{Bail2018Exposure}, examining the relationship between the information individuals are exposed to and subsequent changes in their opinions. For simplicity, we did not divide the overall population into two camps and instead reported the effects on the entire population (as shown in Figure 3d of the Main Text). To further examine the ``backfire'' effect on different camps, we include the breakdown results in Figure~\ref{fig:breakdown_backfire}. We observe that interactions with opposing agents do not consistently reduce polarization levels as expected. Instead, exposure to opposing opinions can occasionally intensify polarization (e.g., the radical green bar in Figure~\ref{fig:breakdown_backfire}a), resembling the backfire effect observed in human societies. Overall, these further analyses are consistent with the results reported in the Main Text, providing more fine-grained insight into \rvss{interaction patterns associated with} echo chamber and backfire effects. 

\begin{figure}[h!]
\centering
\includegraphics[width=0.9\textwidth]{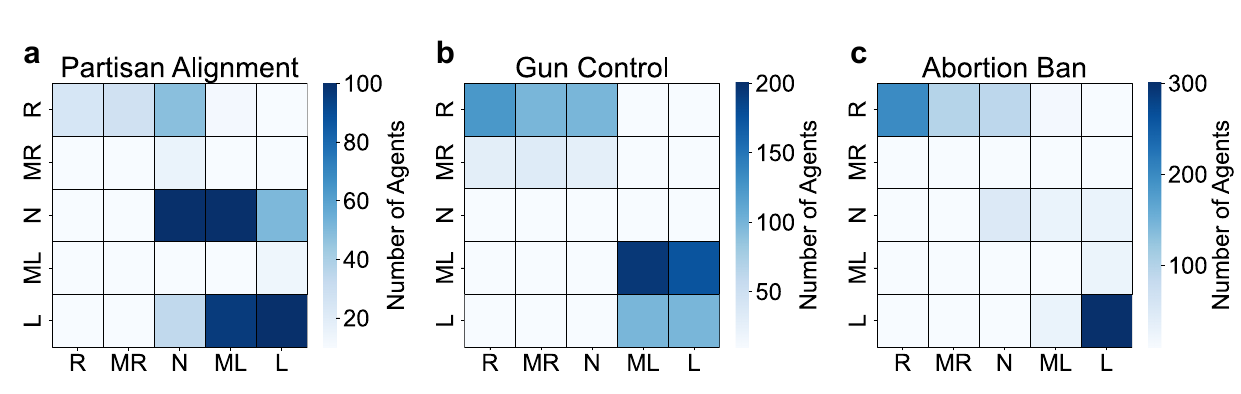}
\caption{\textbf{Illustration of echo chambers following Cinelli et al.~\cite{cinelli2021echo}}, where L, ML, N, MR, and R represent opinions of left, moderate left, neutral, moderate right, and right, respectively.}
\label{fig:echo_chamber}
\end{figure}

\begin{figure}[h!]
\centering
\includegraphics[width=0.9\textwidth]{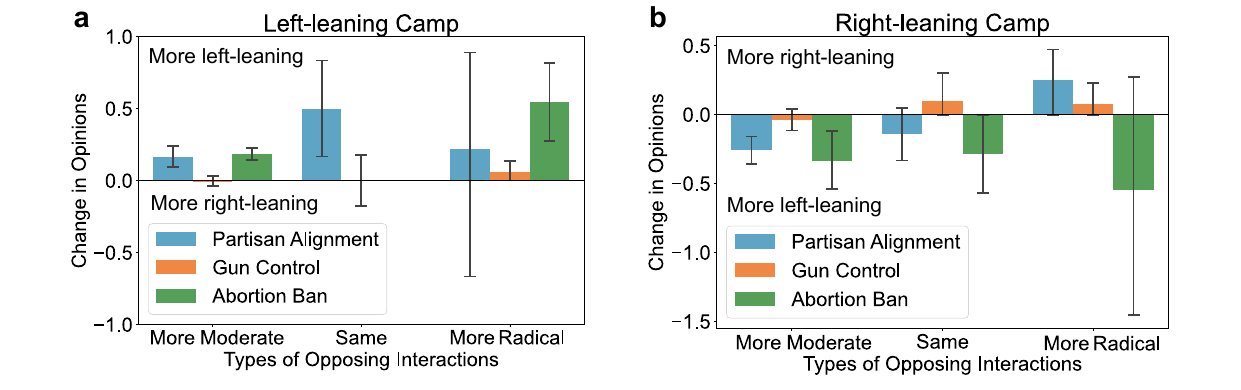}
\caption{\textbf{Backfire effects on a, left-leaning camp, and b, right-leaning camp, following the analysis framework of Bail et al.~\cite{Bail2018Exposure}. Here bars represent the average, and error bars represent the corresponding 95\% confidence intervals (CIs). Sample sizes, corresponding to the number of agent interactions, for bars in (a) are 259, 550, 367, 6, 17, 20, 9, 52, and 11. Sample sizes in (b) are 371, 156, 156, 21, 10, 21, 36, 13, and 11.}}
\label{fig:breakdown_backfire}
\end{figure}

Echo chambers do not universally occur across all individuals in the real world. As noted by Nyhan et al., ``We find that the median Facebook user received a majority of their content from like-minded sources—50.4\%... Just 20.6\% of Facebook users get over 75\% of their exposures from like-minded sources''~\cite{nyhan2023like}, illustrating that while not universal, echo chambers exist for substantial segments of the population. Furthermore, Guess et al. point out that ``Evidence for echo chambers is actually strongest in offline social networks, which can increase exposure to like-minded views and information and amplify partisan messages,''~\cite{guess2018avoiding} highlighting conditions under which echo chambers are more likely to emerge. In our simulations, we find that LLM agents exhibit similar heterogeneity. Specifically, we find that \rvss{patterns of interaction consistent with echo chambers} do not universally occur across all agents, mirroring the heterogeneity observed in real-world populations. As shown in Figure~\ref{fig:echo_chamber_demo}, we find that when discussing partisan alignment, the median proportion of interactions target agents receive from like-minded sources is 48.2\% across the three issues. Moreover, only 29.1\% of agents receive more than 75\% of their interactions from like-minded sources, reflecting the presence of extreme echo chambers for a limited subset of agents. \rvss{These results are qualitatively consistent with patterns reported in empirical studies of human behaviour}, suggesting that echo chambers do not form universally but instead emerge selectively among certain individuals. 


Similarly, the presence and extent of backfire effects remain subjects of debate, and such effects do not appear universally. In Coppock's work, he emphasizes the concept of persuasion in parallel, where exposure to persuasive information leads different groups to shift their attitudes in the same direction, thus questioning backfire effects~\cite{coppock2023persuasion}. He also points out important exceptions, however: ``Political scientists who study American politics have focused on a particular kind of group cue, the party cue, and have found that indeed, such treatments have heterogeneous effects'', emphasizing that backfire effects arise selectively, and are typically triggered by group cues~\cite{coppock2023persuasion}. Moreover, he acknowledges that ``it is difficult to characterize how much of the political information space is filled with group cues versus persuasive information, and furthermore, the distinction between them is not always clear'', highlighting the prevalence of group cues in real-world communication~\cite{coppock2023persuasion}. Similarly, in LLM-driven simulations, agents’ political opinions and associated thoughts, serving as key conditioning factors in communication, naturally provide group cues that influence their interactions and decision-making. \rvss{This interpretation is consistent with Coppock’s framework.} Furthermore, \rvss{behavioural patterns consistent with the backfire effect are observed in only a small portion of agents} (0.82\%–2.12\%), given that agents are rarely situated in environments dominated by opposing opinions. 


\begin{figure}[h!]
\centering
\includegraphics[width=0.9\textwidth]{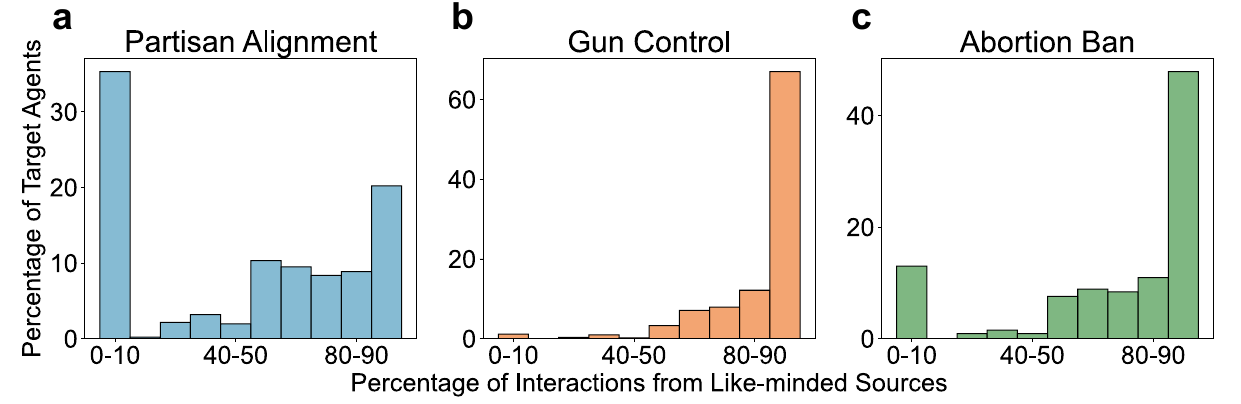}
\caption{\textbf{Distribution of the exposure of target agents to like-minded sources.}}
\label{fig:echo_chamber_demo}
\end{figure}

The degree distribution is a key indicator of network complexity~\cite{barabasi1999emergence, strogatz2001exploring}. 
Therefore, we measure the degree distributions of LLM agents' self-organized social networks (Figure~\ref{fig:degree}). Here we focus on agents' in-degrees, which manifest their popularity in the social network. For example, if an agent has an in-degree of 100, it indicates that 100 agents would like to communicate with this agent. As shown in Figure~\ref{fig:degree}, because the network is randomly initialized with a Watts–Strogatz model~\cite{watts1998collective}, the initial in-degree distribution is characterized by a peak at $k=4$. With an increase in interactions among LLM agents, the degree distribution manifests a power-law tail for large in-degree. This indicates that these networks self-organize into a scale-free state. Moreover, the very existence of a long tail in the in-degree distribution suggests that a small proportion of agents possess unexpectedly greater popularity than others. \rvs{We find that the average in-degree of the top 20 highest in-degree nodes is approximately 9.38, 2.35 times the average. Among these top 20 nodes, 86.7\% belong to the left-leaning camp, suggesting that agents are more likely to maintain social connections with these popular left-leaning individuals.}

\begin{figure}[h!]
\centering
\includegraphics[width=0.79\textwidth]{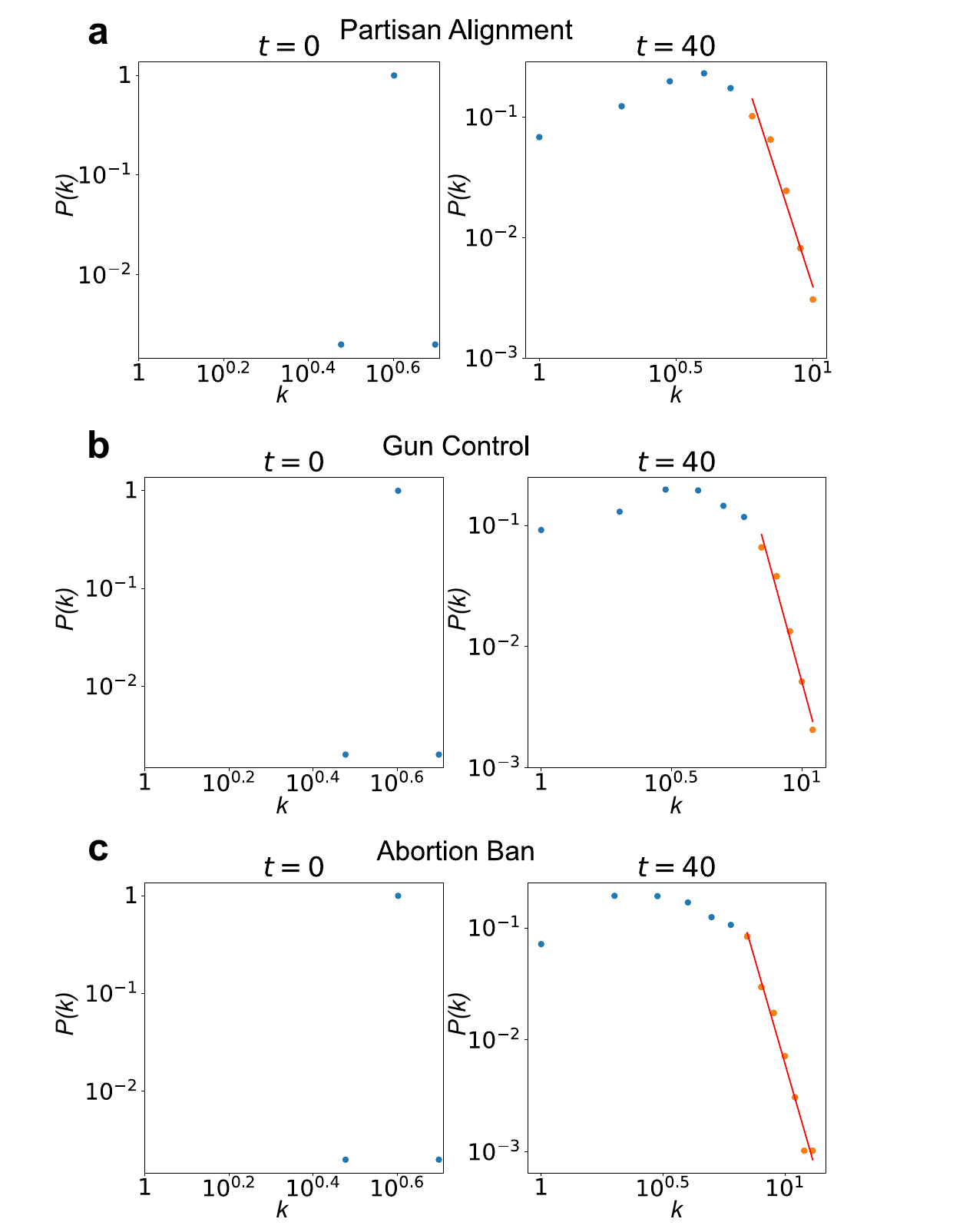}
\caption{\textbf{Degree distribution of LLM agents' self-organized social network.} \textbf{a}, Social network of partisan in the initial ($t=0$) and final ($t=36-40$) states. \textbf{b}, Social network of gun control in the initial ($t=0$) and final ($t=36-40$) states. \textbf{c}, Social network of abortion ban in the initial ($t=0$) and final ($t=36-40$) states.}
\label{fig:degree}
\end{figure}

\clearpage


Moreover, prior observations on LLM agents' self-organized social network raises a new question: what if the organization of social networks deviates from the LLM-driven homophily \rvss{phenomena}? To explore this, we replace the self-organized networks with a random network, where agents always communicate with random ones, and a static network, where agents only communicate with their initial counterparts. As shown in Figure~\ref{fig:randomnetwork}. We discover that in both static and random networks, no balanced polarization pattern forms. Instead, agents with homophilic opinions dominate the overall system, taking up 75\% of the overall agents. This experiment highlights the essential role of self-organized social networks in the polarization of LLM agents.

\begin{figure}[h!]
\centering
\includegraphics[width=0.9\textwidth]{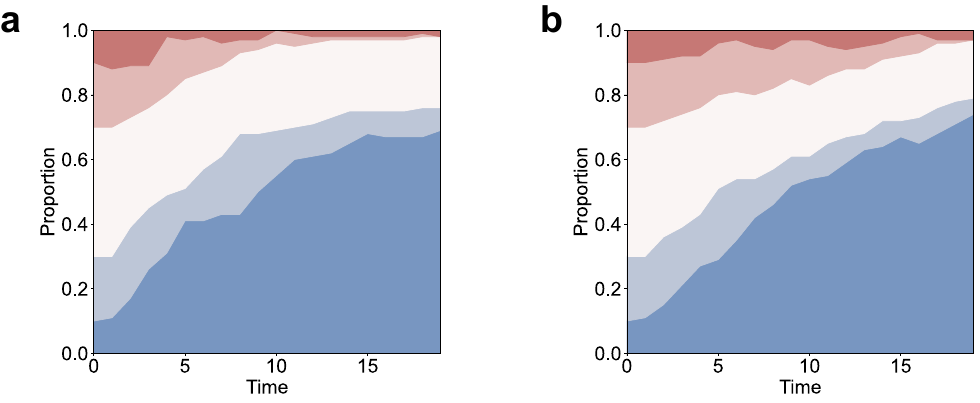}
\caption{\textbf{Opinion dynamics.} \textbf{a}, Opinion dynamics in a static network. \textbf{b}, Opinion dynamics in a random network. In both static and random networks, no balanced polarization pattern forms, with homophilic opinions dominating the system.}
\label{fig:randomnetwork}
\end{figure}

\section{Supplementary Note 5: Results of Other Individual-level Behavioural Patterns}

Besides \rvss{the behavioural patterns examined} in Main Text, we also examine the effects of other social mechanisms widely observed in the real world. In particular, we investigate the other three social mechanisms: (i) Exaggerated misperception, which describes 
the tendency to perceive out-group members as more intensely negative and in-group members as more intensely positive~\cite{lees2020inaccurate,jost2022cognitive}. (ii) Objective illusion, which describes the tendency for people to see those with aligning opinions as more rational and impartial and less biased than others~\cite{schwalbe2020objectivity,jost2022cognitive}. (iii) Stereotyping, which describes the endorsement or acceptance of fixed, categorical, and over-generalized beliefs about the characteristics of a specific social group~\cite{ahler2018parties,jost2022cognitive}. We explicitly \rvss{configure 50\% of the agents via prompting to exhibit behaviours aligned with exaggerated misperception, objective illusion, or stereotyping.}. We rerun the simulations of networked systems with other experimental settings unchanged. As shown in Figure~\ref{fig:other_mechanims}, we find that a higher proportion of agents with these three behavioural patterns increases the polarization level. 


\begin{figure}[h!]
\centering
\includegraphics[width=\textwidth]{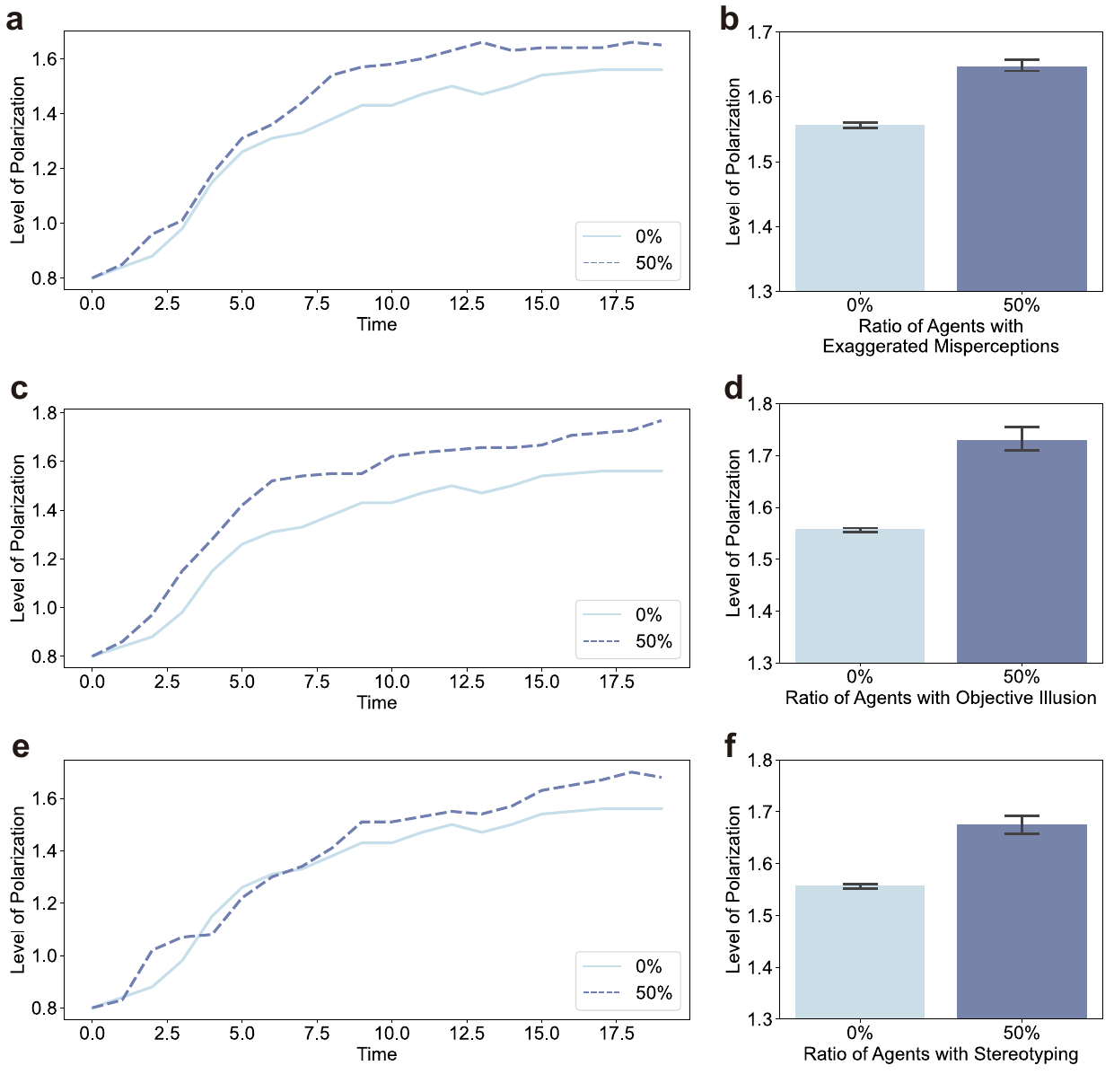}
\caption{\textbf{\rvss{Behavioural patterns} behind the emergence of polarization.} \textbf{a, b}, Effect of exaggerated misperceptions. \textbf{c, d}, Effect of objective illusion. \textbf{e, f}, Effect of stereotyping. Here, a, c, and e show the trend in the level of polarization, and b, d, and f show the average levels of polarization in the last five timesteps ($n=5$, corresponding to five timestep-level polarization values computed from all agents' opinions), where bars represent the average values and error bars represent the corresponding 95\% CIs. When the system consists of more agents with behavioural patterns of exaggerated misperceptions (a and b) or objective illusion (c and d), and stereotyping (e and f), the level of polarization increases. }
\label{fig:other_mechanims}
\end{figure}

\section{Supplementary Note 6: Experiments on Different LLMs}

Results in Main Text are based on the simulations driven by GPT-3.5 Turbo through the public OpenAI API. We also perform similar simulations using other LLMs, including GPT-4o, Llama-3, Claude-3, and ChatGLM. However, Claude-3 hardly responds to political issues with an extremely low response rate of 10\%. Therefore, we exclude it in further experiments. As shown in Figures~\ref{fig:other_llm1} and \ref{fig:other_llm2}, we observe that in all the systems, free-from interactions among LLM agents result in the emergence of polarization. Interestingly, we find that except for Llama-3, all other LLMs, including GPT-3.5 Turbo, GPT-4o, and ChatGLM, show a similar left-skewed tendency. For the Llama-3, the tendency is reversed. This difference lies mainly in the inherent political biases of different LLMs ~\cite{liu2022quantifying,rutinowski2024self,motoki2024more}. Moreover, by introducing the self-regulation strategy, we find that the imbalance between right-leaning and left-leaning camps can be largely alleviated.

\begin{figure}[h!]
\centering
\includegraphics[width=\textwidth]{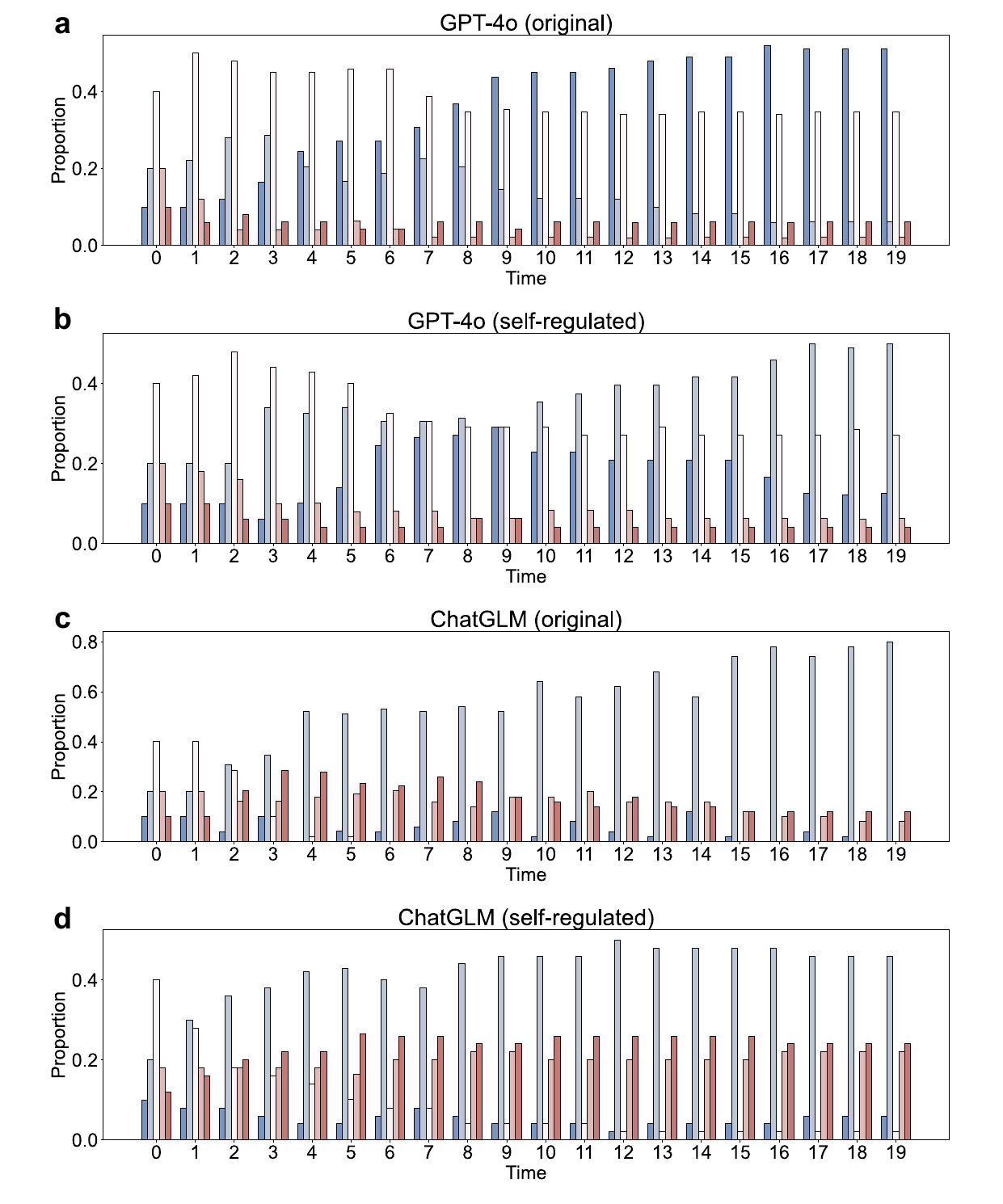}
\caption{\textbf{Opinion dynamics of networked systems driven by various LLMs.} \textbf{a,c}, the original systems driven by GPT-4o and ChatGLM. \textbf{b,d}, the self-regulated systems driven by GPT-4o and ChatGLM.}
\label{fig:other_llm1}
\end{figure}

\begin{figure}[h!]
\centering
\includegraphics[width=\textwidth]{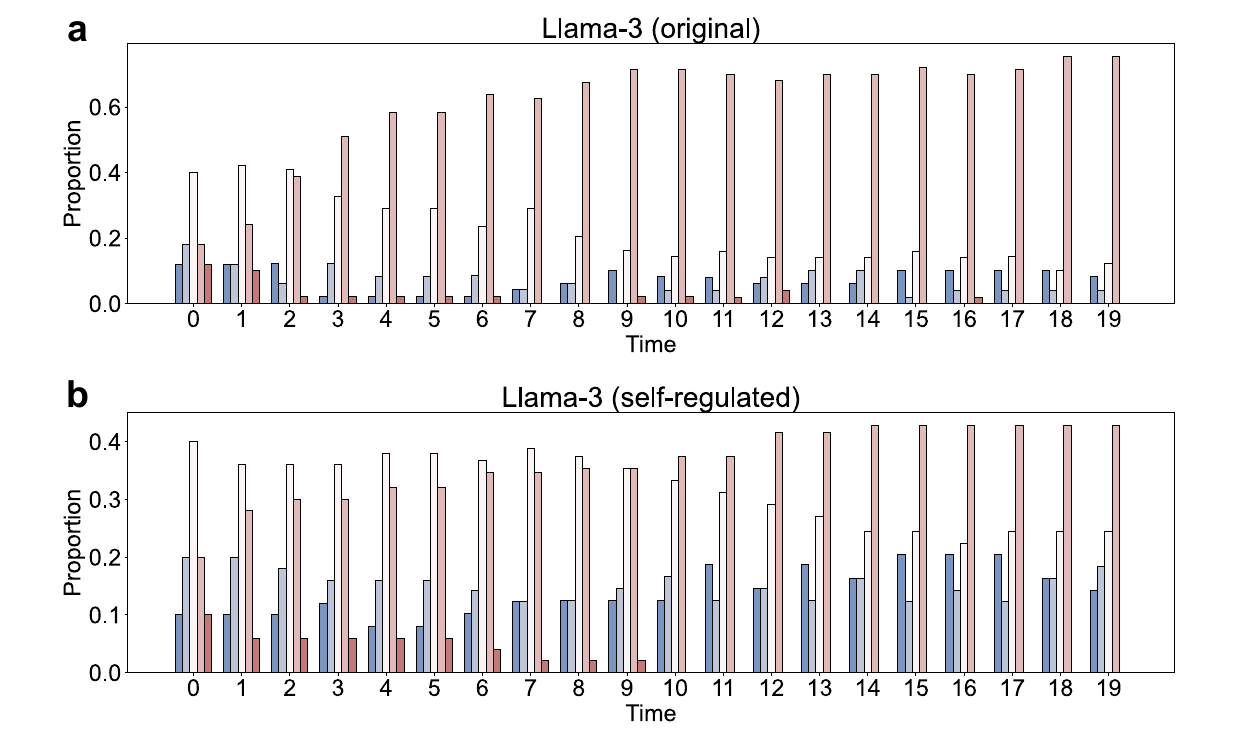}
\caption{\textbf{Opinion dynamics of networks systems driven by various LLMs.}\textbf{a,c}, the original system driven by Llama-3. \textbf{b}, the self-regulated system driven by Llama-3.}
\label{fig:other_llm2}
\end{figure}

We also examine the evolution of social networks developed by LLM agents (Figure~\ref{fig:other_llm3}). We observe that with the increase in interactions, systems driven by various LLMs exhibit a similar homophilic clustering pattern, where agents with homogeneous opinions are increasingly likely to communicate with each other (Figure~\ref{fig:other_llm3}a). Simultaneously, interactions involving opposing agents (Figure~\ref{fig:other_llm3}b) and neutral agents (Figure~\ref{fig:other_llm3}c) are all decreasing, consistent with prior results of GPT-3.5 Turbo (Figure~\ref{fig:networks}).

\begin{figure}[h!]
\centering
\includegraphics[width=\textwidth]{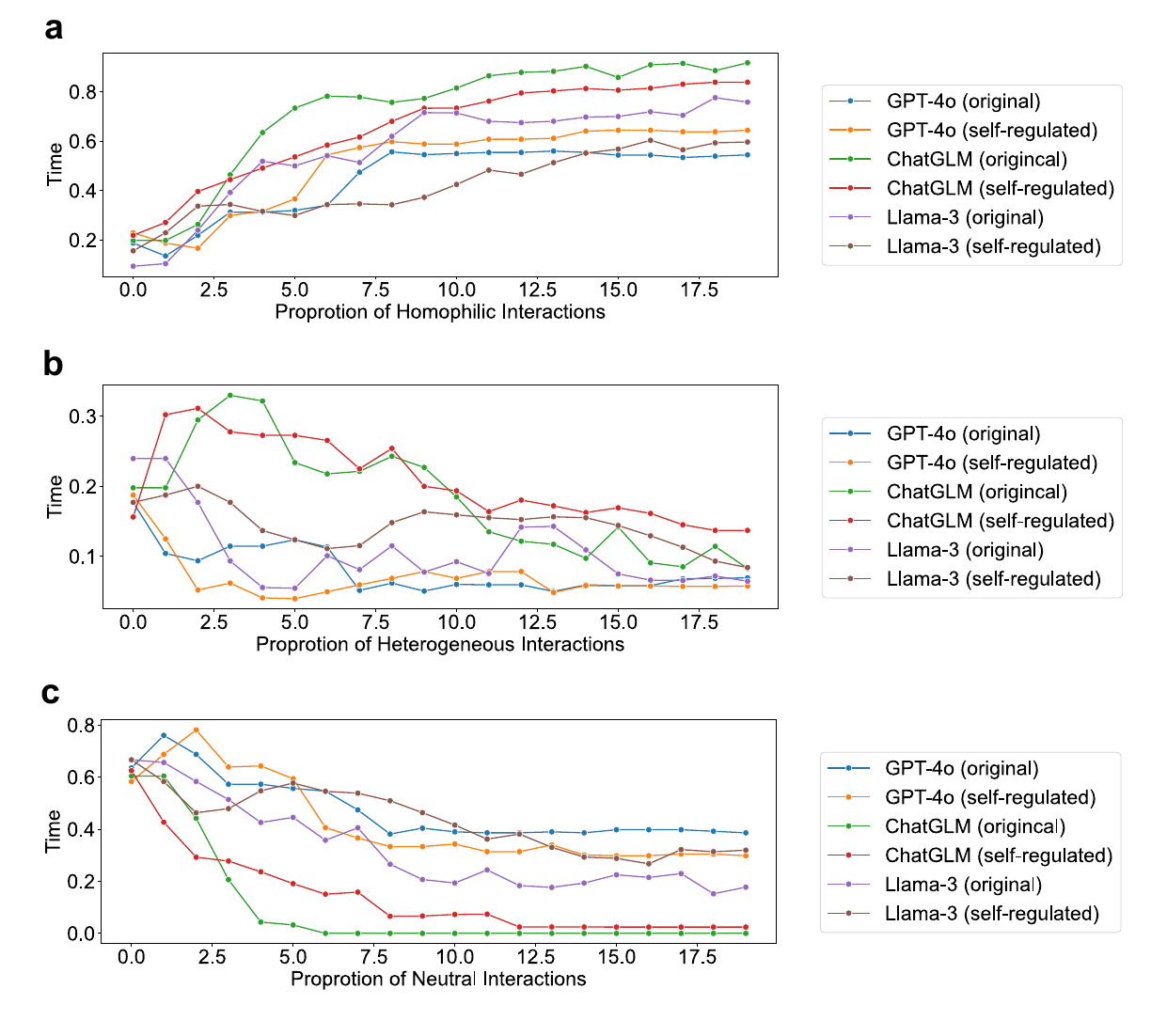}
\caption{\textbf{Evolution of self-organized social network developed by various LLM agents.} \textbf{a}, Changes in the proportion of homophilic interactions over time. \textbf{b}, Changes in the proportion of heterogeneous interactions over time. \textbf{c}, Changes in the proportion of neutral interactions over time. Agents are increasingly likely to interact with those with homophilic opinions while they avoid interaction with others.}
\label{fig:other_llm3}
\end{figure}

Beyond comparing simulations entirely driven by different LLMs, we further explore a more dynamic setting in which agents \rvss{are switched to} a rival language model mid-conversation. Specifically, at the midpoint of simulation, we change the underlying model from GPT-3.5-turbo to the more advanced GPT-4o, allowing agents to adopt a more sophisticated cognitive architecture. As shown in Figures~\ref{fig:mid_orig} and \ref{fig:mid_4o}, this replacement produces only marginal changes: agents still converge to similarly polarized opinion states, and their social networks likewise split into two communities holding opposing views. This suggests that \rvss{switching the underlying language model mid-simulation does not substantially alter the observed polarization patterns in this synthetic system.} We further test a more extreme case by replacing GPT-3.5 with a substantially different rival language model, DeepSeek-V3, developed by a Chinese company and trained on a distinct corpus (Figure~\ref{fig:mid_deepseek}). \rvss{Even under this replacement, results remain similar: neither the overall level of polarization nor the emergent network structure is meaningfully altered.}

\begin{figure}[h!]
\centering
\includegraphics[width=0.75\textwidth]{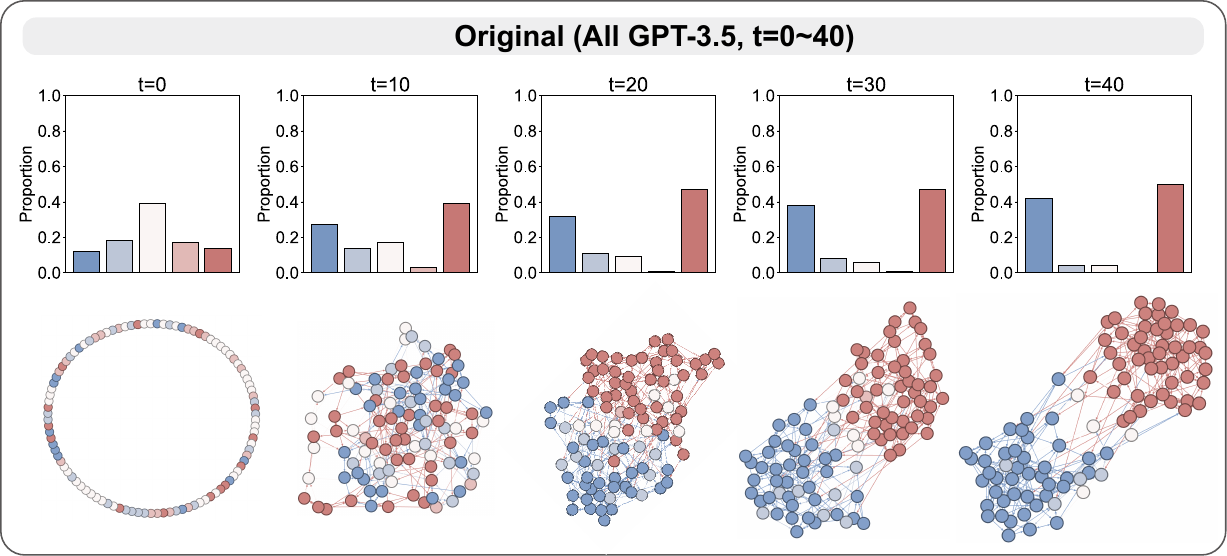}
\caption{\textbf{Evolution of LLM agents' collective opinions and network structures}, where all agents are driven by GPT-3.5 for the entire duration of the simulation and they discuss the political issue of abortion ban.}
\label{fig:mid_orig}
\end{figure}

\begin{figure}[h!]
\centering
\includegraphics[width=0.75\textwidth]{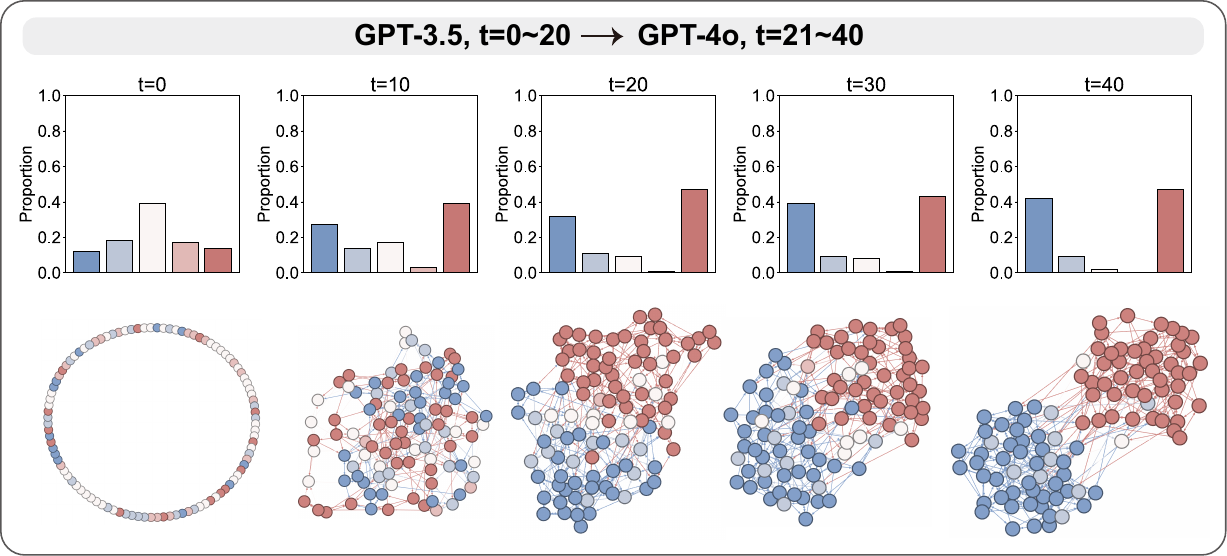}
\caption{\textbf{Evolution of LLM agents' collective opinions and network structures}, where agents are initially driven by GPT-3.5 but defect to GPT-4o amid conversations about the political issue of abortion bans.}
\label{fig:mid_4o}
\end{figure}

\begin{figure}[h!]
\centering
\includegraphics[width=0.75\textwidth]{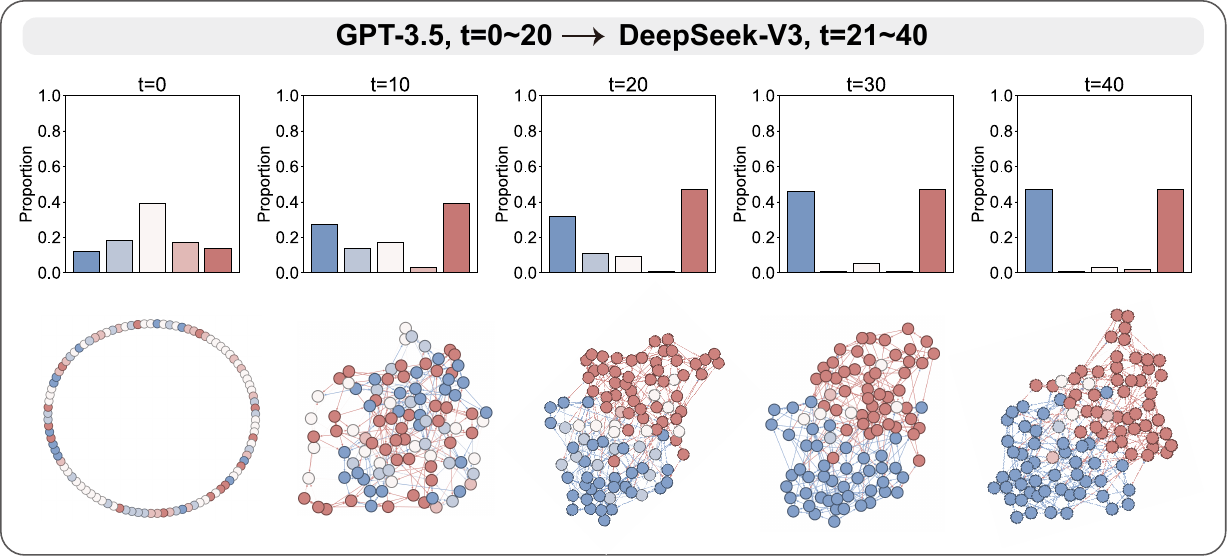}
\caption{\textbf{Evolution of LLM agents' collective opinions and network structures}, where agents are initially driven by GPT-3.5 but defect to DeepSeek-V3 amid conversations about the political issue of abortion bans.}
\label{fig:mid_deepseek}
\end{figure}

Overall, these findings suggest that \rvss{switching the underlying language model mid-conversation} may slightly influence the evolution trajectories of agents’ collective opinions and network structures, but it \rvss{does not substantially alter observed polarization patterns}. A key reason, we believe, lies in the path dependence of these dynamics. By the midpoint of simulation, agents have already developed social networks and collective opinions that reinforce one another’s polarizing tendencies. Their networks have begun to crystallize into two distinct communities, and their opinions are shaped not only by the underlying LLMs, but also by the accumulated history of interactions and surrounding social context. This reflects well-established findings in political psychology, where timing and early exposure are critical to opinion formation and change~\cite{druckman2012source,nickerson1998confirmation,lord1979biased}.  As such, merely replacing the underlying LLMs, without addressing the entrenched social and psychological drivers of polarization, proves insufficient to reverse emergent polarization dynamics.

\section{Supplementary Note 7: Analyses on Different Intervention Strategies}

As discussed in the Main Text, our proposed networked system of LLM agents has the potential to serve \rvss{as a synthetic testbed for the preliminary exploration of hypothetical strategies related to polarization mitigation}. To examine the effectiveness of the proposed intervention strategies and networked system \rvss{in the testbed}, we design five intervention experiments (see the detailed design in SI Section~\ref{sec:intervention_design}). Here we report detailed analyses of different intervention strategies.

\begin{figure}[h!]
\centering
\includegraphics[width=0.9\textwidth]{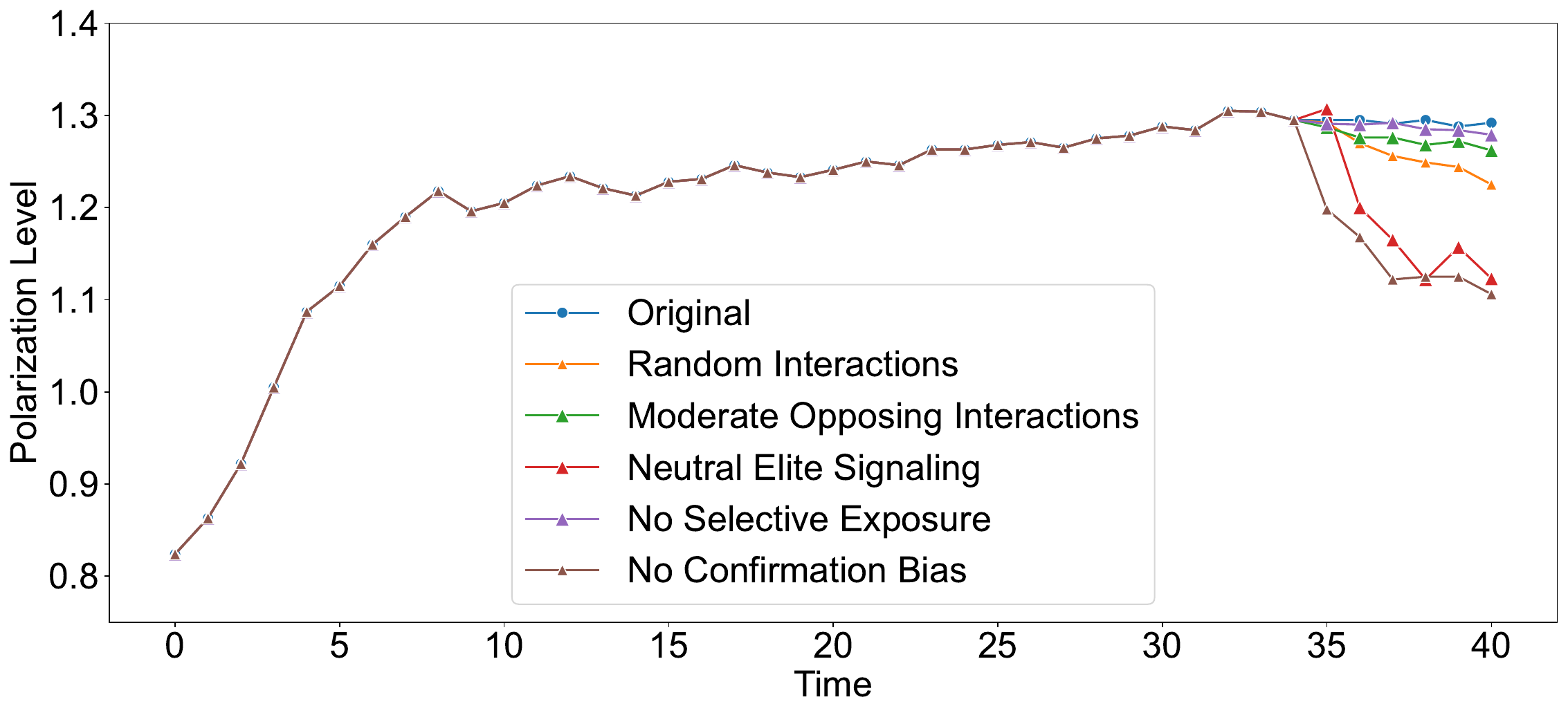}
\caption{\textbf{Trends in polarization levels of networked systems under different intervention strategies.}}
\label{fig:intervention_polarization}
\end{figure}

\begin{figure}[h!]
\centering
\includegraphics[width=\textwidth]{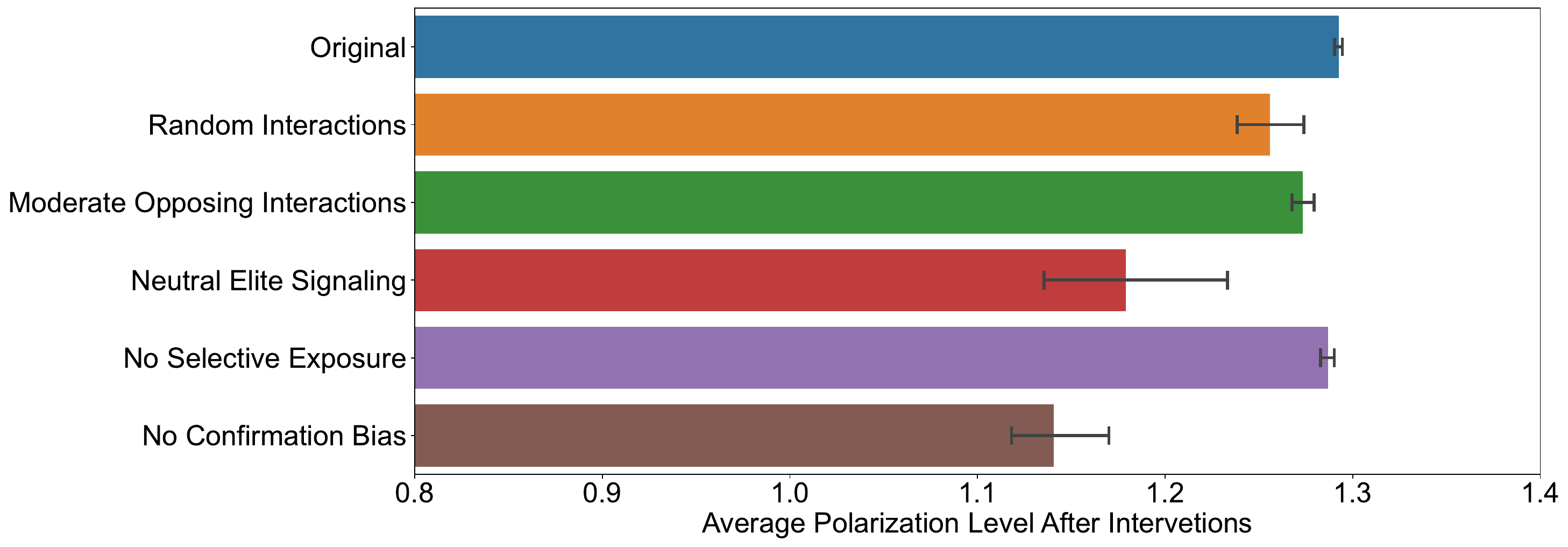}
\caption{\textbf{Effectiveness of different intervention strategies in reducing polarization. Bars show the average values after intervention, and error bars show the corresponding 95\% CIs.}}
\label{fig:intervention_average}
\end{figure}

Figure~\ref{fig:intervention_polarization} shows the trends in polarization systems under different intervention strategies and Figure~\ref{fig:intervention_average} summarizes the average polarization levels after interventions. We observe that all intervention strategies, despite varying effectiveness, can gradually reduce the polarization levels. Among all strategies, we observe that the individual-level interventions of neutral elite signaling and no selective exposure contribute to the greatest reduction \rvss{in this synthetic system}. On the other hand, the network interventions are less effective. This suggests that in an already polarized \rvss{synthetic} system, encouraging free access and open-mindedness to diverse opinions is more effective than directly modifying their social network. This raises a further question: what opinions do these agents form after interventions?

\begin{figure}[h!]
\centering
\includegraphics[width=\textwidth]{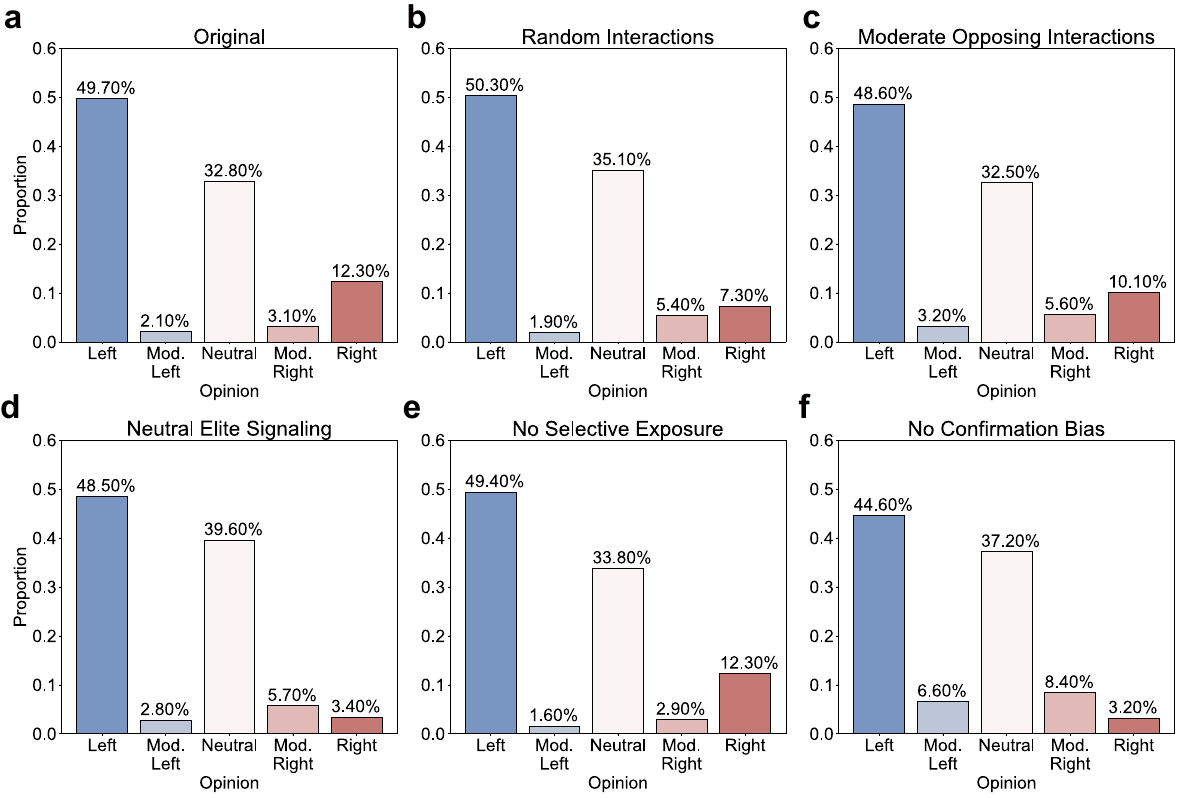}
\caption{\textbf{Opinion distributions of networked systems under different intervention strategies}: \textbf{a}, original; \textbf{b}, random interactions; \textbf{c}, moderate opposing interactions; \textbf{d}, neutral elite signaling; \textbf{e}, no selective exposure; \textbf{f}, no confirmation bias.}
\label{fig:intervention_opinions}
\end{figure}

We further explore the opinion distributions of the intervened systems. As shown in Figure~\ref{fig:intervention_opinions}, we find that different strategies lead to varying changes in opinion distributions. In particular, the strategy of neutral signaling substantially increases the number of agents holding neutral opinions, while the strategy of no confirmation bias leads to more agents with moderate opinions. Moreover, we notice that the strategy of random interaction, despite reducing the overall polarization level, only takes effect on the agents in the right camp, which echoes with prior observations in Figure~\ref{fig:randomnetwork}. This indicates that simply randomizing agents' social relationships could not be an optimal strategy for reducing polarization \rvss{in the synthetic testbed}.

We also investigate how agents change their opinions after receiving interventions (Figure~\ref{fig:intervention_transition}). Here, if the change in opinion is less than 0, it indicates that the agent adopts a more moderate opinion following the intervention. By contrast, if the change value is equal to or greater than 0, it means that the agent adopt an unchanged or more radical opinion. We find that in the original system, which has reached a stable state, the proportions of agents adopting more moderate opinions and those adopting more radical opinions are nearly equal. The introduction of intervention strategies disrupts the stable state: agents are more likely to adopt a more moderate opinion than a more radical one. 


\begin{figure}[h!]
\centering
\includegraphics[width=\textwidth]{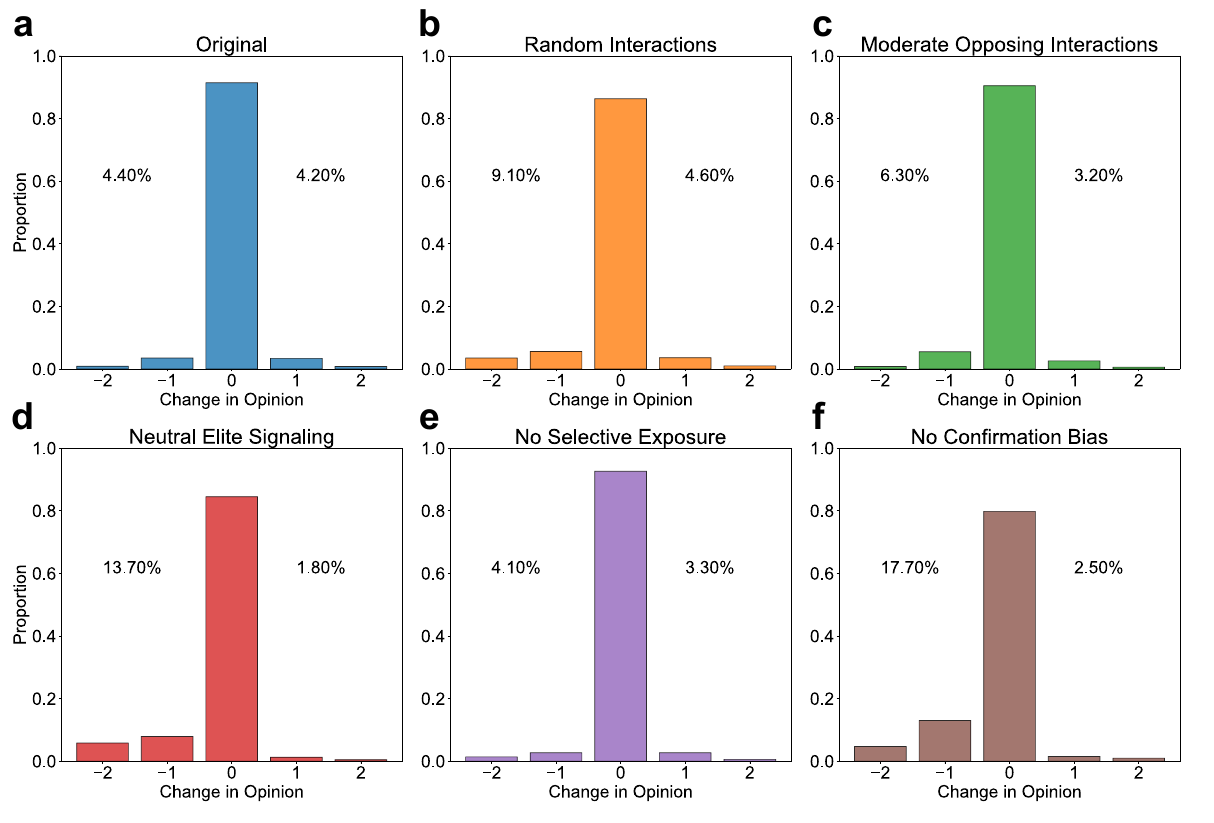}
\caption{\textbf{Probability of opinion change under different intervention strategies}: \textbf{a}, original; \textbf{b}, random interactions; \textbf{c}, moderate opposing interactions; \textbf{d}, neutral elite signaling; \textbf{e}, no selective exposure; \textbf{f}, no confirmation bias.}
\label{fig:intervention_transition}
\end{figure}

After analyzing the effectiveness of these strategies in reducing polarization, one may wonder whether these strategies can foster more inclusive political conversations. As shown in Figure~\ref{fig:trend_network}, we find that all the strategies can effectively reduce the proportions of homophilic interactions, which allows these agents to have the opportunity to interact with those holding diverse opinions. In particular, the interventions on the network level have the greatest impact on the structures. However, the strategies on the individual level gradually shape the network. More interestingly, we find that the strategy of no selective exposure is the least effective, raising questions about the underlying reasons.

\begin{figure}[h!]
\centering
\includegraphics[width=\textwidth]{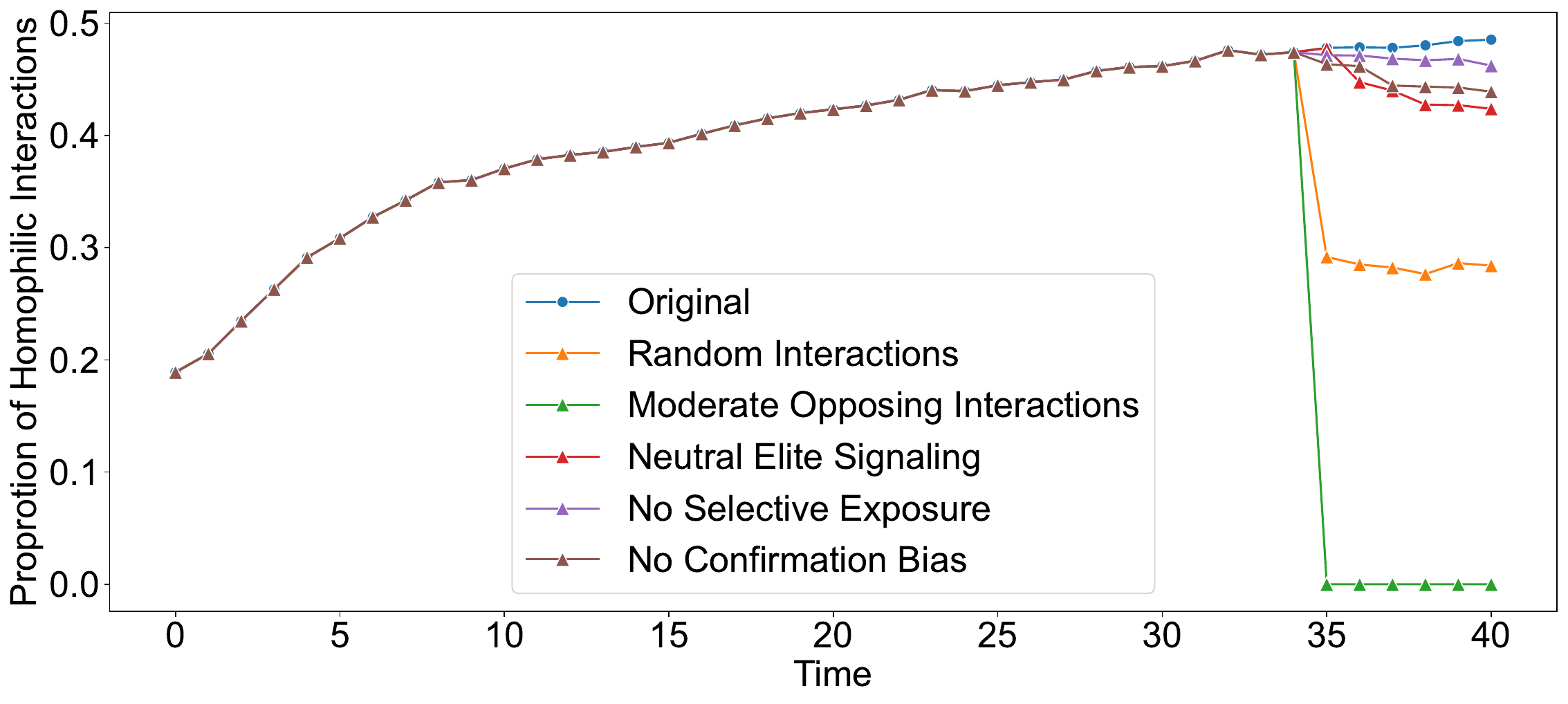}
\caption{\textbf{Trends in homophilic interactions of networked systems under different intervention strategies.}}
\label{fig:trend_network}
\end{figure}

To answer the question, we explore the change in LLM agents' social network by comparing the differences between the current network and that in the former timestep. In particular, we count the number of edges that exist in two networks and the number of edges in the former timestep. By dividing these two numbers, we can measure the difference between these two networks. As shown in Figure~\ref{fig:network_change}, we find that with the evolution of the system, the change in the social network gradually decreases, suggesting the convergence of the system. Moreover, after the intervention of no selective exposure, which requires agents to engage with those holding diverse opinions, the network remains unchanged. This suggests that these agents remain in social relationships dominated by homophilic peers, which not only hinders them from engaging with others holding diverse opinions, but also prevents the formation of less polarized opinions through more diverse interactions.

\begin{figure}[h!]
\centering
\includegraphics[width=\textwidth]{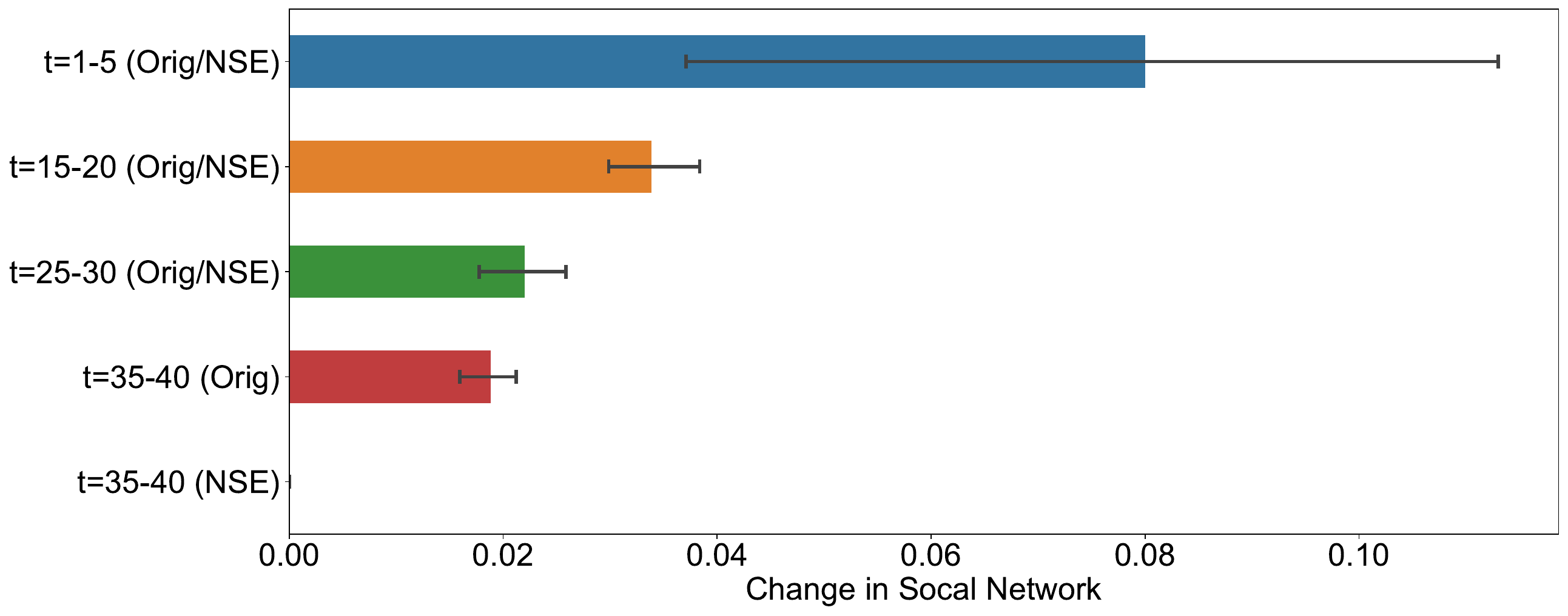}
\caption{\textbf{Comparison of changes in LLM agents' social network}, where we measure the difference between the current social network and that in the former timestep. Here the bars represent the average values ($n=4$ or $5$, corresponding to four or five timestep-level values computed from all agent interactions at each timestep) and the error bars represent the corresponding 95\% CIs.}
\label{fig:network_change}
\end{figure}

Following the design of Groenendyk and Krupnikov~\cite{groenendyk2021motivates}, we have applied a similar open-mindedness intervention to LLM agents. In particular, we first present these agents with a fictitious study linking ``life success'' to the trait of open-mindedness, with ``life success'' operationalized through marital success, income, and IQ. Subsequently, the agents are asked to formulate a theory explaining why open-mindedness might lead to life success. Agents in the control group do not receive the intervention. Following this, both types of agents are exposed only to interactions from the opposing camp, and we observe changes in their opinions to evaluate the effectiveness of the intervention.

In this open-mindedness experiment, 95\% of agents generated a theory supporting the fictitious link between open-mindedness and life success, indicating acceptance of the link — a pattern comparable to that observed in human participants (80\%). Figure~\ref{fig:openminded} presents a comparison of polarization levels in systems subjected to different intervention strategies. Specifically, \textit{Orig} denotes the original system. \textit{Oppose} refers to the intervention condition where agents are only exposed to interactions from the opposing camp. \textit{Open} represents the system in which agents have received the open-mindedness intervention. \textit{Open+Oppose} denotes the condition where agents not only receive the open-mindedness intervention but are also exposed only to interactions from the opposing camp. We observe that the open-mindedness intervention alone leads to a negligible, non-significant reduction in polarization of approximately 1\%. This is because in a polarized system, most agents are already trapped in echo chambers to varying degrees (Figure~\ref{fig:echo_chamber_demo}), limiting their exposure to opposing opinions. As a result, even if these agents become open-minded, they still find it difficult to develop moderate opinions. 

We further increase their exposure to opposing opinions by replacing the peers with whom left-leaning or right-leaning agents interact with, with those from the opposing camp. As shown in Figure~\ref{fig:openminded}, we find that after exposure to opposing opinions, the polarization level of these open-minded agents is significantly reduced (two-sided Student's t-test, \textit{Orig} vs. \textit{Open+Oppose}, $t= 4.18$, $p=.0019<.01$). Moreover, comparing the \textit{Oppose} and \textit{Open+Oppose} conditions reveals that the latter achieves a more substantial reduction in polarization (two-sided Student's t-test, \textit{Oppose} vs. \textit{Open+Oppose}, $t=2.32$, $p=.042<.05$).

\begin{figure}[h!]
\centering
\includegraphics[width=0.9\textwidth]{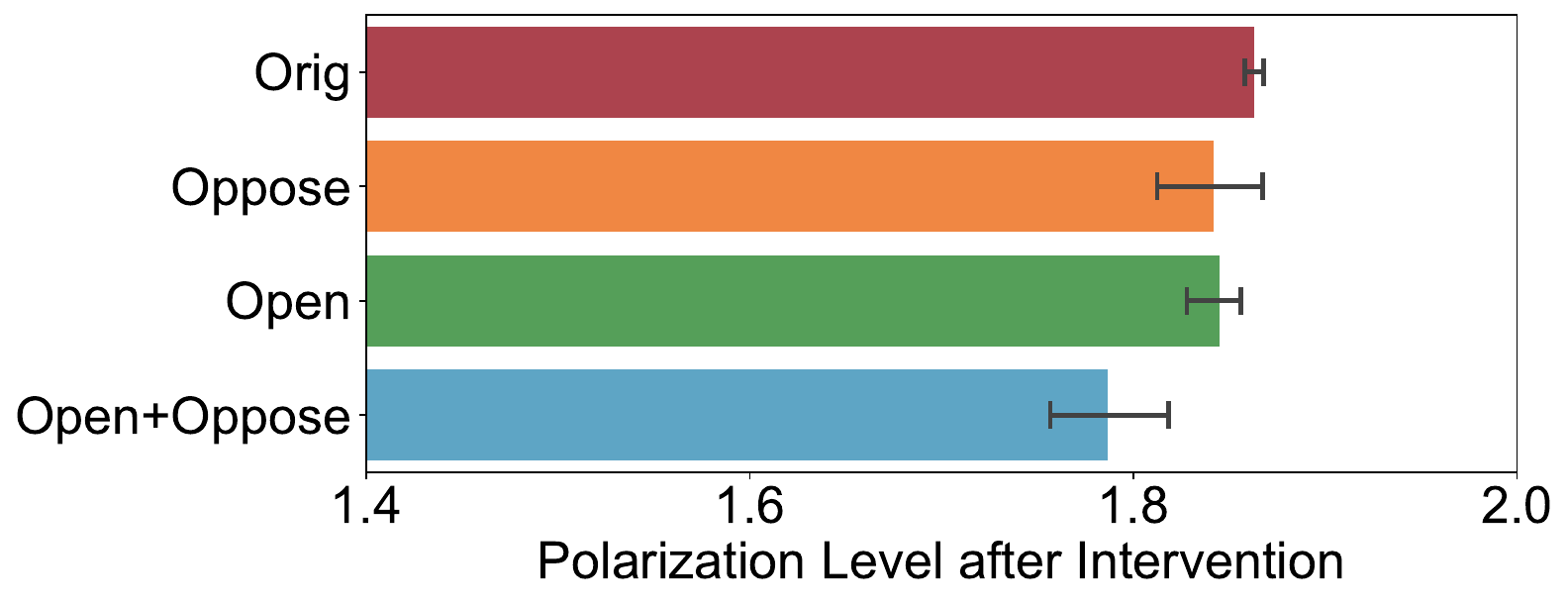}
\caption{\textbf{Comparison of open-mindedness intervention strategies}, where we intervene in systems at t=35 and agents discuss the issue of abortion ban. Here bars show the average values ($n=6$, corresponding to six timestep-level values computed from all agents' opinions) and error bars represent the corresponding 95\% CIs. }
\label{fig:openminded}
\end{figure}

\section{Supplementary Note 8: Analyses on Agents' Social Perceptions and Argument Accumulation Process}

After investigating the collective opinions of LLM agents, one may wonder how these agents perceive each other. Therefore, we examine their perceptions by prompting each agent $i$ to rate its impressions of another agent $j$ and to provide five adjectives that describe agent $j$. The prompts are listed as follows,

\begin{mdframed}

\noindent Assume you are someone who cares about [issue name].

\noindent Towards [issue name], you support [agent $i$'s opinion]

\noindent Your thought is: [agent $i$'s supporting message].

\noindent There is another person who [agent $j$'s opinion].

\noindent That person's thought is: [agent $j$'s supporting message].

\noindent Please rate your impression of that person from 1 to 5, and think of 5 adjectives to describe that person.

\noindent 1 means you have a very negative impression of that person.

\noindent 2 means you have a negative impression of that person.

\noindent 3 means you have a neutral impression of that person.

\noindent 4 means you have a positive impression of that person.

\noindent 5 means you have a very positive impression of that person.

\noindent Respond in JSON format, with keys 'rating' and 'adjectives'. Rating is an integer from 1 to 5, and adjectives are a list of 5 vocabularies.

\end{mdframed}

\begin{figure}[h!]
\centering
\includegraphics[width=\textwidth]{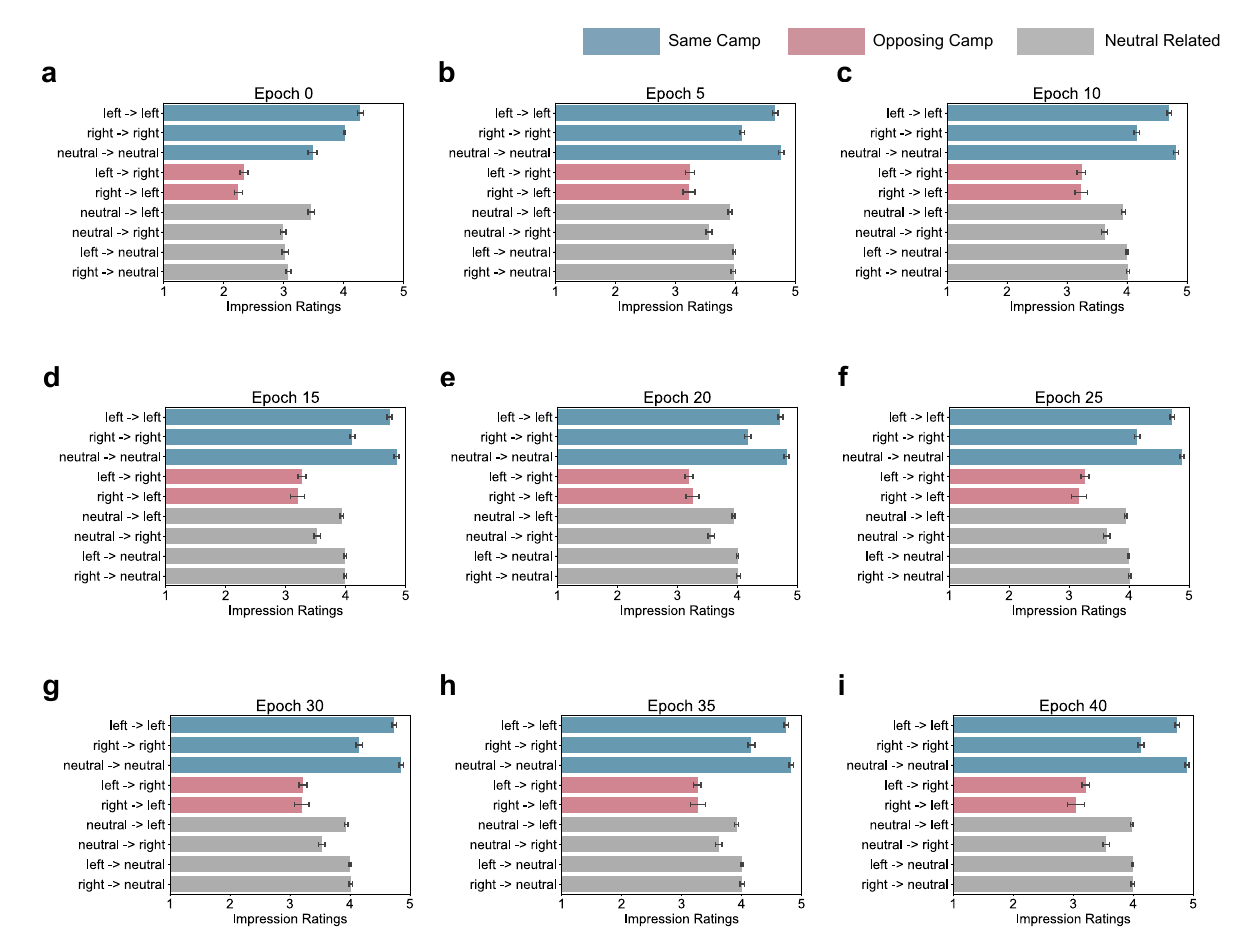}
\caption{\textbf{Impression ratings when agents discuss the issue of partisan alignment in the networked system.} Bars represent mean values, and error bars represent the corresponding 95\% CIs. Each plot includes ratings from 1,000 agents, each evaluating three randomly selected agents: one with similar opinions from the same camp, one with opposing opinions from a different camp, and one with neutral opinions, yielding a total sample size of $n=3,000$.}
\label{fig:politics_rating}
\end{figure}

Specifically, we allow each agent $i$ to rate and describe three randomly selected agents: one with similar opinions from the same camp, one with opposing opinions from a different camp, and one with neutral opinions. We evaluate agents' perceptions of others every five timesteps during the evolution of self-regulated networked systems (Figure 2d in Main Text). As shown in Figures~\ref{fig:politics_rating}-\ref{fig:gun_rating}, we observe that agents are likely to have a better impression of those who share similar opinions from the same camp. By contrast, agents hold the lowest impressions of those in the opposing camp. Moreover, we delve deeper into agents' perceptions by examining both the content and sentiment of their descriptions in a more fine-grained manner. Figures~\ref{fig:politics_dist}-\ref{fig:gun_dist} show the distributions of descriptions across positive, neutral, and negative sentiments. Overall, we find that in most cases, agents tend to adopt positive descriptions when referring to other agents. However, when encountering agents with opposing opinions, they tend to express some negative descriptions. For example, as shown in Figures~\ref{fig:politics_discp}-\ref{fig:gun_discp}, they use adjectives like ``conflicting'', ``rigid'', and ``opinionated'' to describe opposing agents.

\begin{figure}[h!]
\centering
\includegraphics[width=\textwidth]{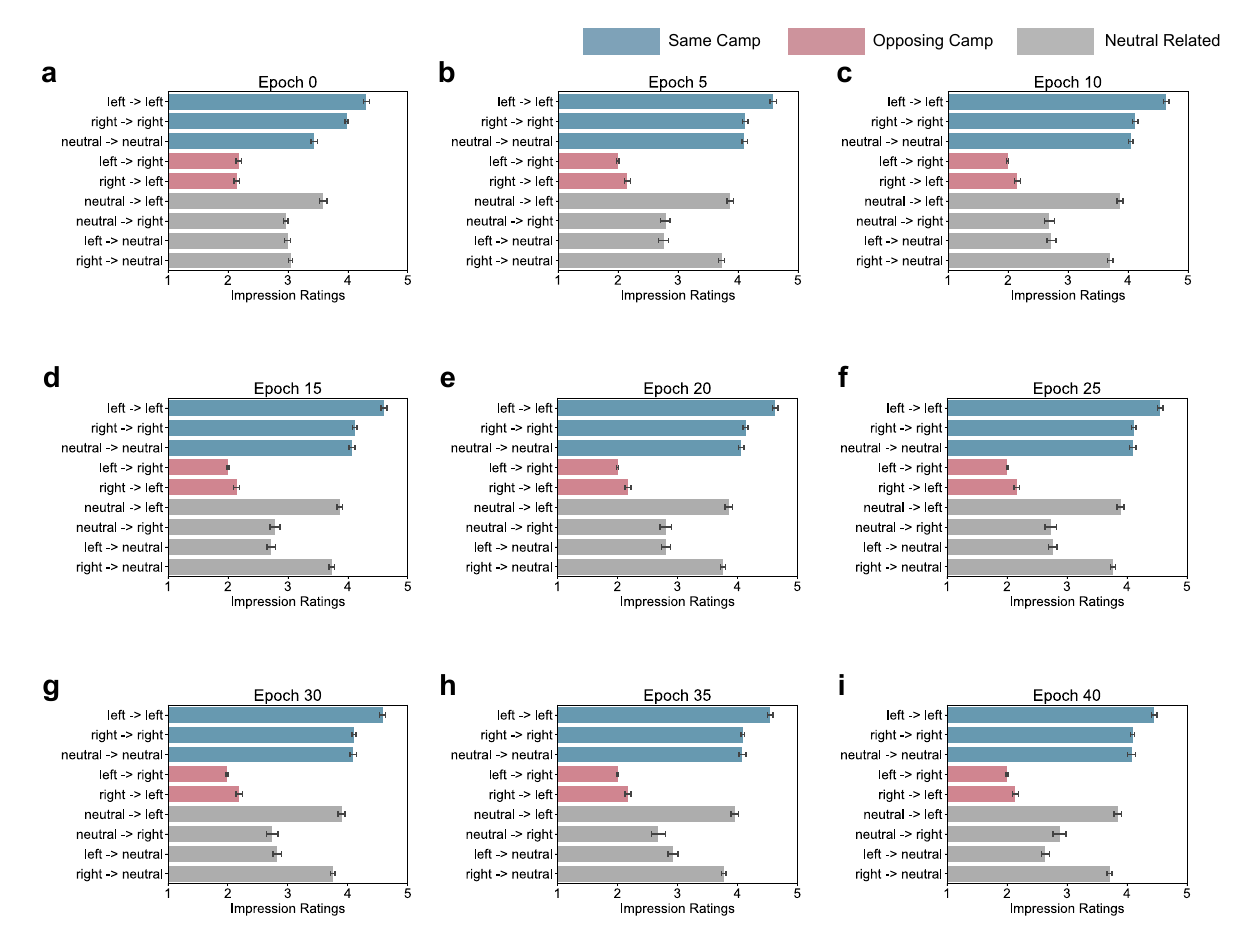}
\caption{\textbf{Impression ratings when agents discuss the issue of abortion ban in the networked system.} Bars represent mean values, and error bars represent the corresponding 95\% CIs. Each plot includes ratings from 1,000 agents, each evaluating three randomly selected agents: one with similar opinions from the same camp, one with opposing opinions from a different camp, and one with neutral opinions, yielding a total sample size of $n=3,000$.}
\label{fig:abortion_rating}
\end{figure}

\begin{figure}[h!]
\centering
\includegraphics[width=\textwidth]{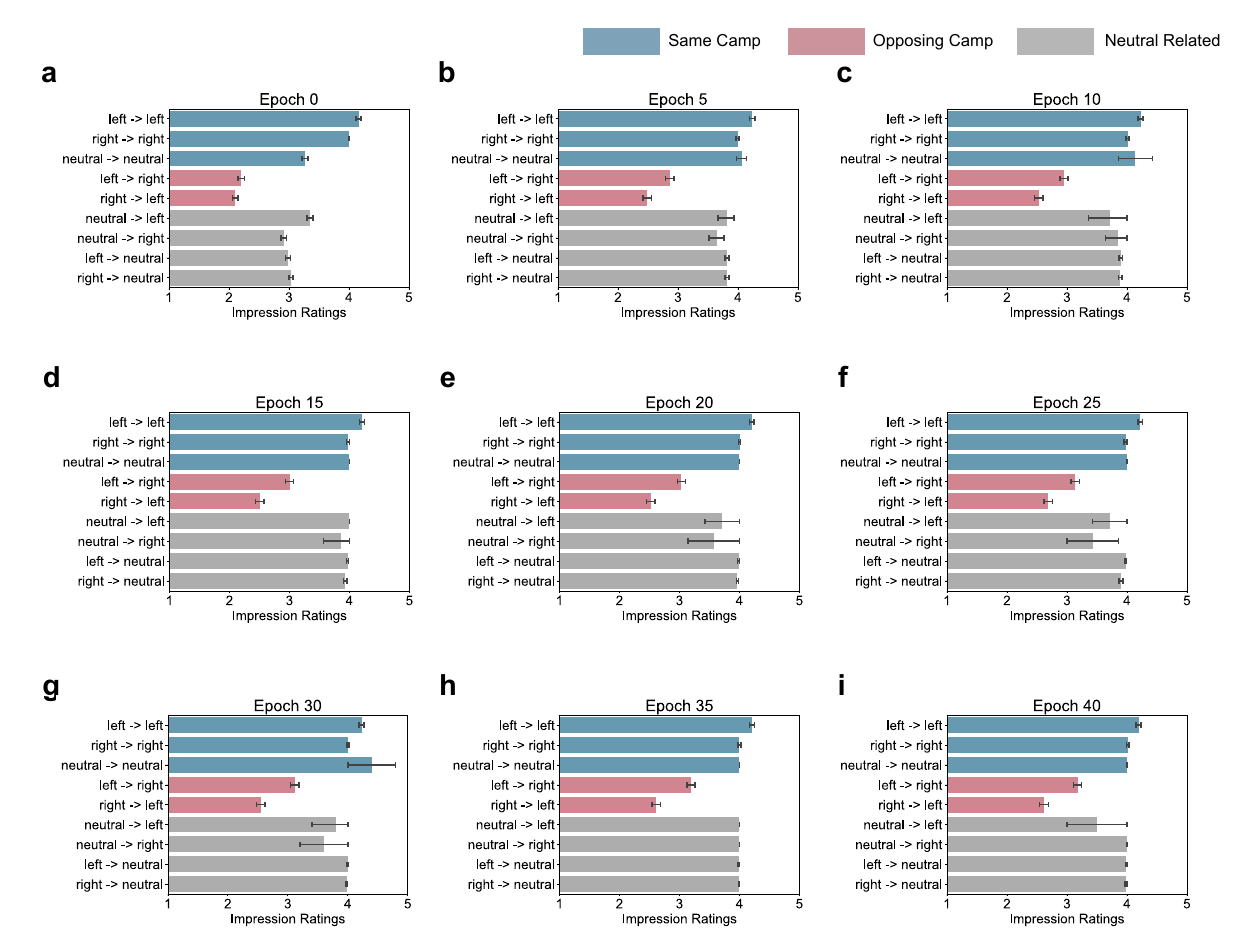}
\caption{\textbf{Impression ratings when agents discuss the issue of gun control in the networked system.} Bars represent mean values, and error bars represent the corresponding 95\% CIs. Each plot includes ratings from 1,000 agents, each evaluating three randomly selected agents: one with similar opinions from the same camp, one with opposing opinions from a different camp, and one with neutral opinions, yielding a total sample size of $n=3,000$.}
\label{fig:gun_rating}
\end{figure}

\begin{figure}[h!]
\centering
\includegraphics[width=\textwidth]{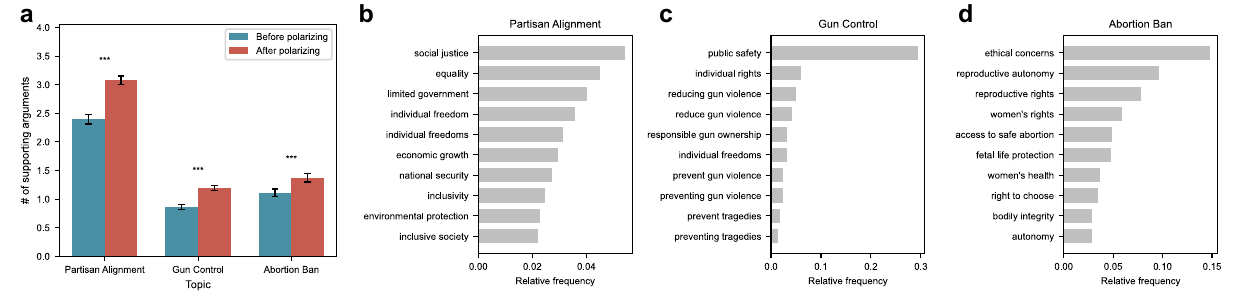}
\caption{\textbf{Accumulation of supporting arguments during polarization.} (a) Average number of self-reported supporting arguments generated by agents before and after polarization across three issues. Bars represent mean values, and error bars represent the corresponding 95\% CIs. Each bar includes responses from 500 randomly sampled agents either before or after they adopted more polarized opinions ($n=500$). Each response contains zero or more self-reported supporting arguments. (b–d) Relative frequency of argument themes appearing in agents' responses.}
\label{fig:argument}
\end{figure}

\begin{figure}[h!]
\centering
\includegraphics[width=\textwidth]{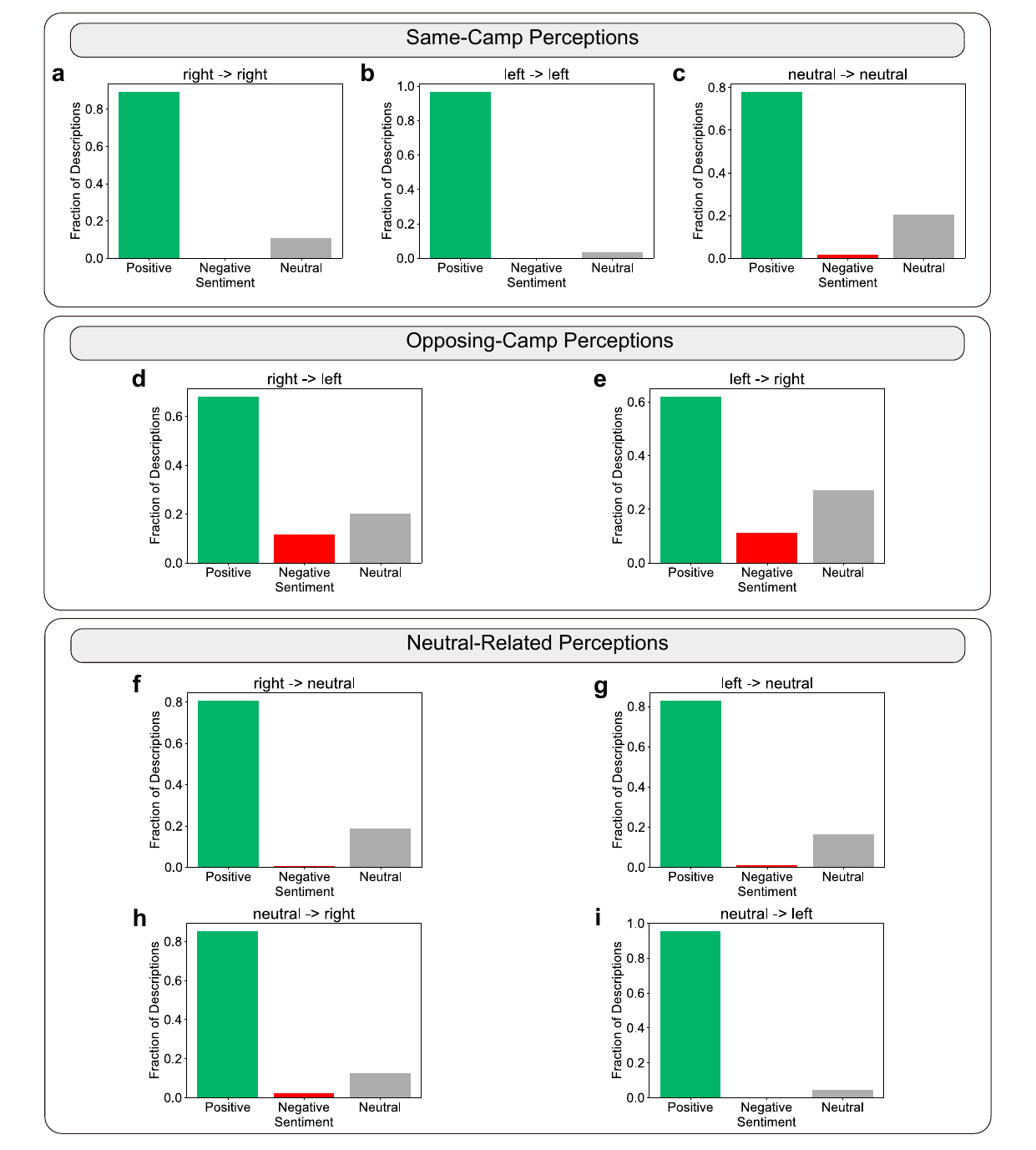}
\caption{\textbf{Distribution of descriptions among different sentiments in the discussion of partisan alignment.}}
\label{fig:politics_dist}
\end{figure}

\begin{figure}[h!]
\centering
\includegraphics[width=\textwidth]{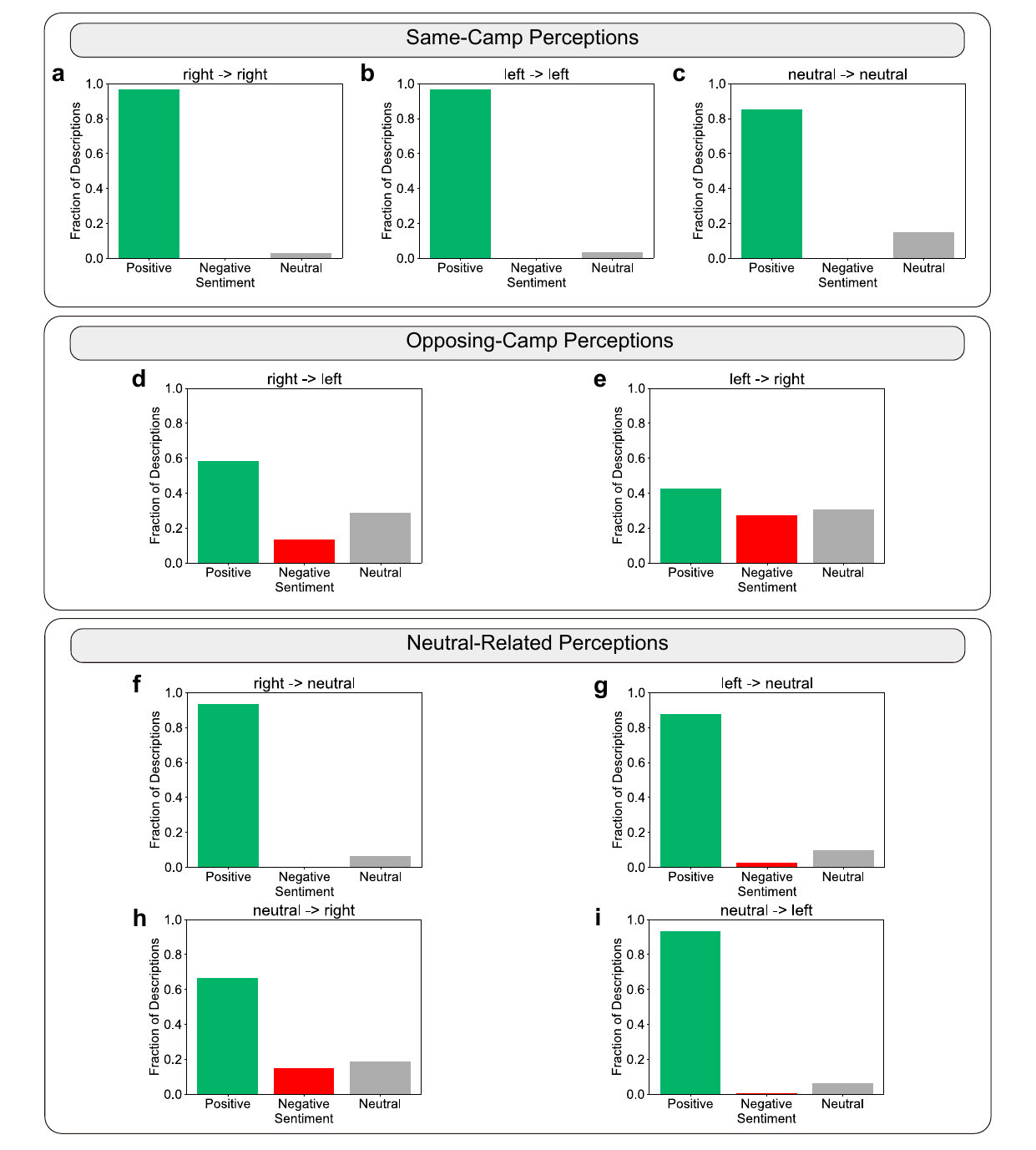}
\caption{\textbf{Distribution of descriptions among different sentiments in the discussion of the abortion ban.}}
\label{fig:abortion_dist}
\end{figure}

\begin{figure}[h!]
\centering
\includegraphics[width=\textwidth]{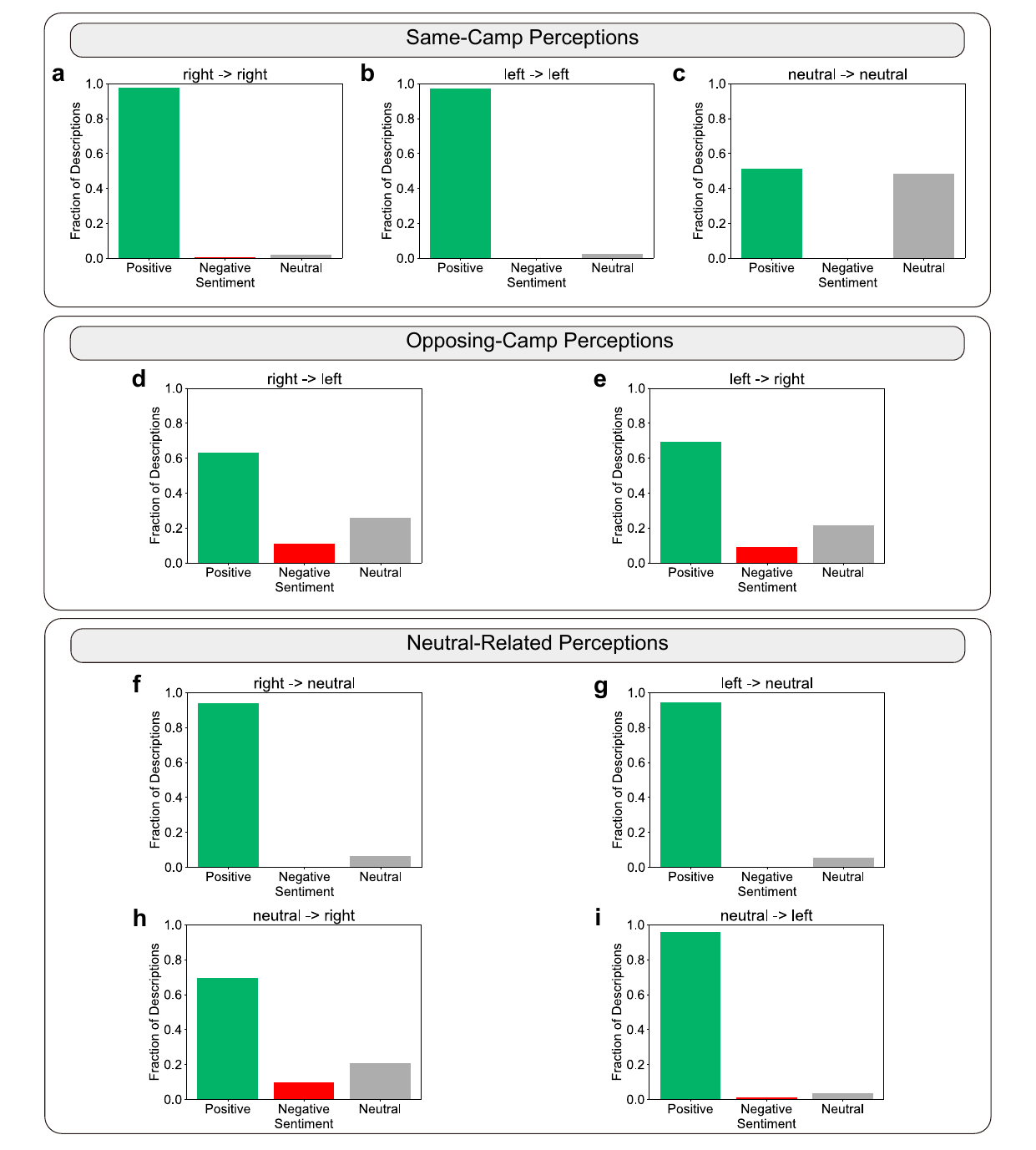}
\caption{\textbf{Distribution of descriptions among different sentiments in the discussion of gun control.}}
\label{fig:gun_dist}
\end{figure}

\begin{figure}[h!]
\centering
\includegraphics[width=\textwidth]{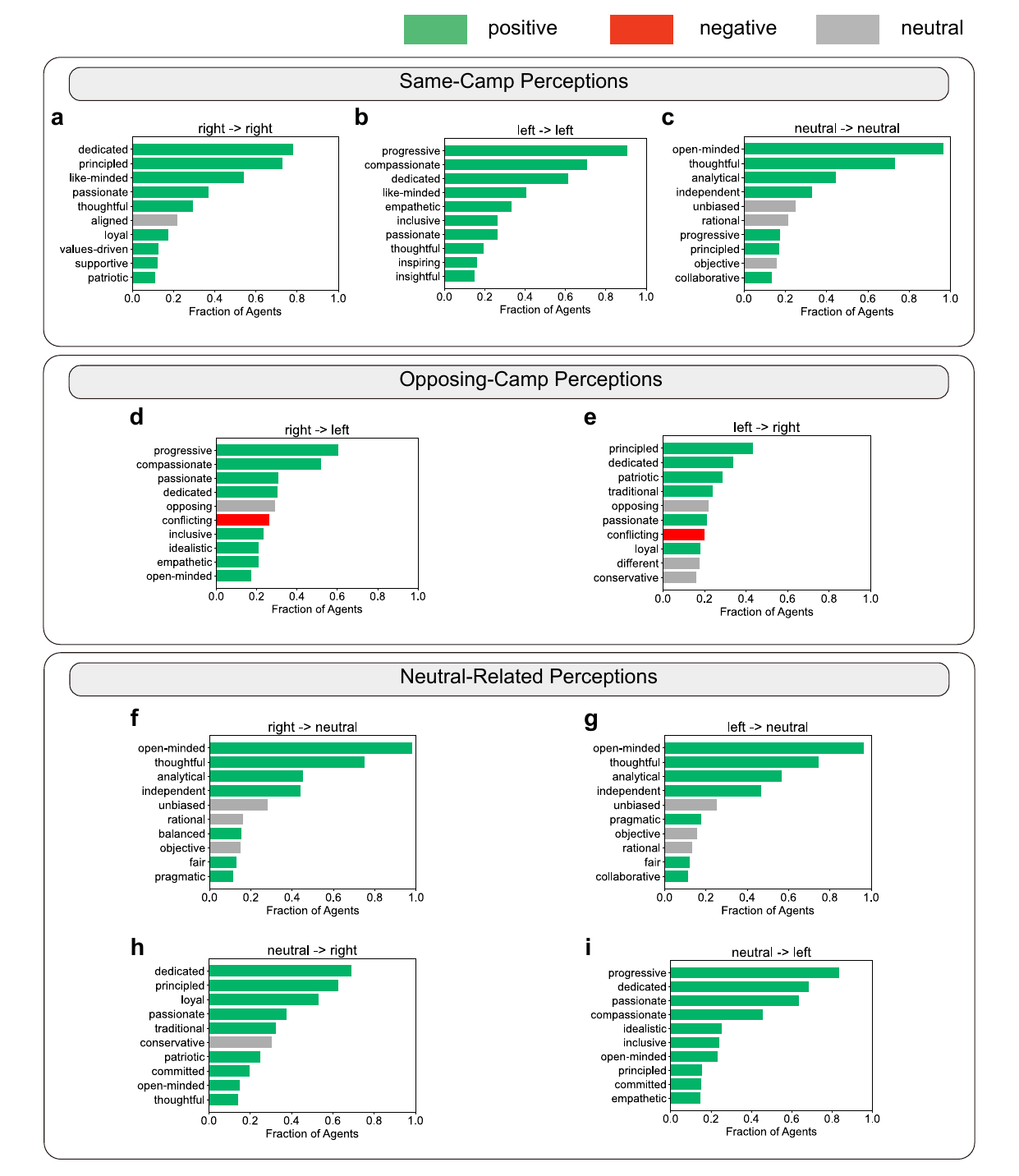}
\caption{\textbf{The 10 most frequently mentioned adjectives in the discussion of partisan alignment.}}
\label{fig:politics_discp}
\end{figure}

\begin{figure}[h!]
\centering
\includegraphics[width=\textwidth]{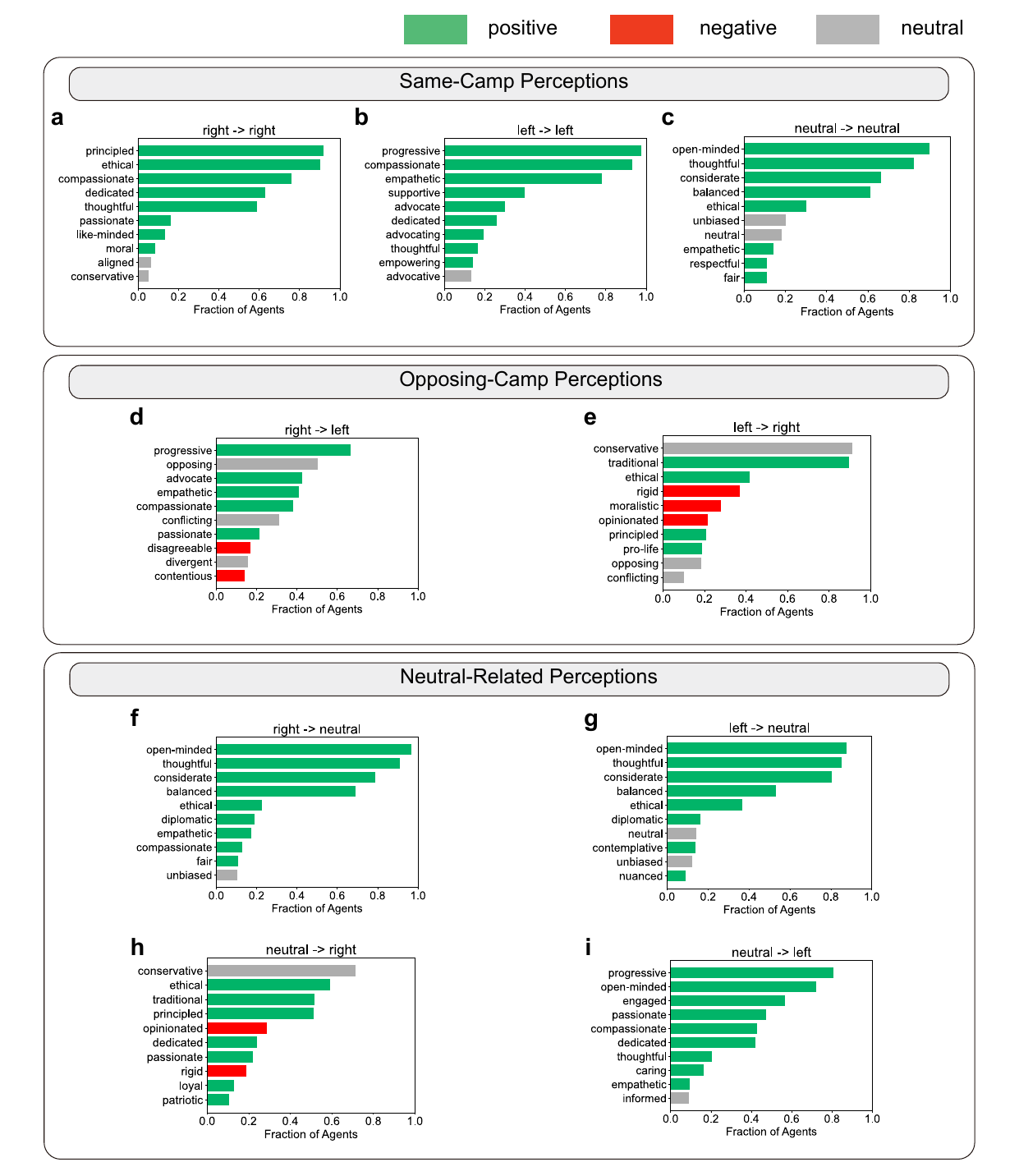}
\caption{\textbf{The 10 most frequently mentioned adjectives in the discussion of the abortion ban.}}
\label{fig:abortion_discp}
\end{figure}

\begin{figure}[h!]
\centering
\includegraphics[width=\textwidth]{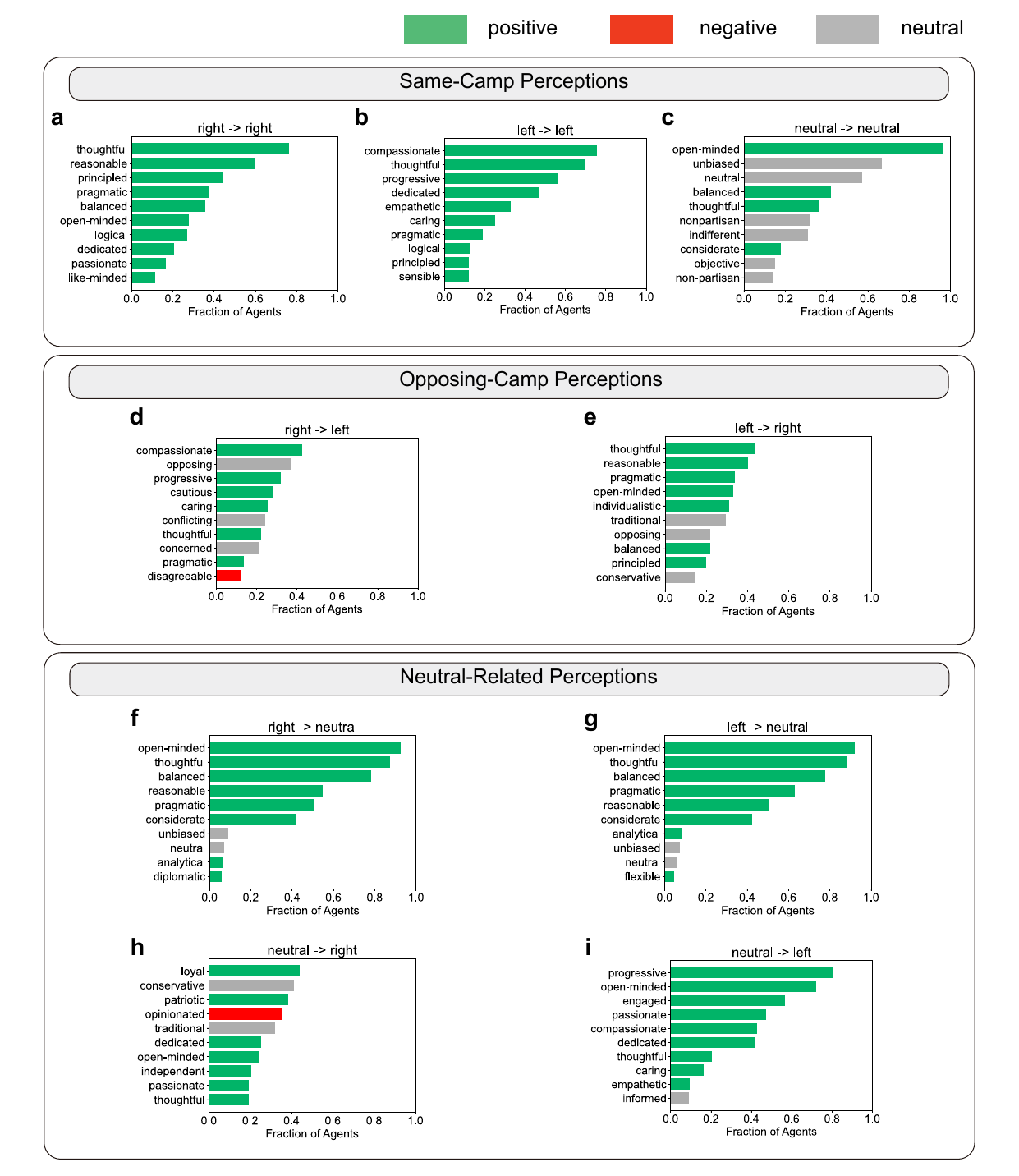}
\caption{\textbf{The 10 most frequently mentioned adjectives in the discussion of gun control.}}
\label{fig:gun_discp}
\end{figure}

After analyzing agents’ social perceptions, we next examine the psychological processes underlying their opinion updates. To clarify, the proposed system does not explicitly assume or hard-code specific mechanisms (e.g., confirmation bias or motivated reasoning) at this stage. Instead, the only assumption is that LLM agents update their opinions based on their current state and the messages received from their socially connected peers. The psychological processes emerge naturally as observable patterns in the outputs produced by the underlying LLMs in response to social inputs.

\rvss{That said, by analyzing how agents update their opinions, we can characterize behavioural patterns, in a manner analogous to how social scientists study human participants}. In line with this, we analyze the self-reported reasons generated by agents during opinion updates, with results summarized in Table~\ref{tab:psychological-processes}. \rvss{The most frequently observed patterns are consistent with} confirmation bias, reflecting agents’ tendency to favor information aligned with their prior beliefs; motivated reasoning, which further strengthens these tendencies through selective interpretation; and moral conviction, which anchors opinions more firmly by framing them as moral standards. Notably, these processes closely mirror well-documented mechanisms in human social psychology, further supporting the plausibility of the agents’ emergent behaviours.

\rvss{To scrutinize the connection between the polarization dynamics observed in our synthetic LLM-agent testbed and the established argument-accumulation lineage in opinion-dynamics research~\cite{mas2013differentiation,flache2017models}, we analyze how agents' self-reported supporting reasons evolve as their opinions become more polarized. Specifically, we compare the supporting reasons generated by agents before and after adopting more polarized opinions. We use a LLM to extract distinct arguments contained in these self-reported reasons. As shown in Figure~\ref{fig:argument}a, agents who become more polarized tend to accumulate a significantly larger number of supporting arguments over time across all three topics. Figure~\ref{fig:argument}b–d further report the most frequently occurring argument themes for each topic after polarization. The distribution of popular arguments varies across topics, reflecting issue-specific reasoning patterns. Overall, these findings provide additional evidence that the results in the synthetic system are consistent with argument-accumulation models.}

\clearpage

\begin{longtable}{
    >{\raggedright\arraybackslash}p{3cm}|
    >{\raggedright\arraybackslash}p{3cm}|
    >{\raggedright\arraybackslash}p{8cm}
}
\caption{\textbf{Psychological process emerges at the opinion update stage.}} \label{tab:psychological-processes} \\

\toprule
\textbf{Topic} & \textbf{Opinion Update Direction} & \textbf{Top 3 Psychological Processes and Their Descriptions}  \\
\midrule
\endfirsthead

\toprule
\textbf{Topic} & \textbf{Opinion Update Direction} & \textbf{Top 3 Psychological Processes and Their Descriptions}  \\
\midrule
\endhead

\midrule 
\endfoot

\bottomrule 
\endlastfoot

\multirow[t]{9}{*}{\textbf{Partisan Alignment}} 
& More Polarized & 
\textbullet\ Confirmation Bias (97\%): Agents selectively focus on information that aligns with their pre-existing beliefs, reinforcing and polarizing their opinions.\\
& & \textbullet\ Motivated Reasoning (57\%): Agents interpret new information in a way that supports their desired conclusions, driven by ideological goals, emotional commitments, or a desire to uphold their political identity.\\
& & \textbullet\ Value Alignment (26\%): Agents update their opinions based on perceived alignment between new information and their core personal values.\\
\cline{2-3}
& Unchanged &
\textbullet\ Confirmation Bias (71\%): Agents selectively focus on information that aligns with their pre-existing beliefs and values, reinforcing their current opinions while ignoring contradictory evidence.\\
& & \textbullet\ Belief Perseverance (44\%): Agents maintain their original opinion despite exposure to new information due to strong prior convictions and deeply ingrained beliefs.\\
& & \textbullet\ Motivated Reasoning (41\%): Agents selectively process information to maintain consistency with their pre-existing beliefs, values, or ideological commitments.\\
\cline{2-3}
& Less Polarized &
\textbullet\ Motivated Reasoning (44\%): Agents selectively process information to align with their pre-existing beliefs and values, reinforcing their political identity.\\
& & \textbullet\ Confirmation Bias (29\%): Agents selectively focus on information that aligns with their pre-existing beliefs, reinforcing their stance while disregarding opposing views.\\
& & \textbullet\ Value Alignment (28\%): Agents update their opinions based on perceived alignment between new information and their core values or political party principles.\\

\midrule
\multirow[t]{9}{*}{\textbf{Abortion Ban}} 
& More Polarized &
\textbullet\ Moral Conviction (84\%): Opinion updates are driven by strong moral or ethical beliefs, leading to more polarized and resistant stances.\\
& & \textbullet\ Confirmation Bias (69\%): Agents selectively focus on information that aligns with their pre-existing beliefs about abortion, reinforcing its polarized opinion.\\
& & \textbullet\ Motivated Reasoning (35\%): Agents selectively process information to align with their pre-existing beliefs and values, reinforcing polarized opinions.\\
\cline{2-3}
& Unchanged &
\textbullet\ Moral Conviction (70\%): Agents' opinions are strongly tied to deeply held moral or ethical principles, making them resistant to change based on new information.\\
& & \textbullet\ Confirmation Bias (65\%): Agents selectively focus on information that aligns with their pre-existing beliefs about reproductive rights, autonomy, or ethical concerns.\\
& & \textbullet\ Belief Perseverance (60\%): Agents maintain their original opinions despite new information due to strong prior beliefs.\\
\cline{2-3}
& Less Polarized &
\textbullet\ Cognitive Dissonance Reduction (81\%): Agents adjust their opinion to reduce discomfort from conflicting beliefs about women's rights and ethical concerns.\\
& & \textbullet\ Value-Based Reasoning (32\%): Agents update their opinions by weighing and prioritizing core values such as autonomy, reproductive rights, and ethical concerns.\\
& & \textbullet\ Motivated Reasoning (22\%): Agents adjust their opinions based on pre-existing values and motivations, seeking moderation or ethical consistency.\\

\midrule
\multirow[t]{9}{*}{\textbf{Gun Control}} 
& More Polarized &
\textbullet\ Motivated Reasoning (91\%): Agents selectively process information to align with pre-existing beliefs or values, reinforcing a polarized stance that prioritizes safety or individual rights.\\
& & \textbullet\ Confirmation Bias (90\%): Agents selectively focus on information that aligns with their pre-existing beliefs about gun control, reinforcing their polarized stance.\\
& & \textbullet\ Moral Conviction (16\%): Agents' opinions are driven by strong moral values like public safety and preventing harm.\\
\cline{2-3}
& Unchanged &
\textbullet\ Confirmation Bias (90\%): Agents selectively focus on information that aligns with their pre-existing beliefs about gun control.\\
& & \textbullet\ Motivated Reasoning (79\%): Agents selectively process new information to align with their pre-existing values and goals.\\
& & \textbullet\ Belief Perseverance (66\%): Agents maintain their original opinions despite exposure to new information.\\
\cline{2-3}
& Less Polarized &
\textbullet\ Cognitive Dissonance Reduction (50\%): Agents adjust their opinions to reduce discomfort from conflicting beliefs about individual rights and public safety.\\
& & \textbullet\ Motivated Reasoning (47\%): Agents adjust their opinion based on desired outcomes (e.g., public safety, reducing gun violence).\\
& & \textbullet\ Social Influence (27\%): Agents adjust their opinions toward moderation in response to perceived societal norms.\\

\end{longtable}

\section{Supplementary Note 9: Experiments on LLMs with Varying Temperatures}

To explore whether the temperature setting — which controls the diversity of LLM-generated outputs — affects the emergence of polarization, we conduct additional simulations with lower (0.5) and higher (1.5) temperature values. Figure~\ref{fig:temperature_1} shows agents' collective opinion dynamics while varying temperatures in their underlying LLMs. We observe that regardless of whether the temperature is lower (Figure~\ref{fig:temperature_1}a) or higher (Figure~\ref{fig:temperature_1}c), these agents spontaneously develop collective opinions through social interaction, leading to the emergence of polarization. Meanwhile, at the network level, temperature changes do not alter the homophily observed in naturalistic LLM social interactions. This indicates that temperature, despite being a key parameter for LLMs, has a limited effect on whether polarization emerges among a collective of interacting agents.

\begin{figure}[h!]
\centering
\includegraphics[width=0.9\textwidth]{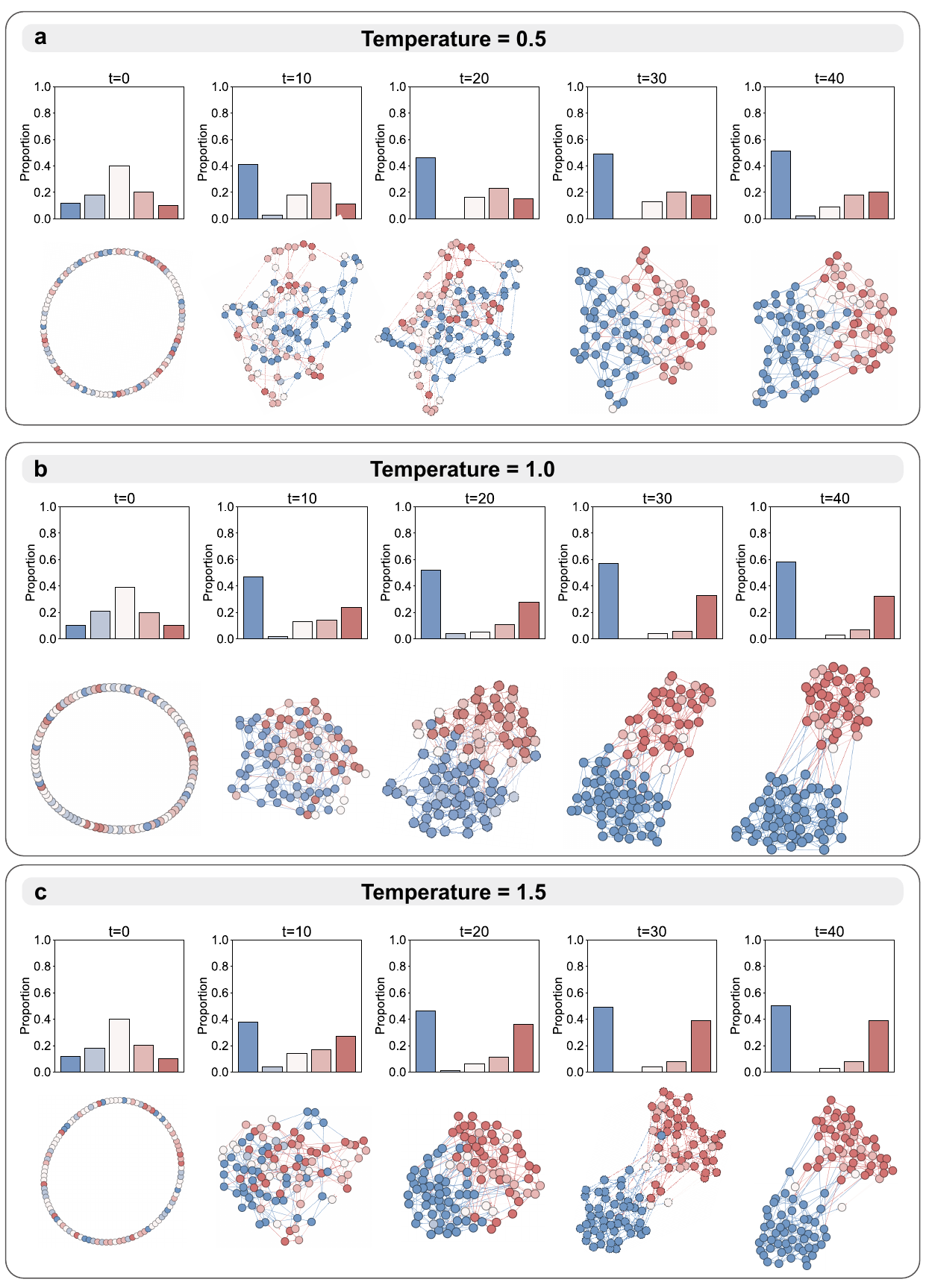}
\caption{\textbf{Evolution of LLM agents' collective opinions and network structures}, where the underlying temperature is \textbf{a,} 0.5, \textbf{b,} 1.0, and \textbf{c,} 1.5.}
\label{fig:temperature_1}
\end{figure}

\begin{figure}[h!]
\centering
\includegraphics[width=0.9\textwidth]{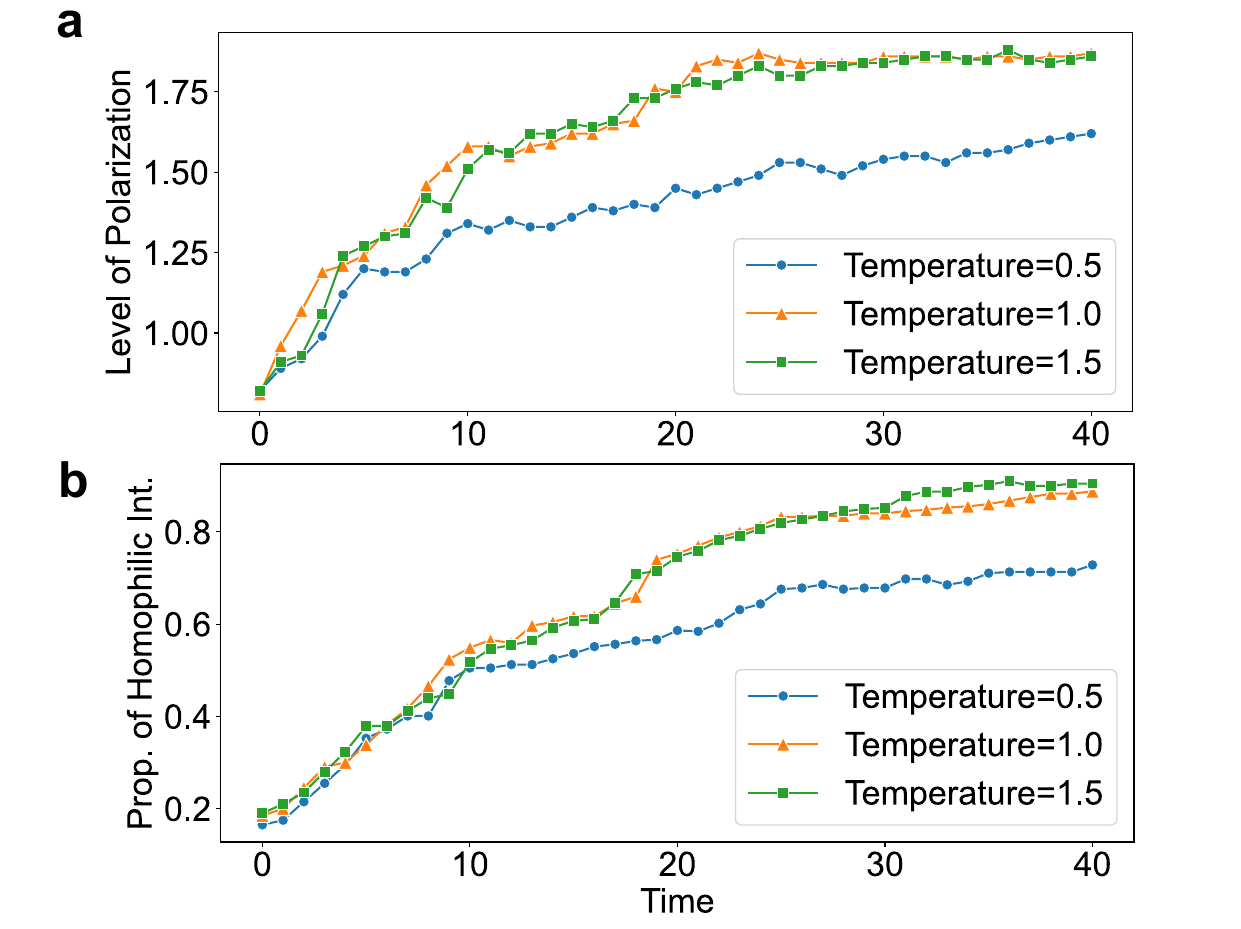}
\caption{\textbf{Changes in a, the level of polarization, and b, the proportion of homophilic interactions over time.}}
\label{fig:temperature_2}
\end{figure}

\begin{figure}[h!]
\centering
\includegraphics[width=0.9\textwidth]{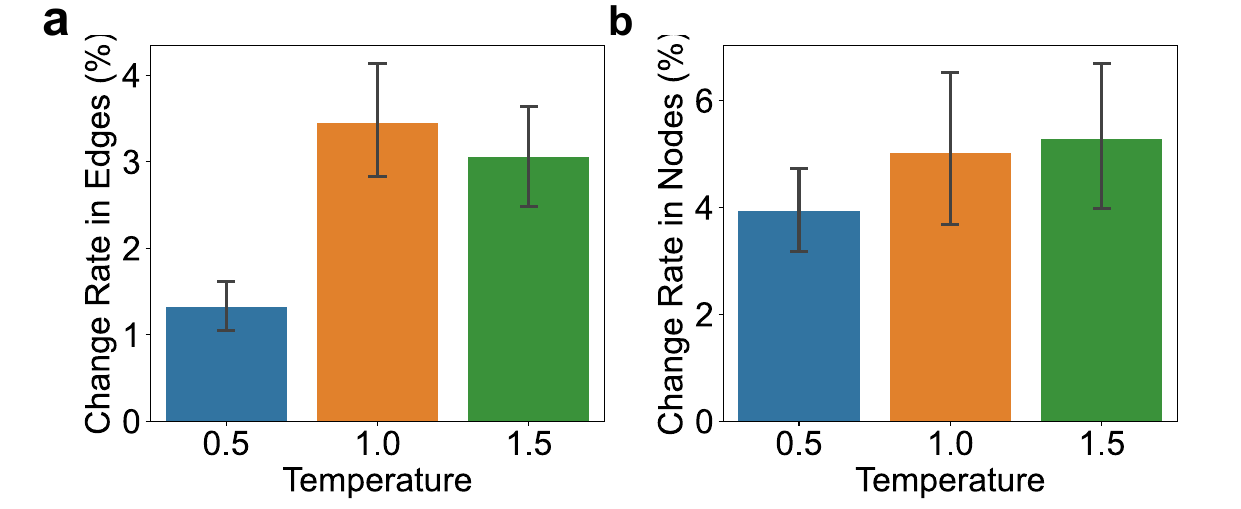}
\caption{\textbf{Effects of temperature on the evolution a, of edges (i.e., social relationships between LLM agents), and b, nodal states (i.e., agents' opinions) in the social network. } Here, bars represent mean values, and error bars represent the corresponding 95\% CIs. Each bar includes 40 samples, with each sample corresponding to the change from one timestep to the previous timestep.}
\label{fig:temperature_3}
\end{figure}
\clearpage

Furthermore, to precisely evaluate temperature effects, we measure both the level of polarization and proportion of homophilic interactions. As presented in Figure~\ref{fig:temperature_2}, our results show that while a higher temperature does not have a noticeable impact, a lower temperature markedly decreases the level of polarization in LLM agents’ opinions and reduces the tendency of agents with similar opinions to cluster. This observation raises an interesting question: how does the diversity of LLM outputs, parameterized by temperature, affect the social behaviours of these agents, and in turn shape the level of polarization?

To answer this question, we further compute the change rate of edges, reflecting how frequently agents switch their communication partners, as well as the change rate of nodal states, indicating how often agents revise their opinions, across different temperature settings. Figure~\ref{fig:temperature_3} illustrates the average values of both change rates over the entire simulation period. Interestingly, we observe that when the temperature is lowered to 0.5, agents exhibit a significantly lower propensity to switch communication partners, instead showing a strong preference for maintaining existing relationships (one-way ANOVA, F(2,117)= 16.90, p << .001; two-sided Student’s t-test, temperature 0.5 vs. 1.5, t=-5.79, p<<.001). Notably, agents driven by LLMs with a temperature of 0.5 adjust only 1.32\% of their edges on average per timestep, which is less than half the rate observed at temperatures of 1.0 or 1.5. Their preference for maintaining existing relationships keeps them within their initial random networks, rather than selectively approaching more homophilic peers (Figure~\ref{fig:temperature_3}b), thereby contributing to the observed reduction in polarization level. Meanwhile, a lower temperature also moderately suppresses agents’ tendency to adopt new opinions, although this effect is not statistically significant (Figure~\ref{fig:temperature_3}b). This tendency, combined with their reluctance to change communication partners, further contributes to the reduction in observed polarization.

\section{Supplementary Note 10: Experiments on LLMs with Enhanced Prompting}

We first explore whether seeding a minority of agents with a markedly different reasoning strategy can temper or amplify polarization. Specifically, we equip 10\% of all agents with chain-of-thought (CoT) prompting, enabling them to think step by step as they establish social relationships, communicate, and form opinions on political issues. These CoT agents are seeded at the beginning of the simulation and randomly distributed across the network. The proportion of CoT agents across different opinions aligns with the overall initialization distribution.

Comparing the evolution of the original system with that of the system seeded with 10\% CoT agents (Figures~\ref{fig:cot1} and \ref{fig:cot2}), we observe a substantial shift: the proportion of agents holding moderate or neutral opinions increases from 8\% to 38\% (Figure~\ref{fig:cot1}), and the overall polarization level drops by 15\% (Figure~\ref{fig:cot2}a). Meanwhile, the network-level division between agents with opposing opinions is largely alleviated, with final modularity scores declining from 0.43 to 0.25 (Figure~\ref{fig:cot2}b; a modularity score above 0.3 typically marks clear community structures). Overall, these observations suggest that even a small minority of agents equipped with a markedly different reasoning strategy and seeded into the system can effectively temper polarization \rvss{in the synthetic testbed}. \rvss{These results illustrate how the proposed synthetic LLM-agent testbed can be used to generate and examine hypotheses about how heterogeneity in agent-level reasoning configurations may shape collective opinion dynamics.}


\begin{figure}[t]
\centering
\includegraphics[width=0.9\textwidth]{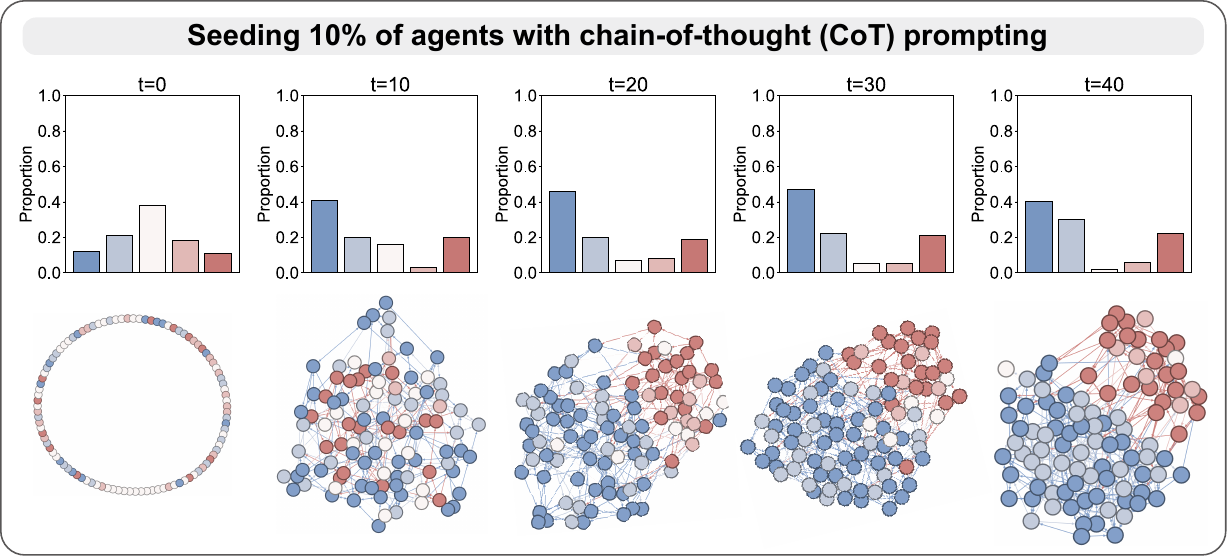}
\caption{\textbf{Evolution of LLM agents' collective opinions and network structures}, where we seed 10\% of agents with chain-of-thought (CoT) prompting. All agents are driven by GPT-3.5 and they discuss the political issue of abortion bans.}
\label{fig:cot1}
\end{figure}

\begin{figure}[h!]
\centering
\includegraphics[width=0.85\textwidth]{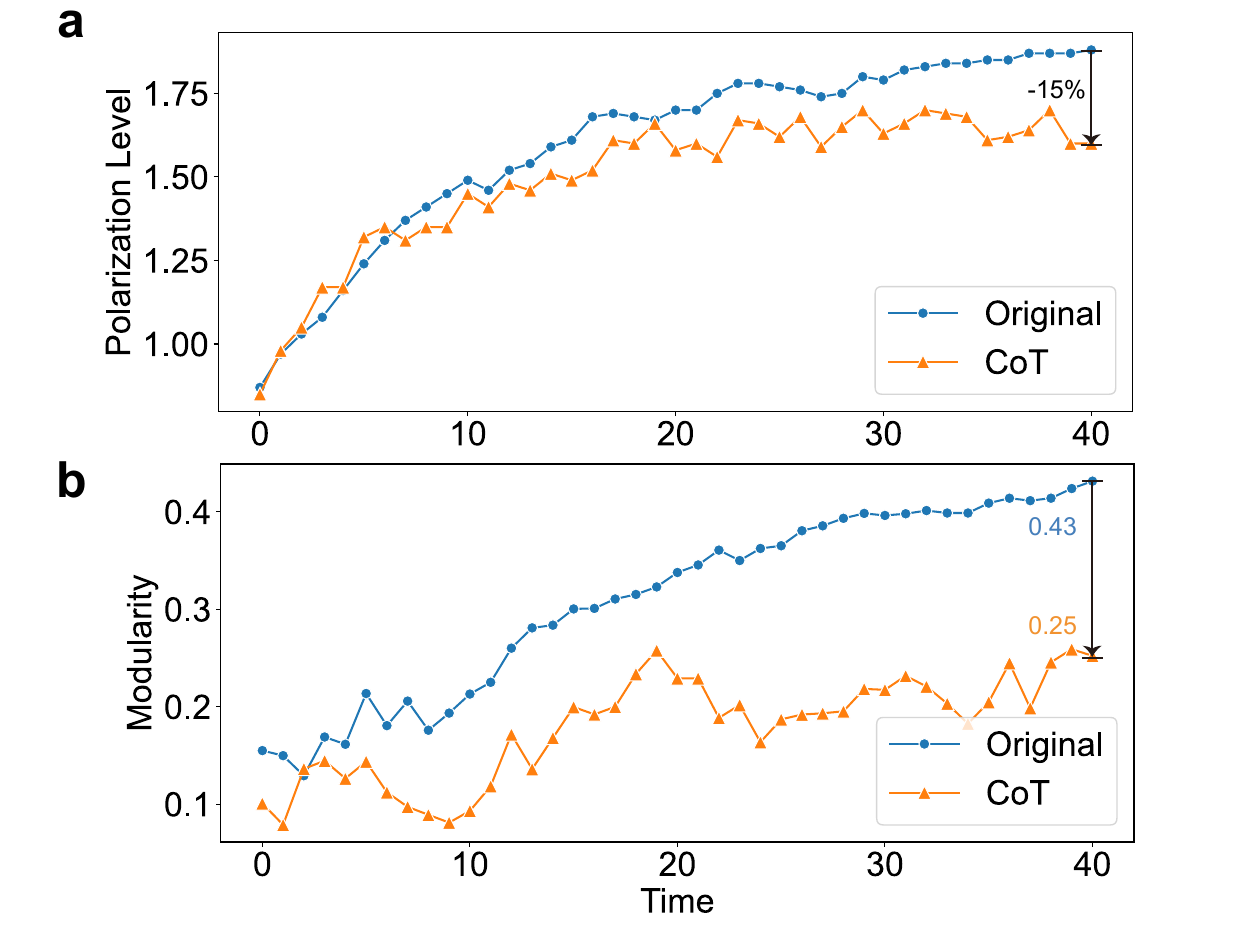}
\caption{\textbf{Comparison between the original system and a system where 10\% of agents are seeded with CoT prompting}, in terms of \textbf{a,} polarization level, and \textbf{b,} network modularity.}
\label{fig:cot2}
\end{figure}

Furthermore, we investigate whether the proposed system can be extended, \rvss{through enhanced prompting, to instantiate agents with subgroup-inspired profiles that reflect heterogeneous behavioural tendencies observed in human populations.}. For example, we have conducted a simple case shown in Figure~\ref{fig:examples_all}, where we use LLMs to generate two representative profiles of human subgroups in the U.S. and incorporate these profiles into the process of forming political opinions. Results show that LLM agents can simulate the two subgroups, capturing their distinctive behavioural patterns. Overall, this extended experiment illustrates that the proposed system remains flexible for modeling diverse human subgroups and their behavioural differences. We note that, in the main experiments of the paper, we intentionally do not insert additional subgroup-specific information into the agents' prompts in order to preserve the spontaneous nature of their behaviour.

\begin{figure}[h!]
\centering
\includegraphics[width=0.85\textwidth]{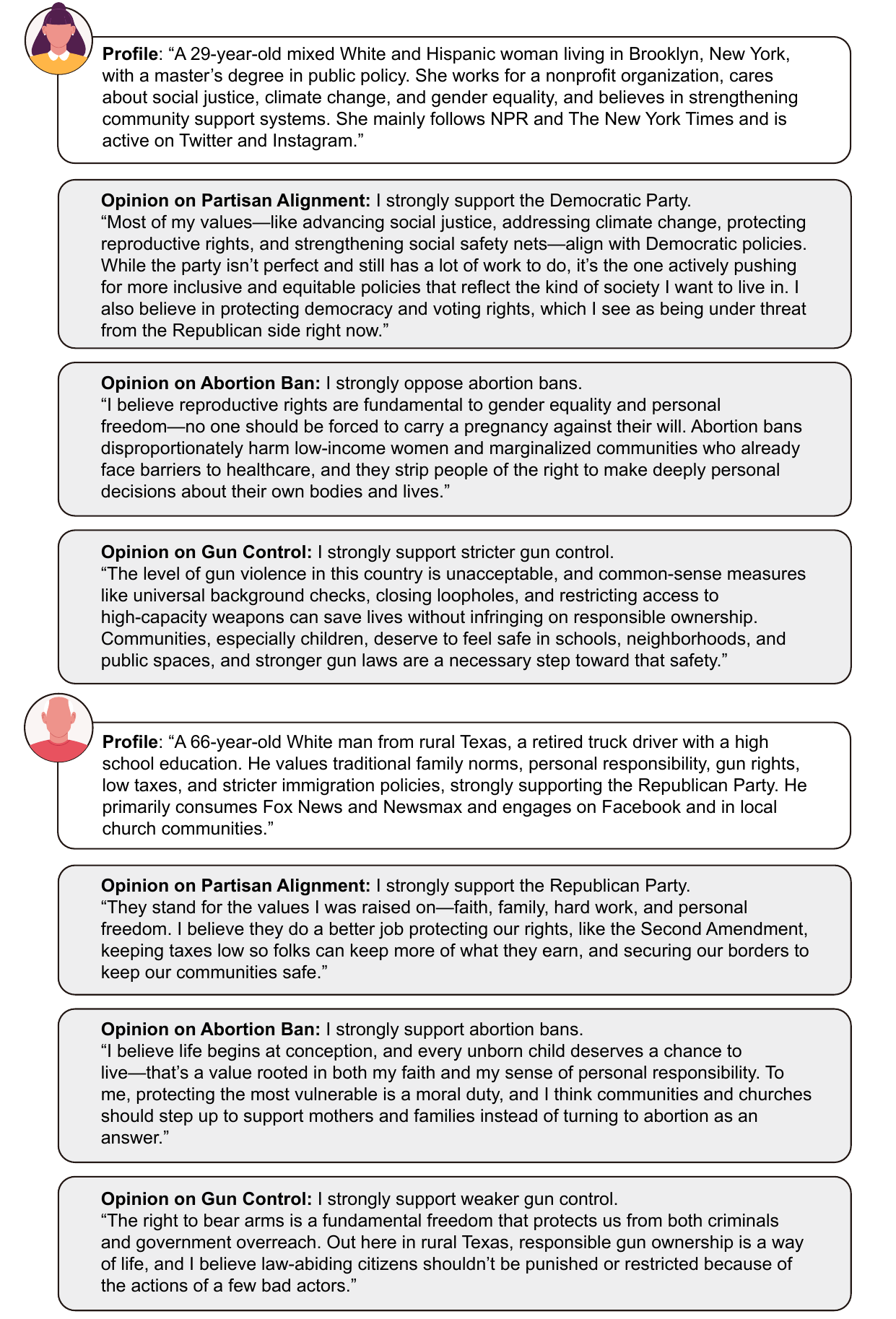}
\caption{\textbf{Two representative profiles of human subgroups in the U.S. and their different behaviours when discussing political issues.} Credit: Profile icons designed by Magnific.}
\label{fig:examples_all}
\end{figure}

\section{Supplementary Note 11: Experiments on the Effect of Increasing Severing Costs}

We examine the sensitivity of our results to the cost of severing social ties. To explore whether even minimal friction can slow the formation of ideological silos, we introduce a simple mechanism in which agents’ decisions to sever existing social ties are blocked with a fixed probability (e.g., 10\%), simulating minimal social friction. Figures~\ref{fig:severing1} and \ref{fig:severing2} show the results under this minimal friction condition, where the cost of severing ties is increased by only 10\%. Notably, while the minimal friction leads to marginal mitigation in polarization levels (Figure~\ref{fig:severing2}a), its effect on suppressing the formation of ideological silos is substantial, as consistently reflected across four different network-level metrics (Figures~\ref{fig:severing2}b–e). This experiment suggests that even minimal social friction can play a nontrivial role in slowing the formation of ideological silos and fostering communication between opposing camps. 


\begin{figure}[h!]
\centering
\includegraphics[width=0.75\textwidth]{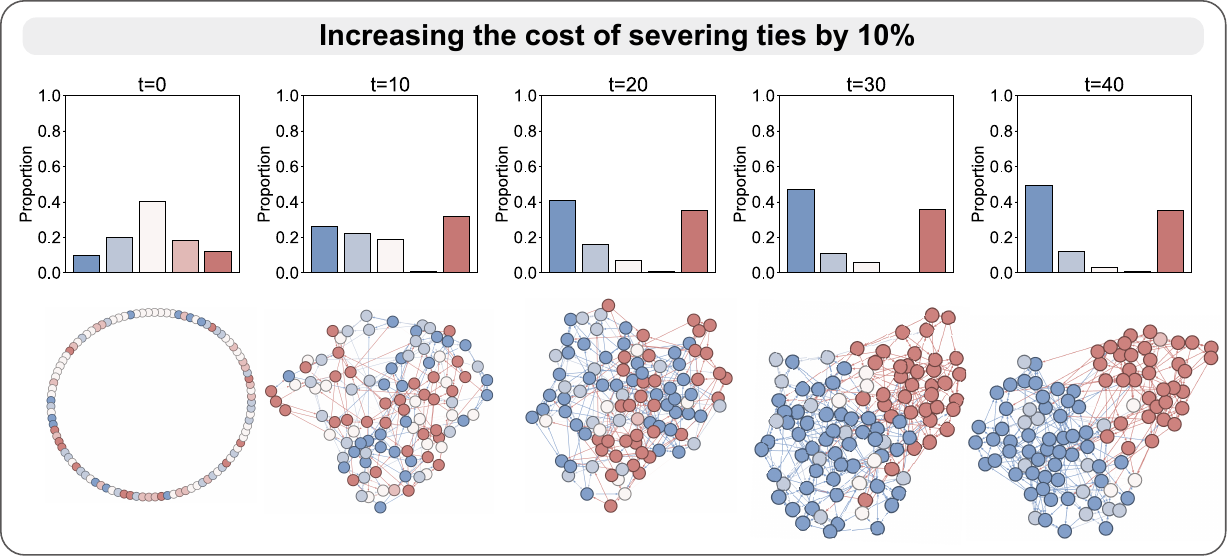}
\caption{\textbf{Evolution of LLM agents' collective opinions and network structures}, where we increase the cost of severing ties by 10\%. All agents are driven by GPT-3.5 and they discuss the political issue of abortion bans.}
\label{fig:severing1}
\end{figure}

\begin{figure}[h!]
\centering
\includegraphics[width=0.72\textwidth]{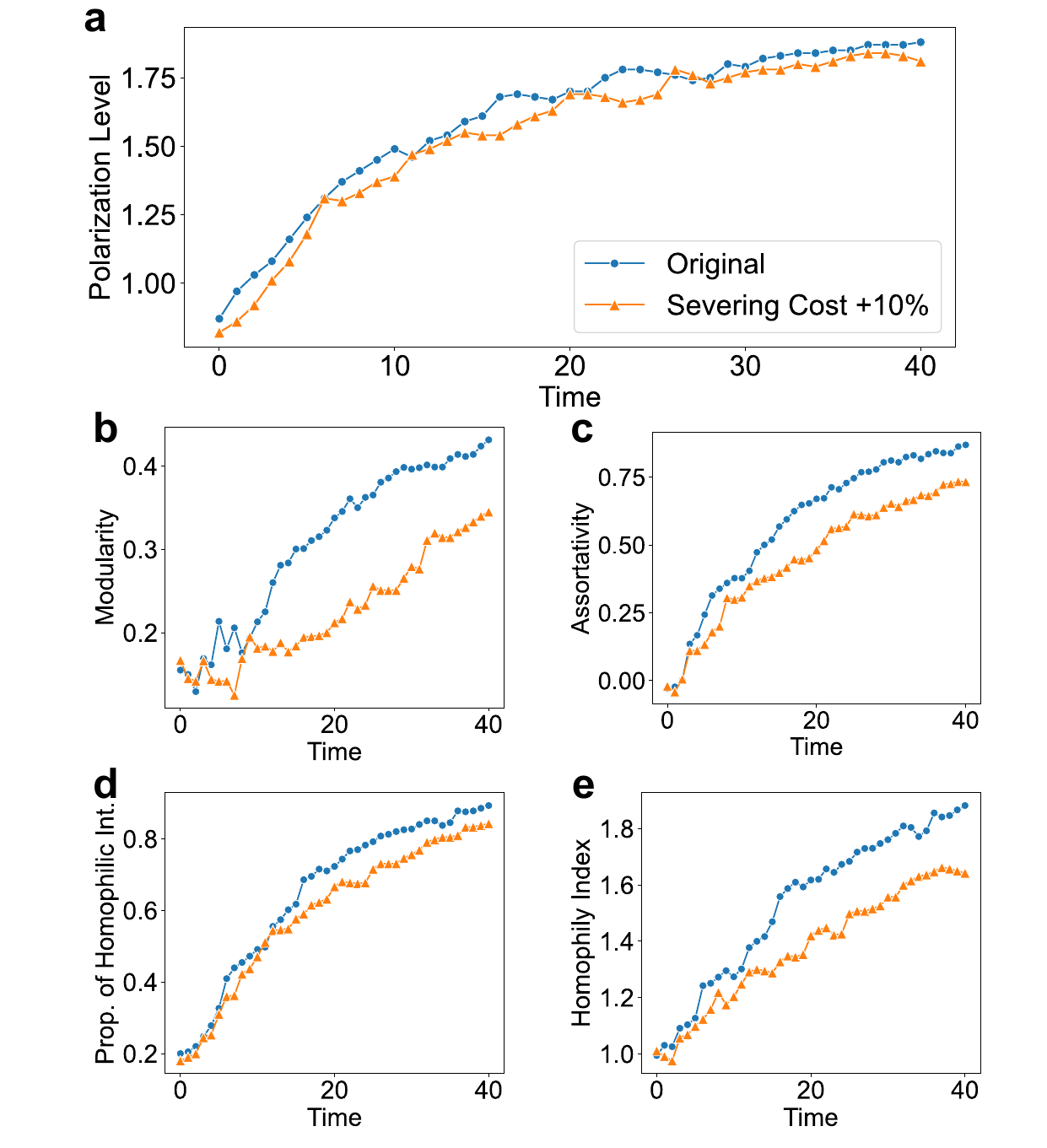}
\caption{\textbf{Comparison between the original system and a system where the cost of severing ties is increased by 10\%}, in terms of \textbf{a,} polarization level, \textbf{b,} network modularity, \textbf{c,} assortativity, \textbf{d,} proportion of homophilic interactions, and \textbf{e,} homophily index.}
\label{fig:severing2}
\end{figure}

\section{Supplementary Note 12: Results of Simulation under Different Initial Conditions}

To assess the robustness and generalizability of the proposed system, we investigate how different initial conditions affect the emergent collective behaviours of LLM agents. Specifically, we focus on two key aspects of initialization: the initial distribution of agents’ opinions and the initial structure of the social network.

We first modify the initial opinion distribution by transitioning from a near-Gaussian distribution of $[0.1, 0.2, 0.4, 0.2, 0.1]$ to a highly centralized distribution of $[0, 0.1, 0.8, 0.1, 0]$. In this setting, a system that initially lacks polarized agents should be less likely to become polarized. Figure~\ref{fig:centralized} shows the evolution of collective opinions in the system. We observe that the system, though taking a longer time to evolve, eventually converges to a polarized opinion distribution. Along with the emergence of opinion polarization, the social network gradually splits into two communities with opposing opinions. To sum up, this experiment demonstrates that even when the \rvss{synthetic} system is initialized in a setting unlikely to produce polarization, long-term free-form social interactions among LLM agents still lead to opinion polarization.

\begin{figure}[h]
\centering
\includegraphics[width=0.75\textwidth]{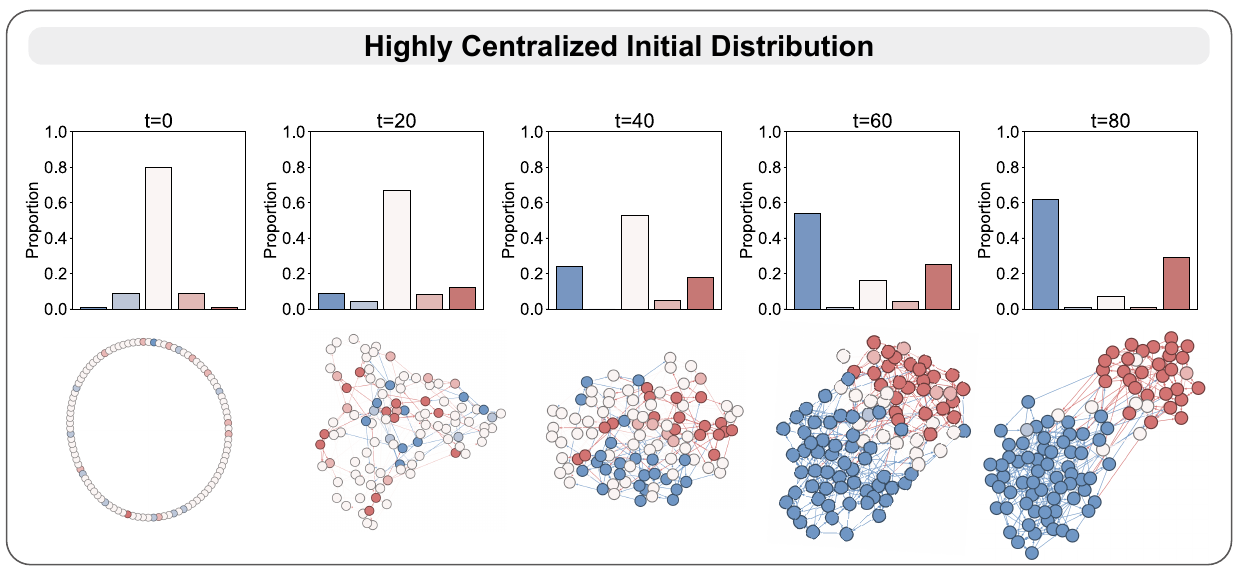}
\caption{\textbf{Evolution of LLM agents' collective opinions and network structure}, where the network is initialized using a highly centralized distribution, and agents discuss the political issue of abortion ban.}
\label{fig:centralized}
\end{figure}

We also conduct experiments to explore whether the initial network affects the emergent collective behaviours of LLM agents. Specifically, we initialize LLM agents’ social network with an Erdős–Rényi model, a Barabási–Albert model, and Watts–Strogatz models with rewiring probabilities of 0.001 and 0.05. In the BA model setup, a subset of agents naturally become highly connected nodes, resembling opinion leaders who exert disproportionate influence on others. Figures~\ref{fig:er}–\ref{fig:ws05} show how these agents self-organize their social networks and collective opinions under different initial conditions. All networks are visualized using the ForceAtlas2 algorithm with identical parameters in Gephi, ensuring direct visual comparability. We find that despite substantial differences in initial network structure, the final outcomes are similar. In all these cases, agents spontaneously organize into two well-defined communities: one that predominantly supports left-leaning opinions and the other right-leaning ones. Moreover, the final opinion distributions converge into a stable polarized pattern, demonstrating that the emergence of polarization and homophilic clustering is not driven by initial configurations, equitable or inequitable. Rather, these patterns are inherently driven by agents' autonomous social interactions. Overall, these results demonstrate the robustness and generalizability of our findings across diverse initial network structures, underscoring that LLM agents, through autonomous interactions, consistently self-organize into polarized communities in ways not determined by initial conditions.

\begin{figure}[h]
\centering
\includegraphics[width=0.75\textwidth]{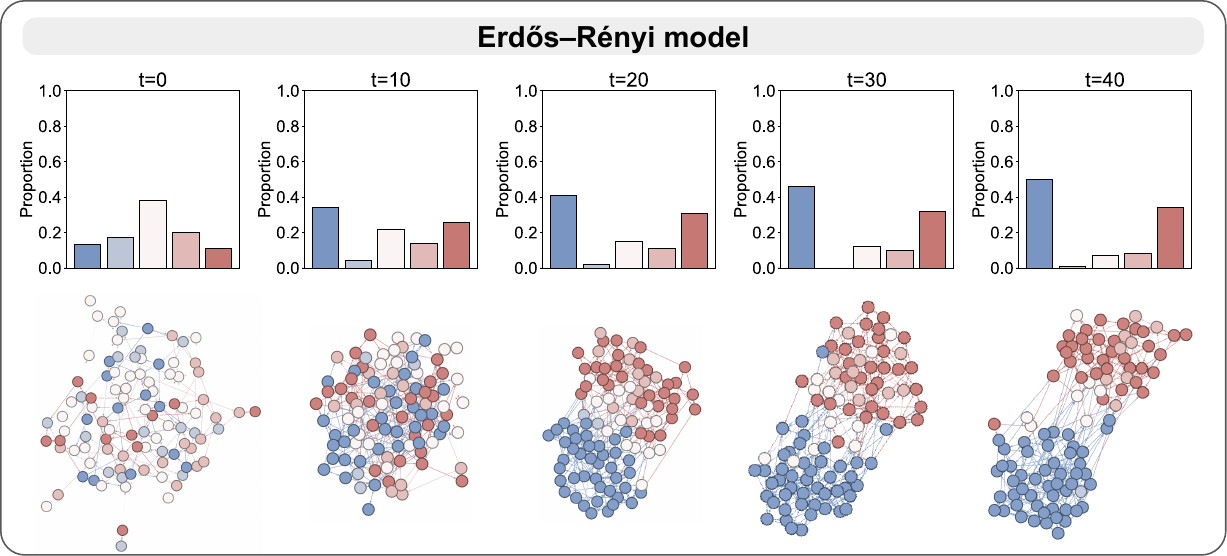}
\caption{\textbf{Evolution of LLM agents' collective opinions and network structures}, where the network is initialized using an Erdős–Rényi model with an average degree of 4, and agents discuss the political issue of abortion ban.}
\label{fig:er}
\end{figure}

\begin{figure}[h]
\centering
\includegraphics[width=0.75\textwidth]{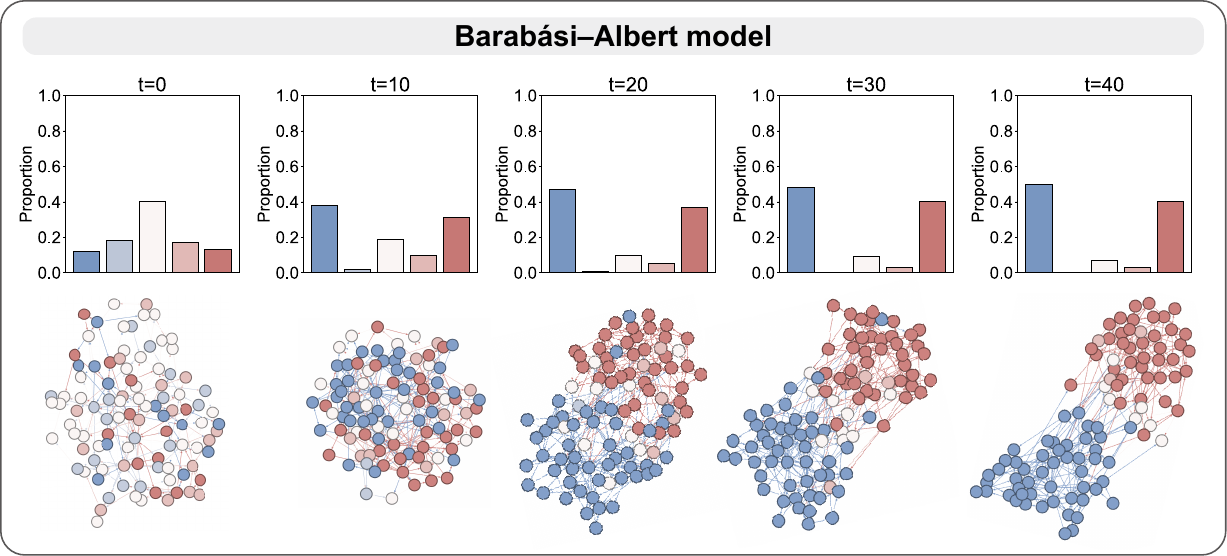}
\caption{\textbf{Evolution of LLM agents' collective opinions and network structures}, where the network is initialized using a Barabási–Albert model with an average degree of 4, and agents discuss the political issue of abortion ban.}
\label{fig:ba}
\end{figure}

\begin{figure}[h]
\centering
\includegraphics[width=0.75\textwidth]{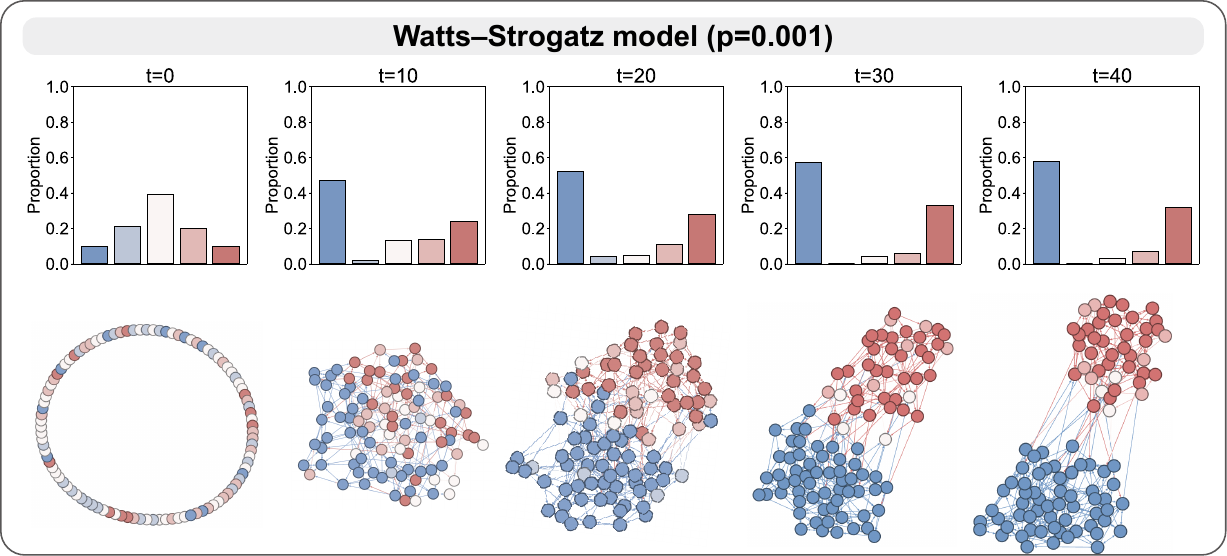}
\caption{\textbf{Evolution of LLM agents' collective opinions and network structures}, where the network is initialized using a Watts–Strogatz model with the rewiring probability of 0.001, and agents discuss the political issue of abortion ban.}
\label{fig:ws001}
\end{figure}

\begin{figure}[h]
\centering
\includegraphics[width=0.75\textwidth]{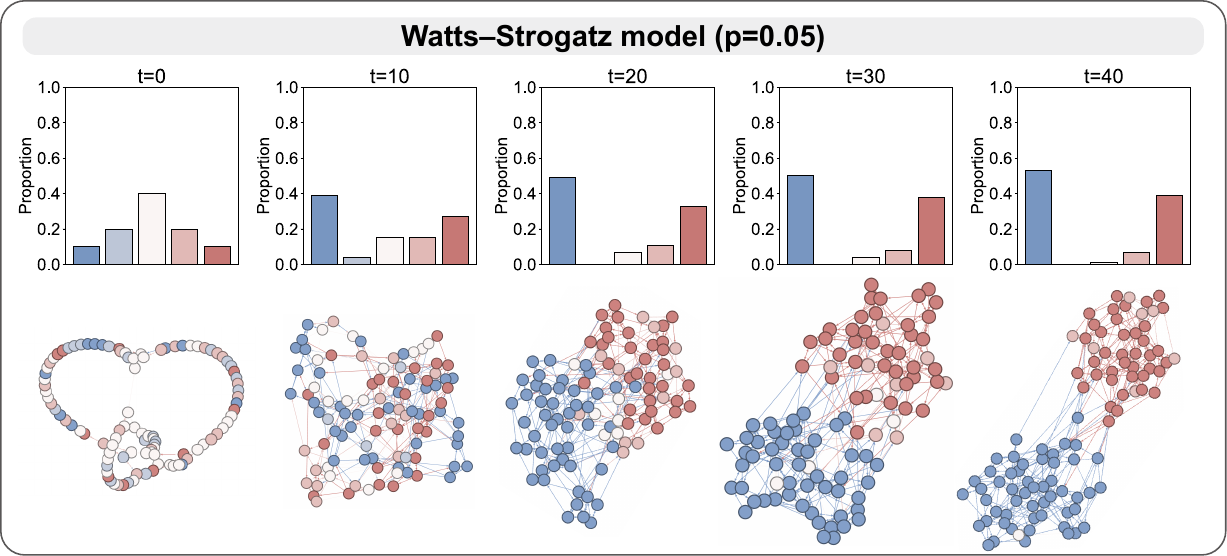}
\caption{\textbf{Evolution of LLM agents' collective opinions and network structures}, where the network is initialized using a Watts–Strogatz model with the rewiring probability of 0.05, and agents discuss the political issue of abortion ban.}
\label{fig:ws05}
\end{figure}

\begin{figure}[h]
\centering
\includegraphics[width=0.75\textwidth]{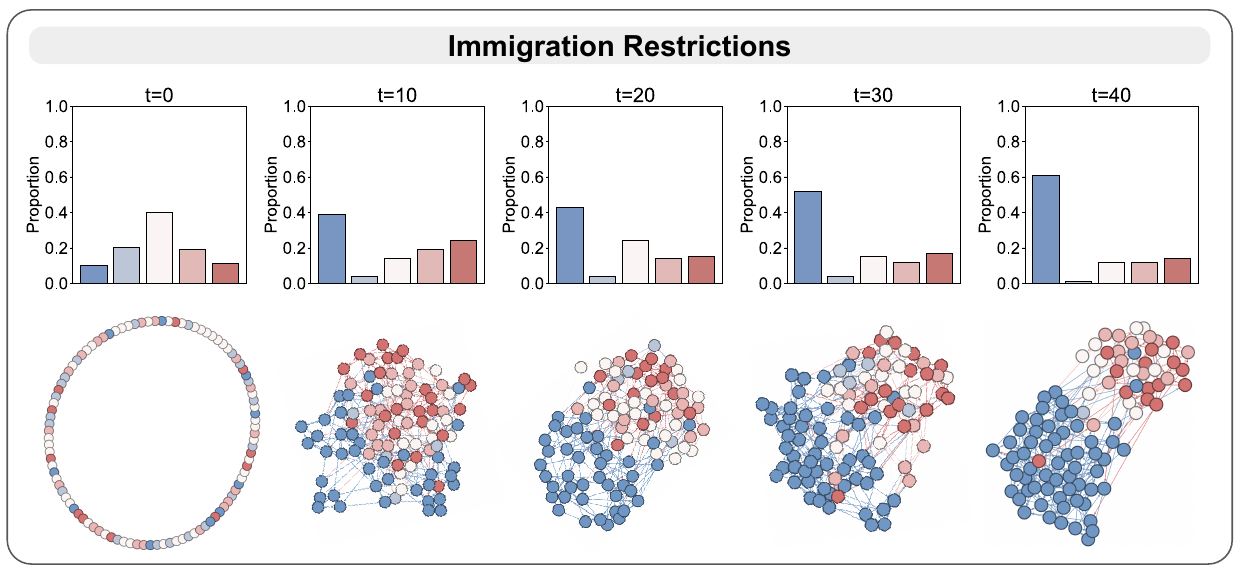}
\caption{\textbf{Evolution of LLM agents' collective opinions and network structures in the discussion of immigration.}}
\label{fig:immigration}
\end{figure}

\section{Supplementary Note 13: Results of Simulation on Different Issues}

To test the system's generalizability across different types of issues, we conduct two additional experiments beyond the political issues previously studied. The first relates to immigration restrictions, a socially relevant and critical issue. The second is the flat Earth theory, a non-political, fact-based issue with clear scientific consensus. These additions help demonstrate the generalizability of the proposed system and its corresponding results across both high-stakes political topics and more neutral or non-controversial domains \rvss{in the synthetic testbed}.

In particular, we illustrate the opinion dynamics when LLM agents discuss immigration restrictions in Figure~\ref{fig:immigration}. We observe that these agents spontaneously develop their collective opinions into a polarized pattern. This pattern emerges consistently across issues that have been shown to trigger polarization observed in human society. This consistency \rvss{suggests that the proposed system can serve as a synthetic LLM-agent testbed for exploring opinion-dynamics hypotheses across different issue domains.}

Furthermore, we wanted to understand how the proposed system would function on non-political, non-controversial issues. To this end, we test the system on a scientific issue – the flat Earth theory – where the topic is grounded in well-established scientific facts, rather than in polarized or contested opinions. As shown in Figure~\ref{fig:flat_earth}, we find that, unlike in political discussions, no polarization phenomenon emerges. Instead, agents rapidly reach a consensus within ten timesteps. These findings demonstrate that our system is not inherently predisposed to generate polarization. Rather, polarization emerges from free-form social interactions among LLM agents when dealing with inherently divisive topics, while it remains stable and convergent in neutral or fact-based domains.


\begin{figure}[h]
\centering
\includegraphics[width=\textwidth]{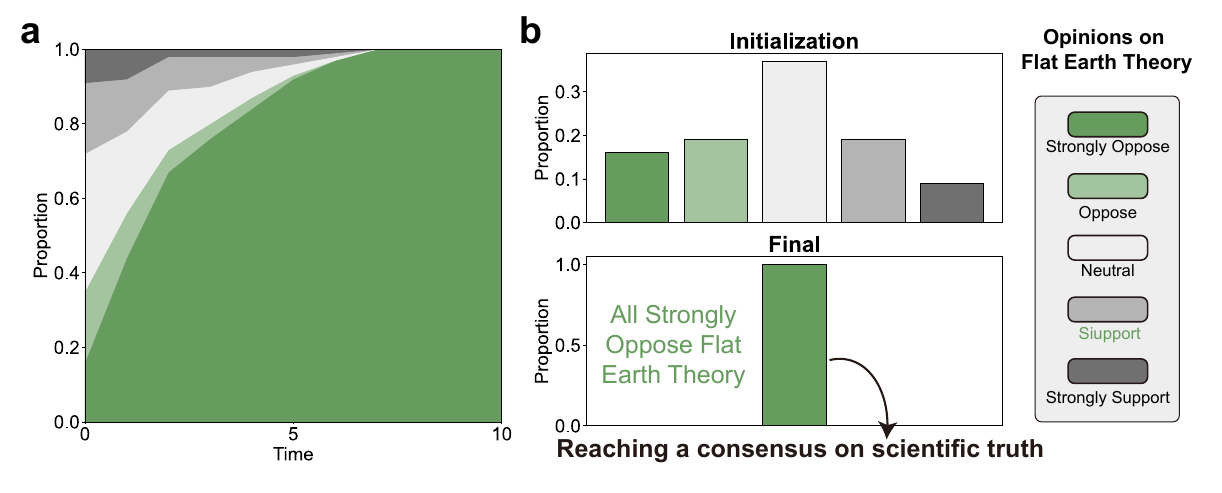}
\caption{\textbf{Opinion dynamics when LLM agents discuss about scientific truth.} After 10 timesteps, all agents are persuaded to strongly oppose the flat Earth theory, reaching consensus.}
\label{fig:flat_earth}
\end{figure}

\section{Supplementary Note 14: Supplementary Notes on Statistical Testing and Robustness Checks}

We assessed the assumptions of each reported frequentist test. For two-sample comparisons, normality was assessed using Shapiro--Wilk tests and equality of variances using Levene's tests; Student's $t$-tests were used when the equal-variance assumption was not violated, whereas Welch's $t$-tests were used otherwise. For tests involving large samples or sample proportions, the normal approximation of the relevant sampling distributions was justified by the central limit theorem. For proportion $z$-tests, formal normality and equal-variance tests were not applicable because the outcomes were binary proportions rather than continuous variables. For one-sample $t$-tests, equal-variance assumptions were not applicable. Where appropriate, we additionally conducted non-parametric robustness checks using Mann--Whitney $U$ tests or Wilcoxon signed-rank tests.

\clearpage
\printbibliography[title={\textrm{References}}]

\end{refsection}
\endgroup


\end{document}